\documentclass[10pt,onecolumn,oneside,a4paper,openany]{article}

\usepackage[dvips]{graphicx}
\usepackage{bbm}
\usepackage{amsfonts}
\usepackage{mathrsfs}
\usepackage{latexsym}

\newcommand{\lo}[1]{^{#1}\!}   
\def\ba{\begin{eqnarray}}
\def\ea{\end{eqnarray}}
\def\be{\begin{equation}}
\def\ee{\end{equation}}

\newtheorem{Theorem}{Theorem}[section]                             
\newtheorem{Definition}{Definition}[section]
\newtheorem{Lemma}{Lemma}[section]

\begin{document}

\vspace{-1cm}
\title{Simplification of the Spectral Analysis of\\ the Volume Operator 
in\\ 
Loop Quantum Gravity}
\date{}
\author{J. Brunnemann\thanks{ jbrunnemann@perimeterinstitute.ca},
T. Thiemann\thanks{tthiemann@perimeterinstitute.ca}\\
\\
Perimeter Institute for Theoretical Physics and University of Waterloo\\
Waterloo, Ontario, Canada}

\maketitle
\abstract{
The Volume Operator plays a crucial role in the definition of the quantum
dynamics of Loop Quantum Gravity (LQG). Efficient calculations for 
dynamical problems of LQG can  therefore be performed only if one has 
sufficient control over the volume
spectrum. While closed formulas for the matrix elements are currently
available in the literature, these are complicated polynomials
in 6j symbols which in turn are given in terms of Racah's formula which 
is too complicated in order to perform even numerical calculations
for the semiclassically important regime of 
large spins. Hence,
so far not even numerically the spectrum could be accessed.

In this article we demonstrate that by means of the Elliot -- Biedenharn
identity one can get rid of all the 6j symbols for any valence of the 
gauge invariant vertex, thus immensely reducing the computational effort.
We use the resulting compact formula to study numerically the spectrum
of the gauge invariant 4 -- vertex. 

The techniques derived in this paper could be of use also for the analysis
of spin -- spin interaction Hamiltonians of many -- particle problems 
in atomic and nuclear physics.
}

\newpage

\tableofcontents

\newpage

\section{Introduction}

The volume operator \cite{1,Geometry II}
plays a pivotal role in the definition of the quantum 
dynamics \cite{TT:QSD I,TT:QSD II,2} of Loop Quantum Gravity 
(LQG) \cite{0}. Since the success of LQG depends on whether the quantum 
dynamics reproduces classical General Relativity (GR) coupled to 
quantum matter in the semiclassical regime it is of outmost importance 
to know as much as possible about the spectrum of the volume operator.

The volume operator has been studied to some extent in the literature
\cite{3,dePietri,dePietriII,TT:vopelm} and it is well known that its 
spectrum is entirely discrete. However, so far only a closed 
formula for its matrix elements has been found. Unfortunately, not only
is the formula for the matrix elements a complicated polynomial in 6j
symbols involving extended sums over intertwiners, in addition the 6j 
symbols themselves are not easy to compute. 
Namely, the only known closed expression for the 6j symbols is Racah's
famous formula which in turn involves fractions of factorials of 
large numbers and sums whose range depends in a complicated way on the 
entries of the 6j symbol. Accordingly, even powerful computer programs 
such as {\it Mathematica} or {\it Maple} run very fast out of memory 
even for moderate values of the spin labels on the edges adjacent to 
vertex in question. For instance, the current authors were not able 
to go beyond $j=3$ when numerically computing the eigenvalues for a gauge 
invariant, four valent vertex, just using the matrix element formulas 
available in the literature. Thus, in order to make progress, analytical 
work is mandatory.

In this paper, which is based on the diploma thesis \cite{4}, we simplify 
the matrix element formula as given in 
\cite{TT:vopelm} tremendously: Using an identity due to Elliot and 
Biedenharn we are able to get rid of all the sums over intertwiners and 
all the 6j symbols in the final 
formula, no matter how large the valence of the vertex is.
The closed expression we obtain is a harmless polynomial of simple roots 
of fractional expressions in the spins and intertwiners, without 
factorials, that label the 
spin network functions in question. We reproduce the closed expression for 
the gauge invariant four -- vertex which has been discovered first by
de Pietri \cite{dePietri}.

This formula should be of interest for a wide range of applications. 
First of all, it opens access to the numerical analysis of dynamical 
questions in canonical LQG. In particular, there is now work in progress 
aiming at extending the spectacular results of \cite{4a} from the 
cosmological minisuperspace truncation to the full theory. Possible first
applications are alluded to in the conclusion section.
Next, the techniques presented here 
could be of use for numerical
investigations of convergence issues of spin foam models, see e.g. 
\cite{5} and references therein. Furthermore, our methods reveal that time 
has come to put LQG calculations on a supercomputer. Finally, it is 
conceivable that 
our formalism is of some use in the physics of many particle spin -- spin
interactions as e.g. in atomic or nuclear physics.\\
\\
The present paper is organized as follows:\\
\\
In section two we review the definition of the volume operator
as derived in \cite{Geometry II} and the closed expression 
for its matrix elements established in \cite{TT:vopelm}. Knowledge of
LQG \cite{1} is not at all necessary for the purpose of this paper which 
can be read also as a paper on the spectral analysis of a specific 
interaction Hamiltonian for a large spin system.

The main result of this paper is contained in the third section where we 
derive the simplification of the matrix elements. At the danger of boring
the reader we display all the intermediate steps. We do this because we 
feel that without these steps the proof,  
which in part is a complicated book keeping problem, cannot be be 
understood. The compact final formula is 
(\ref{Endgueltige Formel fuer das Matrixelement}).

In section four we use our formalism in order to study the gauge invariant 
four vertex. The simplification of the matrix element formula now enables
us to diagonalize the volume operator in a couple of hours for spin 
occupations of up to a $2j_{max}\approx 10^2$. More efficient programming 
and 
compiler -- based programming languages such as {\it Lisp} should be 
able to go significantly higher. Among the ``spectroscopy experiments''
we performed are the investigation of the computational effort, the 
possible existence of a 
volume gap (smallest non -- zero eigenvalue), the spectral density 
distribution and the relative number 
of degenerate (zero volume) configurations. Among the surprises we 
find numerical evidence for a universal density distribution in terms of 
properly rescaled quantities valid at large spin. Next, there is numerical
evidence for the existence of a volume gap at least for the four -- 
valent vertex. Finally, it seems that the 
eigenvalues form distinguishable series just like for the hydrogen atom
which provides a numerical criterion for the question which part of the 
spectrum remains unaffected when removing the finite size ``cut -- off'' 
$j_{max}$. 

In section five we summarize our results and in the appendices we provide
combinatorical and analytical background information which make the paper
hopefully self -- contained.

\newpage

\section{Revision of Known Results}

This section summarizes the definition of the volume operator of LQG and 
reviews the matrix element formula proved in \cite{TT:vopelm}. Readers 
not familiar with LQG can view the volume operator as a specific spin -- 
spin interaction Hamiltonian for a many particle system. After some 
introductory remarks for the benefit of the reader with an LQG 
background we will switch to a corresponding angular momentum 
description immediately which makes knowlegde of LQG unnecessary for the 
purposes of this paper.\\
\\
In LQG typical states are cylindrical functions $f_\gamma$ which are
labelled by
graphs $\gamma$. The graph itself can be thought of a collection 
$E(\gamma)$ of its oriented edges $e$ which intersect in their endpoints
which we call the vertices of $\gamma$. The set of vertices will be 
denoted by $V(\gamma)$. The cylindrical functions $f_\gamma$ depend 
on $SU(2)$ matrices $h_e$ which have the physical interpretation of
holonomies of an $SU(2)-$connection along the edges $e\in E(\gamma)$. 

In \cite{1,Geometry II} the operator describing the volume of a spatial 
region $R$, namely the  Volume operator
$\hat{V}(R)_{\gamma}$ acting on the cylindrical functions over a graph 
$\gamma$ was derived as:
\ba
  \hat{V}(R)_{\gamma}&=&\int\limits_R d^3p\widehat{\sqrt{det(q)(p)_{\gamma}}}
                     =\int\limits_R d^3p~ \hat{V}(p)_{\gamma} 
\ea
where
\ba
   \hat{V}(p)_{\gamma}&=&\ell_P^3 \sum_{v\in V(\gamma)}\delta^3(p,v)~\hat{V}_{v,\gamma}\\
  \label{erste} \hat{V}_{v,\gamma} &=&\sqrt{\Big| \frac{i}{3!\cdot 8}
                   \sum_{{e_I,e_J,e_K \in E(\gamma) \atop {\vspace{2mm}\atop e_I\cap e_J \cap e_K = v}}} 
		   \epsilon (e_I,e_J,e_K)~ q_{IJK}\Big|}\\
	       q_{IJK}&=&\epsilon_{ijk}X^i_IX^j_JX^k_K   
\ea
The sum has to be taken over all vertices $v\in V(\gamma)$ of the graph 
$\gamma$ and at each vertex $v$ over all possible triples $(e_I,e_J,e_K)$ 
of edges of the graph $\gamma$ adjacent to $v$. Here  
$\epsilon (e_I,e_J,e_K)$ is the sign of the cross product of the three 
tangent vectors of the edges $(e_I,e_J,e_K)$ at the vertex $v$ and we have 
assumed without loss of generality that all edges are outgoing from $v$.

The  $X^i_I$ are the right invariant vectorfields on $SU(2)$ 
acting on the holonomy entries of the cylindrical functions. They
satisfy the commutation relation
$[X^i_I,X^j_J]=-2~ \delta_{IJ}~\epsilon^{ijk}~X_I^k$. 
The self-adjoint right invariant vector fields $Y_J^j:=\frac{i}{2} X^j_J$ 
fulfilling $[Y^i_I,Y^j_J]=i~\delta_{IJ}~\epsilon^{ijk}~Y_I^k$ 
are equivalent to the action of angular momentum operators $J_I^i$. 
It is this algebraic property which we use in order to derive the spectral 
properties of the volume operator: It turns out that the Hilbert 
space of LQG reduces on cylindrical functions over a graph 
$\gamma$ to that of an abstract spin system familiar from the theory of 
angular momentum in quantum mechanics. There are as many degrees of 
freedom as there are edges in $\gamma$ and furthermore we can diagonalize 
all the $\hat{V}_{v,\gamma}$ simultaneously as they are obviously mutually
commuting. Hence, in what follows familiarity with LQG is not at all 
necessary, abstractly we are just dealing with an interaction Hamiltonian 
in a many particle spin system.

We can therefore replace:
\ba
       q_{IJK}&=&\left(\frac{2}{i} \right)^3\epsilon_{ijk}J^i_IJ^j_JJ^k_K   
\ea
Using furthermore the antisymmetry of $\epsilon_{ijk}$ and the fact that $[J_I^i,J_J^j]=0$ whenever $I\ne J$ we can restrict the summation in (\ref{erste}) to $I < J<K$ if we simultaneously write a factor $3!$ in front of the sum. The result is:
\ba
  \label{zweite} \hat{V}_{v,\gamma} &=&\sqrt{\Big| 
                   \sum_{I<J<K} 
		   \epsilon (e_I,e_J,e_K)~ \epsilon_{ijk}~J^i_IJ^j_JJ^k_K \Big|}\\
\ea

Now the following identity holds:
\ba
  \epsilon_{ijk}~J^i_I J^j_J J^k_K & = & \frac{i}{4}\big[(J_{IJ})^2,(J_{JK})^2 \big]                                      
\ea
where $J_{IJ}=J_I+J_J$. This relation can be derived by writing down every commutator as
$ \big[ (J_{IJ})^2, (J_{JK})^2\big]=\sum\limits_{i,j=1}^3 \big[(J_I^i+J_J^i)^2,(J_J^j+J_K^j)^2 \big]$, using the
identity $\big[a,bc\big]=\big[a,b\big]c +b\big[a,c \big]$ for the commutator,  using the angular
momentum commutation relations (\ref{CR J}) and the fact that $\big[J_I^i,J_J^j \big]=0$ whenever $I \ne J$.

We may summarize:

\ba
  \label{dritte} \hat{V}_{v,\gamma} &=&\sqrt{\big| Z \cdot\sum_{I<J<K} 
		   \epsilon (e_I,e_J,e_K)~\hat{q}_{IJK} \big|}\\
\ea
where $  \hat{q}_{IJK} :=\big[(J_{IJ})^2,(J_{JK})^2 \big]$ and 
$Z=
\frac{i}{4}$.\\

Unless announced differently we will study the operator 
\ba
  \hat{q}_{IJK} &:=&\big[(J_{IJ})^2,(J_{JK})^2 \big]
\ea
in the following.

\subsection{Matrix Elements in Terms of $3nj$-Symbols}

Now we can apply the recoupling theory of $n$ angular momenta to represent $\hat{q}_{IJK}$ in a recoupling
scheme basis using the definitions (\ref{Def Recoupling scheme}), (\ref{Def Standard basis}), 
(\ref{Def 3nj-symbol}), given in the appendix.

We will do this with respect to the standard basis (\ref{Def Standard basis}), where we can now easily
restrict our calculations to gauge invariant spin network states, by demanding the total angular momentum
$j$ and the total magnetic quantum number $M$ to vanish, that means we will take into account only
recoupling schemes, coupling the outgoing spins at the vertex $v$ to resulting angular momentum 0:
\\
In terms of the recoupling schemes these states are given by:
\be 
  |\vec{g}(IJ)~\vec{j}~j=0~M=0>:=|\vec{g}(IJ)>
\ee
where we introduced an abbreviation, since the quantum numbers $\vec{j}~j=0~M=0$ are the same for every
gauge invariant spin network state with respect to one vertex $v$. 

We will now represent $\hat{q}_{IJK}:=\big[(J_{IJ})^2,(J_{JK})^2 \big]$ in the standard recoupling 
scheme basis of definition (\ref{Def Standard basis}) where 
$|\vec{a}>:=|\vec{a}(12)>,|\vec{a}'>:=|\vec{a}'(12)>$.

The point is that by construction a recoupling scheme basis $|\vec{g}(IJ)>$ diagonalizes the operator
\linebreak $(G_2)^2=(J_{IJ})^2=(J_I+J_J)^2 $ that is 
\be
  (G_2)^2|\vec{g} (IJ)>=g_2(IJ)\big(g_2(IJ)+1 \big)|\vec{g}(IJ)>
\ee      
Furthermore every recoupling scheme $|\vec{g}(IJ)>$ can be expanded in terms of the standard basis via its
expansion coefficients, the $3nj$-symbols given by definition \ref{Def 3nj-symbol} in the appendix.
So it is possible to express
\ba \label{vorformel}
<\vec{a}|\hat{q}_{IJK}|\vec{a}'>
   & = &<\vec{a}|\big[(J_{IJ})^2,(J_{JK})^2 \big]|\vec{a}'> \nonumber\\
   & = &<\vec{a}|(J_{IJ})^2(J_{JK})^2]|\vec{a}'> - <\vec{a}|(J_{JK})^2(J_{IJ})^2|\vec{a}'>\nonumber\\
   & = &  \sum_{\vec{g}(IJ)}g_2(IJ)(g_2(IJ)+1)
   [<\vec{a}|\vec{g}(IJ)><\vec{g}_{IJ}|J_{JK}^2|\vec{a}'>
   -<\vec{a}|J_{JK}^2|\vec{g}_{IJ}><\vec{g}(IJ)|\vec{a}'>]
   \nonumber\\
   &=& 
   \sum_{\vec{g}(IJ),\vec{g}(JK),\vec{g}''(12)}
   g_2(IJ)(g_2(IJ)+1)g_2(JK)(g_2(JK)+1)
   <\vec{g}(IJ)|\vec{g}''><\vec{g}(JK)|\vec{g}''>\times\nonumber\\
   &&\times
   [<\vec{g}(IJ)|\vec{a}><\vec{g}(JK)|\vec{a}'>
   -<\vec{a}(JK)|\vec{g}><\vec{g}(IJ)|\vec{a}'>]\nonumber\\
   & = & 
   \sum_{\vec{g}''(12)} \left[ \sum_{\vec{g}(IJ)} \right. 
   g_2(IJ)(g_2(IJ)+1)<\vec{g}(IJ)|\vec{g}''><\vec{g}(IJ)|\vec{a}>\times\nonumber\\
   &&\times \left. \sum_{\vec{g}(JK)}g_2(JK)(g_2(JK)+1)<\vec{g}(JK)|\vec{g}''><\vec{g}(JK)|\vec{a}'>
   \right] \nonumber\\
   && - \left[ \vec{a} \Longleftrightarrow \vec{a}' \right]
\ea
which is again an antisymmetric matrix possessing purely imaginary eigenvalues (we could alternatively
consider the hermitian version by multiplying all matrix elements by the imaginary unit $i$).  
Here we have inserted suitable recoupling schemes $|\vec{g}(IJ)>,|\vec{g}(JK)>$ diagonalizing 
$(J_{IJ})^2$ and $(J_{JK})^2$ and their expansion in terms of the standard basis $|\vec{g}(12)>$
by using the completeness of the recoupling schemes $|\vec{g}(IJ)>$ for arbitrary $I \ne J $ \footnote{The
summation has to be extended over all possible intermediate recoupling steps $g_2,\ldots,g_{n-1}$
that is $|j_r-j_q|\le g_k(j_q,j_r) \le j_q+j_r$ allowed by theorem \ref{Clebsh Gordan theorem}, given in
the appendix.}:
\be
  \mathbbm{1}=\sum_{\vec{g}(IJ)}|\vec{g}(IJ)><\vec{g}(IJ)| 
\ee

So we have as a first step expressed the matrix elements of $\hat{q}_{IJK}$ in terms of $3nj$-symbols.

\subsection{Closed Expression for the $3nj$-Symbols}
The $3nj$-symbols occurring in (\ref{vorformel}) can be expressed in terms of the individual recouplings implicit in their definition.

\subsubsection{Preparations}

   In \cite{TT:vopelm} the two following lemmas are derived:
  
  \begin{Lemma}{Contraction on Identical Coupling Order} \label{lemma51}\\
      $<\vec{g}(IJ)|\vec{g}'>$\\
      $=<g_2(j_I,j_J),g_3(g_2,j_1),..,g_{I+1}(g_I,j_{I-1}),
      g_{I+2}(g_{I+1},j_{I+1}),..,g_J(g_{J-1},j_{J-1})|$\\
      $~~~~|g''_2(j_1,j_2),g''_3(g''_2,j_3),..,g''_{I+1}(g''_I,j_{I-1}),
      g''_{I+2}(g''_{I+1},j_{I+1}),...,g''_J(g''_{J-1},j_J)>
      \delta_{g_J,g''_J} ... \delta_{g_{n-1},g''_{n-1}}$ 
   \end{Lemma}
    
   \begin{Lemma}{Interchange of Coupling Order} \label{lemma52}\\
      $<g_2'(j_1,j_2),..,g_K'(g_{K-1}',j_K),g_{K+1}'(g_K',j_{K+1}),g_{K+2}'(g_{K+1}',j_{K+2})
      |$\\
      $~~~|g_2(j_1,j_2),..,g_K(g_{K-1},j_K),g_{K+1}(g_K,j_{K+2}),
      g_{K+2}(g_{K+1},j_{K+1})>$\\
      $=<g_{K+1}'(g_K,j_{K+1}),g_{K+2}'(g_{K+1}',j_{K+2})|
      g_{K+1}(g_K,j_{K+2}),g_{K+2}(g_{K+1},j_{K+1})> 
      \delta_{g_2'g_2}\delta_{g_3'g_3}...\delta_{g_K'g_K}\cdot \delta_{g_{K+2}' g_{K+2}}$.
   \end{Lemma}

\subsubsection{Closed Expresion for the $3nj$-Symbols}
   \begin{samepage}
  Now we can reduce out the $3nj$-symbol. In what follows we will not 
explicitly write down the
  $\delta$-expressions occurring by using lemma \ref{lemma51} and lemma \ref{lemma52}, 
  but keep them in mind.\\
  Collecting all the terms mentioned in \cite{TT:vopelm} one obtains the following equation
  for the $3nj$-symbols: \\[5cm]

    \renewcommand{\arraystretch}{2}
   \[ \begin{array}{ccccl} 
      \lefteqn{<\vec{g}(IJ)|\vec{g}'(12)>~=} & & & \nonumber\\
        &= & & \displaystyle\sum\limits_{h_2(j_I,j_1)}
	&<g_2(j_I,j_J),g_3(g_2,j_1)|h_2(j_I,j_1),g_3(h_2,j_J)>\\
       & &\times&\displaystyle\sum\limits_{h_3(h_2,j_2)}
       &<g_3(h_2,j_J),g_4(g_3,j_2)|h_3(h_2,j_2),g_4(h_3,j_J)>\\ 
       & &\vdots &\vdots &~~~~~~~~~~~~~~~~~~~~~~~~~\vdots\\ 
       & &\times&\displaystyle\sum\limits_{h_{I-1}(h_{I-2},j_{I-2})}
       &<g_{I-1}(h_{I-2},j_J),g_I(g_{I-1},j_{I-2})|h_{I-1}(h_{I-2},j_{I_2}),g_I(h_{I-1},j_J)>\\ 
       & & &\times 
       &<g_I(h_{I-1},j_J),g_{I+1}(g_I,j_{I-1})|g'_I(h_{I-1},j_{I-1}),g_{I+1}(g'_I,j_J)>\\          
       & & &\times 
       &<g_{I+1}(g'_I,j_J),g_{I+2}(g_{I+1},j_{I+1})|g'_{I+1}(g'_I,j_{I+1}),g_{I+2}(g'_{I+1},j_J)>\\
       & & &\times 
       &<g_{I+2}(g'_{I+1},j_J),g_{I+3}(g_{I+2},j_{I+2})|g'_{I+2}(g'_{I+1},j_{I+2}),g_{I+3}(g'_{I+2},j_J)>\\                 
       & & &\vdots &~~~~~~~~~~~~~~~~~~~~~~~~~\vdots\\
       & & &\times 
       &<g_{J-1}(g'_{J-2},j_J),g_J(g_{J-1},j_{J-1})|g'_{J-1}(g'_{J-2},j_{J-1}),g_J(g'_{J-1},j_J)>\\                 
       & & &\times 
       &<h_2(j_I,j_1),h_3(h_2,j_2)|g'_2(j_1,j_2),h_3(g'_2,j_I)>\\                
       & & &\times 
       &<h_3(g'_2,j_I),h_4(h_3,j_3)|g'_3(g'_2,j_3),h_4(g'_3,j_I)>\\                
       & & &\times 
       &<h_4(g'_3,j_I),h_5(h_4,j_4)|g'_4(g'_3,j_4),h_5(g'_4,j_I)>\\                
       & & &\vdots &~~~~~~~~~~~~~~~~~~~~~~~~~\vdots\\
       & & &\times 
       &<h_{I-1}(g'_{I-2},j_I),g'_I(h_{I-1},j_{I-1})|g'_{I-1}(g'_{I-2},j_{I-1}),g'_I(g'_{I-1},j_I)>                      
    \end{array} \] \begin{equation}\label{hauptformel}\end{equation}
    \renewcommand{\arraystretch}{1}   
  \end{samepage}

\section{Simplification of the Matrix Elements}

\subsection{$3nj$-Symbols Expressed in Terms of $6j$-Symbols}    

    It is now obvious that we can express (\ref{hauptformel}) via  the $6j$-symbols defined as in
    (\ref{definition 6j symbol})~:

    \ba 
      \lefteqn{<j_{12}(j_1,j_2),j(j_{12},j_3)|j_{23}(j_2,j_3),j(j_1,j_{23})> =}\nonumber\\
      &~~~~~~~~~~~~~~~~~~~~~~~~~~~~ = &
      [(2j_{12}+1)(2j_{23}+1)]^{\frac{1}{2}}(-1)^{j_1+j_2+j_3+j}
      \left\{ \begin{array}{ccc}
      j_1 & j_2 & j_{12}\\
      j_3 & j   & j_{23}
      \end{array} \right\} 
    \ea
    
    Here we have used the fact that $j_1+j_2+j_3+j$ is integer. 
    Now the definition of the $6j$-symbol in terms of Clebsh-Gordon-coefficients (CGC) come into play.
    Because of the properties of the CGC we can change the order of coupling in every recoupling scheme in
    (\ref{hauptformel}) taking care of the minus signs we create:
     \begin{eqnarray*}
       \lefteqn{<j_{12}(j_1,j_2),j(j_{12},j_3)|j_{23}(j_2,j_3),j(j_1,j_{23})>=}\nonumber\\
       &=&(-1)^{j_{12}-j_1-j_2}<j_{12}(j_2,j_1),j(j_{12},j_3)|j_{23}(j_2,j_3),j(j_1,j_{23})>\nonumber\\
       &=&(-1)^{j_{12}-j_1-j_2}(-1)^{j_{23}-j_2-j_3}<j_{12}(j_2,j_1),j(j_{12},j_3)|j_{23}(j_3,j_2),j(j_1,j_{23})>\nonumber\\       
       &=&(-1)^{j_{12}-j_1-j_2}(-1)^{j_{23}-j_2-j_3}(-1)^{j-j_{12}-j_3}
          <j_{12}(j_2,j_1),j(j_3,j_{12})|j_{23}(j_3,j_2),j(j_1,j_{23})>\nonumber\\
       &=&(-1)^{j_{12}-j_1-j_2}(-1)^{j_{23}-j_2-j_3}(-1)^{j-j_{12}-j_3}(-1)^{j-j_1-j_{23}}
          <j_{12}(j_2,j_1),j(j_3,j_{12})|j_{23}(j_3,j_2),j(j_{23},j_1)>	        
     \end{eqnarray*}        
     In this way we are able to change the coupling-order in (\ref{hauptformel}) to get the order 
     required for a translation into the $6j$-symbols. 
     With these preparations we are now able to express (\ref{hauptformel}) in terms of $6j$-symbols:

    \begin{samepage}
    \begin{footnotesize} 
      \[ \begin{array}{ccccr} 
	 \lefteqn{<\vec{g}(IJ)|\vec{g}'(12)>~=} & & & \nonumber\\
           &= & & \displaystyle\sum\limits_{h_2}
	   &(-1)^{-j_I-j_J+g_2}(-1)^{h_2+j_J-g_3}(-1)^{j_I+j_J+j_1+g_3}\sqrt{(2g_2+1)(2h_2+1)}
	   \left\{ \begin{array}{ccc}
                    j_J & j_I & g_2\\
                    j_1 & g_3 & h_2
           \end{array} \right\} \\
	  \\& &\times&\displaystyle\sum\limits_{h_3}
	   &(-1)^{-j_J-h_2+g_3}(-1)^{h_3+j_J-g_4}(-1)^{j_J+h_2+j_2+g_4}\sqrt{(2g_3+1)(2h_3+1)}
	   \left\{ \begin{array}{ccc}
                    j_J & h_2 & g_3\\
                    j_2 & g_4 & h_3
           \end{array} \right\} \\

	  & &\vdots &\vdots &\vdots\hspace{7cm}\vdots~~~~~~~~~~~~\\ 
	  & &\vdots &\vdots &\vdots\hspace{7cm}\vdots~~~~~~~~~~~~\\ 

	  & &\times&\displaystyle\sum\limits_{h_{I-1}}
	  &(-1)^{-j_J-h_{I-2}+g_{I-1}}(-1)^{h_{I-1}+j_J-g_I}(-1)^{j_J+h_{I-2}+j_{I-2}+g_I}\sqrt{(2g_{I-1}+1)(2h_{I-1}+1)}
	   \left\{ \begin{array}{ccc} 
                    j_J & h_{I-2} & g_{I-1}\\
                    j_{I-2} & g_I & h_{I-1}
           \end{array} \right\} \\
	  \\& & &\times 
	  &(-1)^{-j_J-h_{I-1}+g_I}(-1)^{j_J+g'_I-g_{I+1}}(-1)^{j_J+h_{I-1}+j_{I-1}+g_{I+1}}\sqrt{(2g_I+1)(2g'_I+1)}
	   \left\{ \begin{array}{ccc} 
                    j_J & h_{I-1} & g_I\\
                    j_{I-1} & g_{I+1} & g'_I
           \end{array} \right\} \\
	  \\& & &\times 
	  &(-1)^{-j_J-g'_I+g_{I+1}}(-1)^{j_J+g'_{I+1}-g_{I+2}}(-1)^{j_J+g'_I+j_{I+1}+g_{I+2}}\sqrt{(2g_{I+1}+1)(2g'_{I+1}+1)}
	   \left\{ \begin{array}{ccc} 
                    j_J & g'_I & g_{I+1}\\
                    j_{I+1} & g_{I+2} & g'_{I+1}
           \end{array} \right\} \\
	  \\& & &\times 
	  &(-1)^{-j_J-g'_{I+1}+g_{I+2}}(-1)^{j_J+g'_{I+2}-g_{I+3}}(-1)^{j_J+g'_{I+1}+j_{I+2}+g_{I+3}}\sqrt{(2g_{I+2}+1)(2g'_{I+2}+1)}
	   \left\{ \begin{array}{ccc} 
                    j_J & g'_{I+1} & g_{I+2}\\
                    j_{I+2} & g_{I+3} & g'_{I+2}
           \end{array} \right\} \\

	  & & &\vdots &\vdots\hspace{7cm}\vdots~~~~~~~~~~~~\\
	  & & &\vdots &\vdots\hspace{7cm}\vdots~~~~~~~~~~~~\\

	  & & &\times 
	  &(-1)^{-j_J-g'_{J-2}+g_{J-1}}(-1)^{j_J+g'_{J-1}-g_J}(-1)^{j_J+g'_{J-2}+j_{J-1}+g_J}\sqrt{(2g_{J-1}+1)(2g'_{J-1}+1)}
	   \left\{ \begin{array}{ccc} 
                    j_J & g'_{J-2} & g_{J-1}\\
                    j_{J-1} & g_J & g'_{J-1}
           \end{array} \right\} \\
	  \\& & &\times 
	  &(-1)^{j_I+g'_2-h_3}(-1)^{j_I+j_1+j_2+h_3}\sqrt{(2h_2+1)(2g'_2+1)}
	   \left\{ \begin{array}{ccc} 
                    j_I & j_1 & h_2\\
                    j_2 & h_3 & g'_2
           \end{array} \right\} \\
	  \\& & &\times 
	  &(-1)^{-g'_2-j_I+h_3}(-1)^{j_I+g'_3-h_4}(-1)^{j_I+g'_2+j_3+h_4}\sqrt{(2h_3+1)(2g'_3+1)}
	   \left\{ \begin{array}{ccc} 
                    j_I & g'_2 & h_3\\
                    j_3 & h_4 & g'_3
           \end{array} \right\} \\
	  \\& & &\times 
	  &(-1)^{-g'_3-j_I+h_4}(-1)^{j_I+g'_4-h_5}(-1)^{j_I+g'_3+j_4+h_5}\sqrt{(2h_4+1)(2g'_4+1)}
	   \left\{ \begin{array}{ccc} 
                    j_I & g'_3 & h_4\\
                    j_4 & h_5 & g'_4
           \end{array} \right\} \\

	  & & &\vdots &\vdots\hspace{7cm}\vdots~~~~~~~~~~~~\\
	  & & &\vdots &\vdots\hspace{7cm}\vdots~~~~~~~~~~~~\\

	  & & &\times 
	  &(-1)^{-g'_{I-2}-j_I+h_{I-1}}(-1)^{j_I+g'_{I-1}-g_I}(-1)^{j_I+g'_{I-2}+j_{I-1}+g'_I}\sqrt{(2h_{I-1}+1)(2g'_{I-1}+1)}
	   \left\{ \begin{array}{ccc} 
                    j_I & g'_{I-2} & h_{I-1}\\
                    j_{I-1} & g'_I & g'_{I-1}
           \end{array} \right\} \\
       \end{array} \]  
       \end{footnotesize}
       \begin{equation}\label{6j-hauptformel}\end{equation} 
       This is the complete expression of (\ref{hauptformel}) with all the exponents written in detail which 
       are caused by the reordering of the coupling-schemes while bringing them into a form suiteable for
       (\ref{definition 6j symbol}).   
       We want to emphasize that we have the freedom to invert the signs in each of the 
       exponents of
       (\ref{6j-hauptformel}) when convenient for our calculation.
    \end{samepage}

\subsection{The Matrix Elements in Terms of $6j$-Symbols}

    Taking a closer look at (\ref{vorformel}), a basic structure contained in the matrix elements of 
    the volume operator appears:
    \be \label{basic structure}
      \sum_{\vec{g}(IJ)}g_2(IJ)\bigm(g_2(IJ)+1\bigm) <\vec{g}(IJ)|\vec{g}''(12)><\vec{g}(IJ)|\vec{a}(12)>
    \ee          
    Using (\ref{6j-hauptformel})   we now express the $3nj$-symbols occurring in (\ref{basic structure})
    via $6j$-symbols. For
    $<\vec{g}(IJ)|\vec{g}''(12)>$ we use $h_2 \ldots h_{I-1}$ as intermediate summation variables and for
    its $(-1)$~-exponents the sign convention we chose in (\ref{6j-hauptformel}).
    For $<\vec{g}(IJ)|\vec{a}(12)>$ we use $k_2 \ldots k_{I-1}$ as intermediate summation variables 
    and for its $(-1)$~-exponents the negative of every exponent in (\ref{6j-hauptformel}), since every
    exponent is an integer number.
    Writing down carefully all these expressions most of the exponents can be cancelled.    
   \\[15cm] 
   \\
    \begin{samepage}
    The result of this is (using the abbreviation $A(x,y)=\sqrt{(2x+1)(2y+1)}$~):
    
    \begin{footnotesize} 
      \[ \begin{array}{ccccrcr} 
	 \lefteqn{\sum_{\vec{g}(IJ)}g_2(IJ)\bigm(g_2(IJ)+1\bigm) <\vec{g}(IJ)|\vec{g}''(12)>
	               <\vec{g}(IJ)|\vec{a}(12)>~=} 
	 \nonumber\\
	 \nonumber\\
	 \nonumber\\

           \lefteqn{=\sum_{\vec{g}(IJ)}g_2(IJ)\bigm(g_2(IJ)+1\bigm)~\times}
	   \\
	   \\

           & & \times& \displaystyle\sum\limits_{h_2}
	   &(-1)^{h_2} A(g_2,h_2)
	   \left\{ \begin{array}{ccc}
                    j_J & j_I & g_2\\
                    j_1 & g_3 & h_2
           \end{array} \right\} 
	   &\displaystyle\sum\limits_{k_2}
	   &(-1)^{-k_2} A(g_2,k_2)
	   \left\{ \begin{array}{ccc}
                    j_J & j_I & g_2\\
                    j_1 & g_3 & k_2
           \end{array} \right\} \\

	  \\& &\times&\displaystyle\sum\limits_{h_3}
	   &(-1)^{h_3} A(g_3,h_3)
	   \left\{ \begin{array}{ccc}
                    j_J & h_2 & g_3\\
                    j_2 & g_4 & h_3
           \end{array} \right\} 
           &\displaystyle\sum\limits_{k_3}
	   &(-1)^{-k_3} A(g_3,k_3)
	   \left\{ \begin{array}{ccc}
                    j_J & k_2 & g_3\\
                    j_2 & g_4 & k_3
           \end{array} \right\}\\

	  & &\vdots &\vdots &\vdots~~~~~~~~~~~~&\vdots &\vdots~~~~~~~~~~~~\\ 
	  & &\vdots &\vdots &\vdots~~~~~~~~~~~~&\vdots &\vdots~~~~~~~~~~~~\\

	  & &\times&\displaystyle\sum\limits_{h_{I-1}}
	  &(-1)^{h_{I-1}} A(g_{I-1},h_{I-1})
	   \left\{ \begin{array}{ccc} 
                    j_J & h_{I-2} & g_{I-1}\\
                    j_{I-2} & g_I & h_{I-1}
           \end{array} \right\} 
           &\displaystyle\sum\limits_{k_{I-1}}	   
	   &(-1)^{-k_{I-1}} A(g_{I-1},k_{I-1})
	   \left\{ \begin{array}{ccc} 
                    j_J & k_{I-2} & g_{I-1}\\
                    j_{I-2} & g_I & k_{I-1}
           \end{array} \right\} \\

	  \\& & &\times 
	  &(-1)^{g''_I} A(g_I,g''_I)
	   \left\{ \begin{array}{ccc} 
                    j_J & h_{I-1} & g_I\\
                    j_{I-1} & g_{I+1} & g''_I
           \end{array} \right\} 
	  & &(-1)^{-a_I} A(g_I,a_I)
	   \left\{ \begin{array}{ccc} 
                    j_J & k_{I-1} & g_I\\
                    j_{I-1} & g_{I+1} & a_I
           \end{array} \right\} \\

	  \\& & &\times 
	  &(-1)^{g''_{I+1}} A(g_{I+1},g''_{I+1})
	   \left\{ \begin{array}{ccc} 
                    j_J & g''_I & g_{I+1}\\
                    j_{I+1} & g_{I+2} & g''_{I+1}
           \end{array} \right\} 
	  & &(-1)^{-a_{I+1}} A(g_{I+1},a_{I+1})
	   \left\{ \begin{array}{ccc} 
                    j_J & a_I & g_{I+1}\\
                    j_{I+1} & g_{I+2} & a_{I+1}
           \end{array} \right\} \\

	  \\& & &\times 
	  &(-1)^{g''_{I+2}} A(g_{I+2},g''_{I+2})
	   \left\{ \begin{array}{ccc} 
                    j_J & g''_{I+1} & g_{I+2}\\
                    j_{I+2} & g_{I+3} & g''_{I+2}
           \end{array} \right\} 
	  & & (-1)^{-a_{I+2}} A(g_{I+2},a_{I+2})
	   \left\{ \begin{array}{ccc} 
                    j_J & a_{I+1} & g_{I+2}\\
                    j_{I+2} & g_{I+3} & a_{I+2}
           \end{array} \right\} \\

	  & & &\vdots &\vdots~~~~~~~~~~~~& &\vdots~~~~~~~~~~~~\\
	  & & &\vdots &\vdots~~~~~~~~~~~~& &\vdots~~~~~~~~~~~~\\

	  & & &\times 
	  &(-1)^{g''_{J-1}} A(g_{J-1},g''_{J-1})
	   \left\{ \begin{array}{ccc} 
                    j_J & g''_{J-2} & g_{J-1}\\
                    j_{J-1} & g_J & g''_{J-1}
           \end{array} \right\} 
	  & &(-1)^{-a_{J-1}} A(g_{J-1},a_{J-1})
	   \left\{ \begin{array}{ccc} 
                    j_J & a_{J-2} & g_{J-1}\\
                    j_{J-1} & g_J & a_{J-1}
           \end{array} \right\} \\

	  \\& & &\times 
	  &(-1)^{g''_2} A(h_2,g''_2)
	   \left\{ \begin{array}{ccc} 
                    j_I & j_1 & h_2\\
                    j_2 & h_3 & g''_2
           \end{array} \right\} 
	  & &(-1)^{-a_2} A(k_2,a_2)
	   \left\{ \begin{array}{ccc} 
                    j_I & j_1 & k_2\\
                    j_2 & k_3 & a_2
           \end{array} \right\} \\

	  \\& & &\times 
	  &(-1)^{h_3+g''_3} A(h_3,g''_3)
	   \left\{ \begin{array}{ccc} 
                    j_I & g''_2 & h_3\\
                    j_3 & h_4 & g''_3
           \end{array} \right\}
	  & &(-1)^{-k_3-a_3} A(k_3,a_3)
	   \left\{ \begin{array}{ccc} 
                    j_I & a_2 & k_3\\
                    j_3 & k_4 & a_3
           \end{array} \right\} \\

	  \\& & &\times 
	  &(-1)^{h_4+g''_4} A(h_4,g''_4)
	   \left\{ \begin{array}{ccc} 
                    j_I & g''_3 & h_4\\
                    j_4 & h_5 & g''_4
           \end{array} \right\} 
	  & &(-1)^{-k_4-a_4} A(k_4,a_4)
	   \left\{ \begin{array}{ccc} 
                    j_I & a_3 & k_4\\
                    j_4 & k_5 & a_4
           \end{array} \right\} \\

	  & & &\vdots &\vdots~~~~~~~~~~~~& &\vdots~~~~~~~~~~~~\\
	  & & &\vdots &\vdots~~~~~~~~~~~~& &\vdots~~~~~~~~~~~~\\

	  & & &\times 
	  &(-1)^{h_{I-1}+g''_{I-1}} A(h_{I-1},2g'_{I-1})
	   \left\{ \begin{array}{ccc} 
                    j_I & g''_{I-2} & h_{I-1}\\
                    j_{I-1} & g''_I & g'_{I-1}
           \end{array} \right\}
	  & &(-1)^{-k_{I-1}-a_{I-1}} A(k_{I-1},a_{I-1})
	   \left\{ \begin{array}{ccc} 
                    j_I & a_{I-2} & k_{I-1}\\
                    j_{I-1} & a_I & a_{I-1}
           \end{array} \right\} \\
       \end{array} \]  
    \end{footnotesize} 
    \\
    \be \label{basic structure in 6j symbols}\ee
    \end{samepage}
    Up to now we have only made a translation between different notations. The reason for doing such 
    an amount
    of writing will become clear soon: Using identities between the $6j$-symbols it is possible to derive a
    much shorter closed expression for the matrix elements by evaluating step by step all the summations in
    (\ref{basic structure in 6j symbols})~.

\subsection{A Useful Identity} 
   
   Before we can start the evaluation we want to derive an identity, which will be essential.
   We want to evaluate the following sum:
   \be \label{F1}
     F(j_{12},j'_{12}) := 
     \sum_{j_{23}}(2j_{23}+1)~j_{23}(j_{23}+1)~
     	   \left\{ \begin{array}{ccc} 
                    j_1 & j_2 & j_{12}\\
                    j_3 & j_4 & j_{23}
           \end{array} \right\} 
	   \left\{ \begin{array}{ccc} 
                    j_1 & j_2 & j'_{12}\\
                    j_3 & j_4 & j_{23}
           \end{array} \right\} 
   \ee   
   There exist numerous closed expressions for $6j$-symbols, whose entries have special relations. They are
   much more manageable than the general expression given in the appendix (\ref{A1}) (Racah formula). One of them reads as
   \cite{Edmonds}, p.130:
   \be
     \left\{ \begin{array}{ccc} 
              a & b & c\\
              1 & c & b
     \end{array} \right\}
     = (-1)^{a+b+c+1} \frac{2[b(b+1)+c(c+1)-a(a+1)]}
                           {[2b(2b+1)(2b+2)2c(2c+1)(2c+2)]^{\frac{1}{2}}}                           
   \ee
   Using the shorthand $X(b,c)=2b(2b+1)(2b+2)2c(2c+1)(2c+2)$ and the fact that $a+b+c$ is integer, 
   we can rewrite the equation to obtain
   \be \label{F2}
     a(a+1)= (-1)^{a+b+c}
     \left\{ \begin{array}{ccc} 
              a & b & c\\
              1 & c & b
     \end{array} \right\} X(b,c)^{\frac{1}{2}}+\big[(b+1)+c(c+1) \big]                     
   \ee 
   Putting $a=j_{23}$ and inserting (\ref{F2}) into (\ref{F1}) one finds for $F(j_{12},j'_{12})$ for any
   $b,c~$:
   
   \ba \label{F3}
     F(j_{12},j'_{12}) 
     & = & \frac{1}{2}(-1)^{b+c}X(b,c)^{\frac{1}{2}}
           \overbrace{
           \sum_{j_{23}}(-1)^{j_{23}}~(2j_{23}+1)  	     
	     \left\{ \begin{array}{ccc} 
                      j_{23} & b & c \\
                           1 & c & b
             \end{array} \right\} 
	     \left\{ \begin{array}{ccc} 
                      j_1 & j_2 & j_{12}\\
                      j_3 & j_4 & j_{23}
             \end{array} \right\} 
	     \left\{ \begin{array}{ccc} 
                      j_1 & j_2 & j'_{12}\\
                      j_3 & j_4 & j_{23}
             \end{array} \right\} 
	     }^{I}\nonumber\\    
     & & +\frac{b(b+1)+c(c+1)}{(2j_{12}+1)} \underbrace{\sum_{j_{23}}(2j_{12}+1)(2j_{23}+1)
     	     \left\{ \begin{array}{ccc} 
                      j_1 & j_2 & j_{12}\\
                      j_3 & j_4 & j_{23}
             \end{array} \right\} 
	     \left\{ \begin{array}{ccc} 
                      j_1 & j_2 & j'_{12}\\
                      j_3 & j_4 & j_{23}
             \end{array} \right\} 
	     }_{\displaystyle{\delta_{j_{12}j'_{12}}}}    
   \ea
   where we have used the orthogonality relation (\ref{orthogonality relation}) for the $6j$-symbols.
   Let us take a closer look at the three $6j$-symbols of $I$ on the right hand side of (\ref{F3}).
   We now apply some permutations to the rows and columns of the first $6j$-symbol within $I$ 
   which leave this $6j$-symbol invariant, see 
   (\ref{symmetry1}), (\ref{symmetry2}). 
   After that we have for $I$:
   \ba
     I & = & 
      \sum_{j_{23}}(-1)^{j_{23}}~(2j_{23}+1)  	     
	 \left\{ \begin{array}{ccc} 
                  b & b & 1 \\
                  c & c & j_{23}
         \end{array} \right\} 
	 \left\{ \begin{array}{ccc} 
                  j_1 & j_2 & j_{12}\\
                  j_3 & j_4 & j_{23}
         \end{array} \right\} 
	 \left\{ \begin{array}{ccc} 
                  j_1 & j_2 & j'_{12}\\
                  j_3 & j_4 & j_{23}
         \end{array} \right\} \nonumber  
    \ea
    Now we fix $b=j_1$, $c=j_4$ and can evaluate:  
    \ba
      I 
      & = & \sum_{j_{23}}(-1)^{j_{23}}~(2j_{23}+1)  	     
	     \left\{ \begin{array}{ccc} 
                      j_1 & j_1 & 1 \\
                      j_4 & j_4 & j_{23}
             \end{array} \right\} 
	     \left\{ \begin{array}{ccc} 
                      j_1 & j_2 & j_{12}\\
                      j_3 & j_4 & j_{23}
             \end{array} \right\} 
	     \left\{ \begin{array}{ccc} 
                      j_1 & j_2 & j'_{12}\\
                      j_3 & j_4 & j_{23}
             \end{array} \right\} \nonumber \\  
       & = & (-1)^{-(j_{12}+j_1+j_1+j_3+j_2+1+j_4+j'_{12}+j_4)}
	     \left\{ \begin{array}{ccc} 
                      j_{12} & j_1 & j_2 \\
                      j_1 & j'_{12} & 1
             \end{array} \right\} 
	     \left\{ \begin{array}{ccc} 
                      1 & j_{12} & j'_{12}\\
                      j_3 & j_4 & j_4
             \end{array} \right\} \nonumber
    \ea       
    Here we have used the Elliot-Biedenharn-identity (\ref{EBI}) .
    Inserting this back into (\ref{F3}) yields: 
    \ba \label{F4}
     F(j_{12},j'_{12}) 
     & = & \frac{1}{2}(-1)^{j_1+j_4}(-1)^{-(j_{12}+j_1+j_1+j_3+j_2+1+j_4+j'_{12}+j_4)}
           X(j_1,j_4)^{\frac{1}{2}}
	     \left\{ \begin{array}{ccc} 
                      j_{12} & j_1 & j_2 \\
                      j_1 & j'_{12} & 1
             \end{array} \right\} 
	     \left\{ \begin{array}{ccc} 
                      1 & j_{12} & j'_{12}\\
                      j_3 & j_4 & j_4
             \end{array} \right\} 
	     \nonumber\\    
     &   & +\frac{j_1(j_1+1)+j_4(j_4+1)}{(2j_{12}+1)}\displaystyle{\delta_{j_{12}j'_{12}}} \nonumber
     \ea
      \\
     Using that, by definition of the $6j$-symbols (\ref{definition 6j symbol}),
     $(j_{12}+j_1+j_2)$ and $(j_3+1+j_4+j'_{12})$ are integer numbers, we can invert their common sign in 
     the exponent of the $(-1)$ . 
     After summing up all the terms in the exponents and performing some permutations on the arguments of
     the $6j$-symbols according to (\ref{symmetry1}), (\ref{symmetry2})
     we obtain the final result for $F(j_{12},j'_{12})$:\\\\
     \fbox{\parbox{15.5cm}{
     \ba \label{F5}
         \lefteqn{F(j_{12},j'_{12}):=\sum_{j_{23}}(2j_{23}+1)~j_{23}(j_{23}+1)~
     	   \left\{ \begin{array}{ccc} 
                    j_1 & j_2 & j_{12}\\
                    j_3 & j_4 & j_{23}
           \end{array} \right\} 
	   \left\{ \begin{array}{ccc} 
                    j_1 & j_2 & j'_{12}\\
                    j_3 & j_4 & j_{23}
           \end{array} \right\}~=}\hspace{2cm} \nonumber\\
      & ~~~~~~~=~ & \frac{1}{2}(-1)^{j_1+j_2+j_3+j_4+j_{12}+j'_{12}+1}
           X(j_1,j_4)^{\frac{1}{2}}
	     \left\{ \begin{array}{ccc} 
                      j_2 & j_1 & j_{12} \\
                      1 & j'_{12} & j_1
             \end{array} \right\} 
	     \left\{ \begin{array}{ccc} 
                      j_3 & j_4 & j_{12}\\
                        1 & j'_{12} & j_4
             \end{array} \right\} 
	     \nonumber\\    
     &   & +~\frac{j_1(j_1+1)+j_4(j_4+1)}{(2j_{12}+1)}\displaystyle{\delta_{j_{12}j'_{12}}}                 
     \ea }}
     \\ \\
     with $X(j_1,j_4)=2j_1(2j_1+1)(2j_1+2)2j_4(2j_4+1)(2j_4+2)$. 
     \begin{description}
       \item[Remark] By the integer / positivity requirements of the factorials occurring in the definition
       of the $6j$-symbols of (\ref{F4}), see (\ref{integer conditions}) we can read off restrictions for
       $j_{12},j'_{12}$, namely the selection rules
       \[ j'_{12} = \left\{ \begin{array}{l}
                          j_{12}-1\\ 
			  j_{12}\\
			  j_{12}+1
		        \end{array}
	            \right.
	\]               
     \end{description} 
     
\subsection{Precalculation} 
   After these preparations we can now go into 'medias res':
   We will carry out all the summations in (\ref{basic structure in 6j symbols}).
   Step by step we will only write down (with reordered prefactors) the terms containing the actual summation variable, suppressing all
   the other terms and sums in (\ref{basic structure in 6j symbols}).\\
   \\
   We start with the summation over $\vec{g}(IJ)$ (using again the 
shorthand
   $A(x,y)=\sqrt{(2x+1)(2y+1)}$~and the fact, that $A(g_2,h_2)\cdot A(g_2,k_2)=A(h_2,k_2)\cdot (2g_2+1)$.
   Additionally we will frequently use the integer conditions (\ref{integer conditions})~):
   
   \begin{itemize}
     \item Summation over $g_2$:
        \begin{footnotesize}
        \ba
         \lefteqn{
	   A(h_2,k_2)(-1)^{h_2-k_2}\sum_{g_2}g_2(g_2+1)(2g_2+1)
     	   \left\{ \begin{array}{ccc} 
                    j_J & j_I & g_2\\
                    j_1 & g_3 & h_2
           \end{array} \right\} 
	   \left\{ \begin{array}{ccc} 
                    j_J & j_I & g_2\\
                    j_1 & g_3 & k_2
           \end{array} \right\}~=} \nonumber\\ \nonumber\\ \nonumber\\
          & \stackrel{\ref{symmetry1},\ref{symmetry2}}{=} &	
	   A(h_2,k_2)(-1)^{h_2-k_2}\sum_{g_2}g_2(g_2+1)(2g_2+1)
     	   \left\{ \begin{array}{ccc} 
                    j_I & j_1 & h_2\\
                    g_3 & j_J & g_2
           \end{array} \right\} 
	   \left\{ \begin{array}{ccc} 
                    j_I & j_1 & k_2\\
                    g_3 & j_J & g_2
           \end{array} \right\} \nonumber \\
          & \stackrel{\ref{F4}}{=} & 
	   A(h_2,k_2)(-1)^{h_2-k_2} 
	   \biggm[ \frac{1}{2}(-1)^{\overbrace{j_I+j_1+h_2}^{integer}+g_3+j_J+1+k_2} X(j_I,j_J)^{\frac{1}{2}} 
      	   \left\{ \begin{array}{ccc} 
                    j_1 & j_I & h_2\\
                     1  & k_2 & j_I
           \end{array} \right\} 
	   \left\{ \begin{array}{ccc} 
                    g_3 & j_J & h_2\\
                     1  & k_2 & j_J
           \end{array} \right\} \nonumber \\
	   & & \hspace{3cm} +\frac{j_I(j_I+1)+j_J(j_J+1)}{2h_2+1}\delta_{h_2k_2}  \biggm] \nonumber \\
          & = & \underbrace{
	   A(h_2,k_2)\frac{1}{2}(-1)^{-j_I-j_1+j_J+1} X(j_I,j_J)^{\frac{1}{2}}
      	   \left\{ \begin{array}{ccc} 
                    j_1 & j_I & h_2\\
                     1  & k_2 & j_I
           \end{array} \right\} }_{M_2} 
	   (-1)^{g_3}
	   \left\{ \begin{array}{ccc} 
                    g_3 & j_J & h_2\\
                     1  & k_2 & j_J
           \end{array} \right\} 
	   +\underbrace{\bigm[j_I(j_I+1)+j_J(j_J+1)\bigm]}_{N}\delta_{h_2k_2} \nonumber   
	\ea
	\end{footnotesize}
     \item Summation over $g_3$:
        \begin{footnotesize}
        \ba
         \lefteqn{
	   A(h_3,k_3)(-1)^{h_3-k_3}\sum_{g_3}\biggm[
	   M_2(-1)^{g_3}
	   \left\{ \begin{array}{ccc} 
                    g_3 & j_J & h_2\\
                     1  & k_2 & j_J
	   \end{array} \right\}  + N\delta_{h_2k_2} \biggm]
	   (2g_3+1) 
	   \left\{ \begin{array}{ccc} 
                    j_J & h_2 & g_3\\
                    j_2 & g_4 & h_3
           \end{array} \right\} 
	   \left\{ \begin{array}{ccc} 
                    j_J & k_2 & g_3\\
                    j_2 & g_4 & k_3
           \end{array} \right\}~=} \nonumber\\ \nonumber\\ \nonumber\\
          & = &	
	   A(h_3,k_3)(-1)^{h_3-k_3}\biggm\{M_2\sum_{g_3}(-1)^{g_3}(2g_3+1)
     	   \left\{ \begin{array}{ccc} 
                    g_3 & j_J & h_2\\
                     1  & k_2 & j_J
           \end{array} \right\} 
     	   \left\{ \begin{array}{ccc} 
                    j_J & h_2 & g_3\\
                    j_2 & g_4 & h_3
           \end{array} \right\} 
	   \left\{ \begin{array}{ccc} 
                    j_J & k_2 & g_3\\
                    j_2 & g_4 & k_3
           \end{array} \right\} \nonumber \\
	  &   & \hspace{3cm} +
	    N\delta_{h_2k_2}\sum_{g_3}(2g_3+1) 
     	   \left\{ \begin{array}{ccc} 
                    j_J & h_2 & g_3\\
                    j_2 & g_4 & h_3
           \end{array} \right\} 
	   \left\{ \begin{array}{ccc} 
                    j_J & k_2 & g_3\\
                    j_2 & g_4 & k_3
           \end{array} \right\} \biggm\} \nonumber \\
          & \stackrel{\ref{symmetry1},\ref{symmetry2}}{=} & 
	   A(h_3,k_3)(-1)^{h_3-k_3}\biggm\{M_2\sum_{g_3}(-1)^{g_3}(2g_3+1)
     	   \left\{ \begin{array}{ccc} 
                    j_J & j_J & 1 \\
                    h_2 & k_2 & g_3
           \end{array} \right\} 
     	   \left\{ \begin{array}{ccc} 
                    j_J & h_3 & g_4\\
                    j_2 & g_3 & h_2
           \end{array} \right\} 
	   \left\{ \begin{array}{ccc} 
                    j_J & g_4 & k_3\\
                    j_2 & k_2 & g_3
           \end{array} \right\} \nonumber \\
	  & & \hspace{3cm} +
	    \frac{N}{(2k_3+1)}\delta_{h_2k_2}\sum_{g_3}(2k_3+1)(2g_3+1) 
     	   \left\{ \begin{array}{ccc} 
                    j_J & g_4 & h_3\\
                    j_2 & h_2 & g_3
           \end{array} \right\} 
	   \left\{ \begin{array}{ccc} 
                    j_J & g_4 & k_3\\
                    j_2 & h_2 & g_3
           \end{array} \right\}  \biggm\} \nonumber \\
          & \stackrel{\ref{EBI},\ref{orthogonality relation}}{=} & 
           A(h_3,k_3)(-1)^{h_3-k_3}\biggm\{M_2(-1)^{-(h_3+j_J+j_2+\overbrace{j_J+g_4+1+k_3}^{integer}
	                                            +\overbrace{h_2+k_2}^{integer})}
     	   \left\{ \begin{array}{ccc} 
                    h_3 & j_J & g_4\\
                    j_J & k_3 &  1
           \end{array} \right\} 
	   \left\{ \begin{array}{ccc} 
                     1  & h_3 & k_3\\
                    j_2 & k_2 & h_2
           \end{array} \right\} \nonumber \\ 
	  &   & \hspace{3cm} +
	    \frac{N}{(2k_3+1)}\delta_{h_2k_2}\delta_{h_3k_3} \biggm\} \nonumber \\
          & \stackrel{\ref{integer conditions},\ref{symmetry1},\ref{symmetry2}}{=} & 
           M_2\underbrace{
	   A(h_3,k_3)(-1)^{-j_2+h_2+k_2+1}
      	   \left\{ \begin{array}{ccc} 
                    j_2 & h_2 & h_3\\
                     1  & k_3 & k_2
           \end{array} \right\} }_{M_3} 
	   (-1)^{g_4}
	   \left\{ \begin{array}{ccc} 
                    g_4 & j_J & h_3\\
                     1  & k_3 & j_J
           \end{array} \right\} 
	   +N\delta_{h_2k_2}\delta_{h_3k_3} \nonumber   
	\ea
	\end{footnotesize}
	Note: that is in principle the same term with the same index-order as we got from the summation 
	over $g_2$.
	\item Generally we have for the summation over $g_i$ for $4\le i \le I$:
        \begin{footnotesize}
        \ba
         \lefteqn{
	   A(h_i,k_i)(-1)^{h_i-k_i}\sum_{g_i} \Biggm\{ \biggm[
	   M_2 \ldots M_{i-1}(-1)^{g_i}
	   \left\{ \begin{array}{ccc} 
                    g_i & j_J     & h_{i-1}\\
                     1  & k_{i-1} & j_J
	   \end{array} \right\}  + N\delta_{h_2k_2} \ldots \delta_{h_{i-1}k_{i-1}} \biggm] \times}\nonumber\\
          \lefteqn{  
	   \hspace{4cm} \times~(2g_i+1) 
	   \left\{ \begin{array}{ccc} 
                    j_J     & h_{i-1} & g_i\\
                    j_{i-1} & g_{i+1} & h_i
           \end{array} \right\} 
	   \left\{ \begin{array}{ccc} 
                    j_J     & k_{i-1} & g_i\\
                    j_{i-1} & g_{i+1} & k_i
           \end{array} \right\} \Biggm\}~=} \nonumber\\ \nonumber\\ \nonumber\\
          & = &	
	   A(h_i,k_i)(-1)^{h_i-k_i} \times \nonumber\\
	  & & 
	   \times \biggm\{M_2 \ldots M_{i-1}\sum_{g_i}(-1)^{g_i}(2g_i+1)
     	   \left\{ \begin{array}{ccc} 
                    g_i & j_J     & h_{i-1}\\
                     1  & k_{i-1} & j_J
           \end{array} \right\} 
     	   \left\{ \begin{array}{ccc} 
                    j_J     & h_{i-1} & g_i\\
                    j_{i-1} & g_{i+1} & h_i
           \end{array} \right\} 
	   \left\{ \begin{array}{ccc} 
                    j_J     & k_{i-1} & g_i\\
                    j_{i-1} & g_{i+1} & k_i
           \end{array} \right\} \nonumber \\
	  &   & \hspace{3cm} +
	    N\delta_{h_2k_2} \ldots \delta_{h_{i-1}k_{i-1}}\sum_{g_i}(2g_i+1) 
     	   \left\{ \begin{array}{ccc} 
                    j_J     & h_{i-1} & g_i\\
                    j_{i-1} & g_{i+1} & h_i
           \end{array} \right\} 
	   \left\{ \begin{array}{ccc} 
                    j_J     & k_{i-1} & g_i\\
                    j_{i-1} & g_{i+1} & k_i
           \end{array} \right\}  \biggm\} \nonumber \\
          & \stackrel{\ref{symmetry1},\ref{symmetry2}}{=} & 
	   A(h_i,k_i)(-1)^{h_i-k_i}\times \nonumber \\
	  &  & 
	   \times \biggm\{M_2 \ldots M_{i-1}\sum_{g_i}(-1)^{g_i}(2g_i+1)
     	   \left\{ \begin{array}{ccc} 
                    j_J     & j_J     &  1 \\
                    h_{i-1} & k_{i-1} & g_i
           \end{array} \right\} 
     	   \left\{ \begin{array}{ccc} 
                    j_J     & h_i & g_{i+1}\\
                    j_{i-1} & g_i & h_{i-1}
           \end{array} \right\} 
	   \left\{ \begin{array}{ccc} 
                    j_J     & g_{i+1} & k_i\\
                    j_{i-1} & k_{i-1} & g_i
           \end{array} \right\} \nonumber \\
	  & & \hspace{3cm} +
	    \frac{N}{(2k_i+1)}\delta_{h_2k_2} \ldots \delta_{h_{i-1}k_{i-1}}\sum_{g_i}(2k_i+1)(2g_i+1) 
     	   \left\{ \begin{array}{ccc} 
                    j_J     & h_i & g_{i+1}\\
                    j_{i-1} & g_i & h_{i-1}
           \end{array} \right\} 
	   \left\{ \begin{array}{ccc} 
                    j_J     & g_{i+1} & k_i\\
                    j_{i-1} & k_{i-1} & g_i
           \end{array} \right\}  \biggm\}\nonumber \\
          & \stackrel{\ref{EBI},\ref{orthogonality relation}}{=} & 
           A(h_i,k_i)(-1)^{h_i-k_i} \times \nonumber\\
	  &  &
	   \times \biggm\{M_2 \ldots M_{i-1}
            (-1)^{-(h_i+j_J+j_{i-1}+\overbrace{j_J+g_{i+1}+1+k_i}^{integer}
                                   +\overbrace{h_{i-1}+k_{i-1}}^{integer})} 
	   \left\{ \begin{array}{ccc} 
                    h_i & j_J & g_{i+1}\\
                    j_J & k_i &  1
           \end{array} \right\} 
	   \left\{ \begin{array}{ccc} 
                     1      & h_i     & k_i\\
                    j_{i-1} & k_{i-1} & h_{i-1}
           \end{array} \right\}  \nonumber \\
          &   & \hspace{3cm}
	    + N\delta_{h_2k_2} \ldots \delta_{h_{i-1}k_{i-1}}\delta_{h_ik_i}  \biggm\}\nonumber \\
          & \stackrel{\ref{integer conditions},\ref{symmetry1},\ref{symmetry2}}{=} & 
           M_2\ldots M_{i-1} \underbrace{
	   A(h_i,k_i)(-1)^{-j_{i-1}+h_{i-1}+k_{i-1}+1}
      	   \left\{ \begin{array}{ccc} 
                    j_{i-1} & h_{i-1} & h_i\\
                       1    & k_i     & k_{i-1}
           \end{array} \right\} }_{M_i} 
	   	   (-1)^{g_{i+1}}
	   \left\{ \begin{array}{ccc} 
                    g_{i+1} & j_J & h_i\\
                      1     & k_i & j_J
           \end{array} \right\} \nonumber\\
          &   & \hspace{3cm}
	   +N\delta_{h_2k_2} \ldots \delta_{h_{i-1}k_{i-1}}\delta_{h_ik_i} \nonumber   
	\ea
	\end{footnotesize}
	\item For the summation over $g_{I+1}$ \\
	      we have the same terms as above with a slight difference in one index (underlined):
        \begin{footnotesize}
        \ba
         \lefteqn{
	   A(h_{I+1},k_{I+1})(-1)^{h_{I+1}-k_{I+1}}\sum_{g_{I+1}} \Biggm\{ \biggm[
	   M_2 \ldots M_{I}(-1)^{g_{I+1}}
	   \left\{ \begin{array}{ccc} 
                    g_{I+1} & j_J & h_I\\
                     1  & k_I & j_J
	   \end{array} \right\}  
	   + N\delta_{h_2k_2} \ldots \delta_{h_Ik_I} \biggm] \times}\nonumber\\
          \lefteqn{  
	   \hspace{4cm} \times~(2g_{I+1}+1) 
	   \left\{ \begin{array}{ccc} 
                    j_J     & h_I & g_{I+1}\\
                    \underline{j_{I+1}} & g_{I+2} & h_{I+1}
           \end{array} \right\} 
	   \left\{ \begin{array}{ccc} 
                    j_J     & k_I & g_{I+1}\\
                    \underline{j_{I+1}} & g_{I+2} & k_{I+1}
           \end{array} \right\} \Biggm\}~=} \nonumber\\ \nonumber\\ \nonumber\\
          & = &
           M_2\ldots M_I \underbrace{
	   A(h_{I+1},k_{I+1})(-1)^{-j_{I+1}+h_I+k_I+1}
      	   \left\{ \begin{array}{ccc} 
                    j_{I+1} & h_I     & h_{I+1}\\
                       1    & k_{I+1} & k_I
           \end{array} \right\} }_{\tilde{M}_{I+1}} 
	   	   (-1)^{g_{I+2}}
	   \left\{ \begin{array}{ccc} 
                    g_{I+2} & j_J & h_{I+1}\\
                      1     & k_{I+1} & j_J
           \end{array} \right\} \nonumber\\
          &   & \hspace{3cm}
	   +N\delta_{h_2k_2} \ldots \delta_{h_Ik_I}\delta_{h_{I+1}k_{I+1}} \nonumber   
	\ea
	\end{footnotesize}
	\item Summation over $g_i$ for $I+2\le i \le J-1$:
        \begin{samepage}
        \begin{footnotesize}
        \ba
         \lefteqn{
	   A(h_i,k_i)(-1)^{h_i-k_i}\sum_{g_i} \Biggm\{ \biggm[
	   M_2 \ldots M_I\tilde{M}_{I+1} \ldots \tilde{M}_{i-1}(-1)^{g_i}
	   \left\{ \begin{array}{ccc} 
                    g_i & j_J     & h_{i-1}\\
                     1  & k_{i-1} & j_J
	   \end{array} \right\}  
	   + N\delta_{h_2k_2} \ldots \delta_{h_{i-1}k_{i-1}} \biggm] \times}\nonumber\\
          \lefteqn{  
	   \hspace{4cm} \times~(2g_i+1) 
	   \left\{ \begin{array}{ccc} 
                    j_J & h_{i-1} & g_i\\
                    \underline{j_i} & g_{i+1} & h_i
           \end{array} \right\} 
	   \left\{ \begin{array}{ccc} 
                    j_J & k_{i-1} & g_i\\
                    \underline{j_i} & g_{i+1} & k_i
           \end{array} \right\} \Biggm\}~=} \nonumber\\ \nonumber\\ \nonumber\\
          & = &
           M_2\ldots M_I\tilde{M}_{I+1}\ldots \tilde{M}_{i-1} \underbrace{
	   A(h_i,k_i)(-1)^{-j_i+h_{i-1}+k_{i-1}+1}
      	   \left\{ \begin{array}{ccc} 
                    j_{i-1} & h_{i-1} & h_i\\
                       1    & k_i     & k_{i-1}
           \end{array} \right\} }_{\tilde{M}_i} 
	   	   (-1)^{g_{i+1}}
	   \left\{ \begin{array}{ccc} 
                    g_{i+1} & j_J & h_i\\
                      1     & k_i & j_J
           \end{array} \right\} \nonumber\\
          &   & \hspace{3cm}
	   +N\delta_{h_2k_2} \ldots \delta_{h_{i-1}k_{i-1}}\delta_{h_ik_i} \nonumber   
	\ea
	\end{footnotesize}
	\end{samepage}
   \end{itemize}

   Here we keep the following notation in mind:\\\\
   \fbox{\parbox{15.5cm}{
   \[\begin{array}{rclcrclcl}
      h_1     &  =    & j_I                     &  &k_1    &  =    &j_I  
                                                &\hspace{1cm}& J\le i \le N:~~g_i=g''_i=a_i\\
      h_2     &  =    & h_2(j_I,j_1)            &  &k_2    &  =    &k_2(j_I,j_1)\\      
      h_3     &  =    & h_3(h_2,j_2)            &  &k_3    &  =    &k_3(k_2,j_2)\\
              &\vdots &                         &\hspace{1cm} &       &\vdots &            \\
      h_{I-1} &  =    & h_{I-1}(h_{I-2},j_{I-2})&  &k_{I-1}&  =    &k_{I-1}(k_{I-2},j_{I-2})\\
      h_I     &  =    & g''_I                   &  &k_I    &  =    &a_I\\
      h_{I+1} &  =    & g''_{I+1}               &  &k_{I+1}&  =    &a_{I+1}\\
              &\vdots &                         &  &       &\vdots &            \\
      h_{J-1} &  =    & g''_{J-1}               &  &k_{J-1}&  =    &a_{J-1}    
   \end{array} \]     
   \be \label{kopplungsschemata} \ee}}\\\\\\
   We have now carried out completely the summation over $\vec{g}(IJ)$ in 
   (\ref{basic structure in 6j symbols}).
   After this we write down the remaining terms of this summation and the terms of 
   (\ref{basic structure in 6j symbols}) which have not taken part in the summation yet. 
   So we end up with (do not get confused about the notation $A(x,y)=\sqrt{(2x+1)(2y+1)}$ whereas $A$, $A_i$
   are abbreviations form certain terms):
   \begin{footnotesize}
   \ba
     \lefteqn{\sum_{\vec{g}(IJ)}g_2(g_2+1)<\vec{g}(IJ)|\vec{g}''><\vec{g}(IJ)|\vec{a}>~=}\hspace{1cm} \nonumber\\
     &=&\sum_{h_2 \ldots h_{I-1} \atop k_2 \ldots k_{I-1}} 
        \Biggm[\overbrace{\prod_{n=2}^{I}M_n \prod_{m=I+1}^{J-1}\tilde{M}_m 
	\times  (-1)^{a_J} \left\{ \begin{array}{ccc} 
                                  a_J & j_J     & g''_{J-1}\\
                                    1 & a_{J-1} & j_J
                            \end{array} \right\} }^{part~I}
      	+\overbrace{N \prod_{n=2}^{J-1}\delta_{h_n k_n}}^{part~II}  \Biggm] \times ~remaining~terms \nonumber \\
     &=&\sum_{h_2 \ldots h_{I-1} \atop k_2 \ldots k_{I-1}} 
    \Biggm[\frac{1}{2}(-1)^{j_J-j_I-\sum_{n=1}^{I-1}j_n-\sum_{m=I+1}^{J-1}j_m}X(j_I,j_J)^{\frac{1}{2}}
    \times \nonumber\\
     & & \begin{array}{llclr} 
	    &\times \overbrace{(-1)^{+1}A(h_2,k_2)\left\{ \begin{array}{ccc} 
                                	   j_1 & j_I & h_2\\
                                	    1  & k_2 & j_I
                        	  \end{array} \right\}}^{A}  
	    &\displaystyle\prod\limits_{n=3}^{I}&\overbrace{A(h_n,k_n)(-1)^{h_{n-1}+k_{n-1}+1}
                        	 \left\{ \begin{array}{ccc} 
                                	  j_{n-1}& h_{n-1} & h_n\\
                                	     1   & k_n     & k_{n-1}
                         	 \end{array} \right\}}^{A_1\leftrightarrow n=3,
				                        A_2\leftrightarrow n=4,
							A_3\leftrightarrow n=5,
							\ldots,
							A_{I-2} \leftrightarrow n=I} &\times\\  
	    &\times  & \displaystyle\prod\limits_{n=I+1}^{J-1}& A(g''_n,a_n)(-1)^{g''_{n-1}+a_{n-1}+1}
                        	 \left\{ \begin{array}{ccc} 
                                	  j_n & g''_{n-1} & g''_n\\
                                	    1 &  a_n      & a_{n-1}
                         	 \end{array} \right\} &\times \\ 
	    &\times ~(-1)^{a_J}   			 
               	 \left\{ \begin{array}{ccc} 
                      	  a_J & j_J     & g''_{J-1}\\
                      	    1 & a_{J-1} & j_J
               	 \end{array} \right\}  &
                + & \big[ j_I(j_I+1)+j_J(j_J+1)\big] \displaystyle\prod\limits_{n=2}^{J-1}\delta_{h_nk_n}  
		  \Biggm] \hspace{1cm}&\times \nonumber \\
         \end{array} \nonumber\\
     & & \begin{array}{crr}
	    \times 
	      &\underbrace{(-1)^{g''_2} A(h_2,g''_2)
	      \left\{ \begin{array}{ccc} 
                       j_I & j_1 & h_2\\
                       j_2 & h_3 & g''_2
              \end{array} \right\}}_{B_1} 
	      &\underbrace{(-1)^{-a_2} A(k_2,a_2)
	      \left\{ \begin{array}{ccc} 
                       j_I & j_1 & k_2\\
                       j_2 & k_3 & a_2
              \end{array} \right\}}_{C_1} \\
              \times
              &\underbrace{(-1)^{h_3-k_3+g''_3} A(h_3,g''_3)
	      \left\{ \begin{array}{ccc} 
                       j_I & g''_2 & h_3\\
                       j_3 & h_4 & g''_3
              \end{array} \right\}}_{B_2}
	      &\underbrace{(-1)^{-a_3} A(k_3,a_3)
	      \left\{ \begin{array}{ccc} 
                       j_I & a_2 & k_3\\
                       j_3 & k_4 & a_3
              \end{array} \right\}}_{C_2} \\
              \times
	      &\underbrace{(-1)^{h_4-k_4+g''_4} A(h_4,g''_4)
	      \left\{ \begin{array}{ccc} 
                       j_I & g''_3 & h_4\\
                       j_4 & h_5 & g''_4
              \end{array} \right\}}_{B_3} 
	      &\underbrace{(-1)^{-a_4} A(k_4,a_4)
	      \left\{ \begin{array}{ccc} 
                       j_I & a_3 & k_4\\
                       j_4 & k_5 & a_4
              \end{array} \right\}}_{C_3} \\

             \vdots&\vdots~~~~~~~~~~~~& \vdots~~~~~~~~~~~~\\
	     \vdots&\vdots~~~~~~~~~~~~& \vdots~~~~~~~~~~~~\\
             \times
	      &\underbrace{(-1)^{h_{I-1}-k_{I-1}+g''_{I-1}} A(h_{I-1},2g'_{I-1})
	      \left\{ \begin{array}{ccc} 
                       j_I & g''_{I-2} & h_{I-1}\\
                       j_{I-1} & g''_I & g'_{I-1}
              \end{array} \right\}}_{B_{I-2}}
	      &\underbrace{(-1)^{-a_{I-1}} A(k_{I-1},a_{I-1})
	      \left\{ \begin{array}{ccc} 
                       j_I & a_{I-2} & k_{I-1}\\
                       j_{I-1} & a_I & a_{I-1}
              \end{array} \right\}}_{C_{I-2}} \nonumber
         \end{array} 	 		    
   \ea  
   \end{footnotesize}
   \be \label{zwischenschritt}\ee
   
   After noting this intermediate result we finally have to execute the remaining summations of
   (\ref{zwischenschritt}), namely the summations over $h_2,\ldots ,h_{I-1}$ and $k_2,\ldots ,k_{I-1}$ (leaving
   out the signs $(-1)^{g''_i-a_i}$~, since they will be cancelled, as we will see, due to the occurrence
   of $\delta_{g''_ia_i}$-terms in the following calculations).\\
   
   First we do this summation for the \fbox{$\bf part~I$ of (\ref{zwischenschritt})}.

   \begin{itemize}
   \item first step
      \subitem summation over $h_2$: 
        \begin{footnotesize}
	\ba
	  \lefteqn{\sum_{h_2}	      
	     \overbrace{(-1)^{+1}A(h_2,k_2)       
	                     \left\{ \begin{array}{ccc} 
                                     j_1 & j_I & h_2 \\
                                      1  & k_2 & j_I
                                     \end{array} \right\}}^{A}
	     \overbrace{(-1)^{h_2+k_2+1}A(h_3,k_3)
	                     \left\{ \begin{array}{ccc} 
                                     j_2 & h_2 & h_3 \\
                                      1  & k_3 & k_2
                                     \end{array} \right\}}^{A_1}
	     \overbrace{A(h_2,g''_2)
	                     \left\{ \begin{array}{ccc} 
                                     j_I & j_1 & h_2 \\
                                     j_2 & h_3 & g''_2
                                     \end{array} \right\}}^{B_1} = }\nonumber\\\nonumber\\
             &\hspace{0cm} \stackrel{\ref{symmetry1},\ref{symmetry2}}{=}&
	          (-1)^{k_2}A(k_2,g''_2)A(h_3,k_3) 
		          \underbrace{ \sum_{h_2}(-1)^{h_2}(2h_2+1)
	                     \left\{ \begin{array}{ccc} 
                                      j_1 & k_2 & j_I \\
                                      1  & j_I & h_2
                                     \end{array} \right\}
	                     \left\{ \begin{array}{ccc} 
                                     k_2 & k_3 & j_2 \\
                                     h_3 & h_2 &  1
                                     \end{array} \right\}
	                     \left\{ \begin{array}{ccc} 
                                     j_1 & j_2 & g''_2 \\
                                     h_3 & j_I & h_2
                                     \end{array} \right\} } 
	                   _{\begin{array}{l}
			       \stackrel{\ref{EBI}}{=} 
				(-1)^{-(\overbrace{k_3+j_I+1+g''_2}^{integer}+k_2+j_1+h_3+j_2)}
	                        \left\{ \begin{array}{ccc} 
                                         k_3 & k_2   & j_2 \\
                                         j_1 & g''_2 & j_I
                                        \end{array} \right\}
	                        \left\{ \begin{array}{ccc} 
                                         j_I & k_3 & g''_2 \\
                                         h_3 & j_I &  1
                                     \end{array} \right\} \\ 
			       \stackrel{\ref{integer conditions}}{=}
			        (-1)^{k_3+j_I+1+g''_2-k_2-j_1-h_3-j_2-j_I}
	                        \left\{ \begin{array}{ccc} 
                                         k_3 & k_2   & j_2 \\
                                         j_1 & g''_2 & j_I
                                        \end{array} \right\}
	                        \left\{ \begin{array}{ccc} 
                                         j_I & k_3 & g''_2 \\
                                         h_3 & j_I &  1
                                     \end{array} \right\} 
		             \end{array}} \nonumber\\ \nonumber\\
             &\hspace{0cm} = &
	                     \underbrace{
                             (-1)^{g''_2+1+k_3-j_1-j_2-h_3}A(h_3,k_3) 
	                     \left\{ \begin{array}{ccc} 
                                     j_I & k_3 & g''_2 \\
                                     h_3 & j_I &  1
                                     \end{array} \right\} }_{D_1}
			     \underbrace{	     
	                     A(k_2,g''_2)
			     \left\{ \begin{array}{ccc} 
                                     k_3 & k_2   & j_2 \\
                                     j_1 & g''_2 & j_I
                                     \end{array} \right\} }_{E_1} \nonumber
 	\ea
        \end{footnotesize} \be \ee
      \subitem summation over $k_2$: 
        \begin{footnotesize}
	\ba 
	  \lefteqn{
	  \sum_{k_2}	      
            \overbrace{A(k_2,g''_2)
			     \left\{ \begin{array}{ccc} 
                                     k_3 & k_2   & j_2 \\
                                     j_1 & g''_2 & j_I
                                     \end{array} \right\} }^{E_1}
	     \overbrace{A(k_2,a_2)       
	                     \left\{ \begin{array}{ccc} 
                                     j_I & j_1 & k_2 \\
                                     j_2 & k_3 & a_2
                                     \end{array} \right\}}^{C_1} = } \nonumber\\ \nonumber \\
             &\hspace{3cm} \stackrel{\ref{symmetry1},\ref{symmetry2}}{=}&
	                     \frac{A(a_2,g''_2)}{(2a_2+1)} 
			     \underbrace{\sum_{k_2}(2k_2+1)(2a_2+1)
	                        \left\{ \begin{array}{ccc} 
                                        j_1 & j_2 & g''_2 \\
                                        k_3 & j_I & k_2
                                        \end{array} \right\}
	                	\left\{ \begin{array}{ccc} 
                                	j_1 & j_2 & a_2 \\
                                	k_3 & j_I & k_2
                                	\end{array} \right\} }
			      _{\stackrel{\ref{orthogonality relation}}{=}~\delta_{g''_2a_2}} \nonumber\\
             &\hspace{3cm} =& \delta_{g''_2a_2} \nonumber
 	\ea
        \end{footnotesize} \be \label{summation k2} \ee \\[1cm]

   \item second step
      \subitem summation over $h_3$: 
        \begin{footnotesize}
	\ba
	  \lefteqn{\sum_{h_3}	      
	     \overbrace{(-1)^{g''_2+1+k_3-j_1-j_2-h_3}A(h_3,k_3)       
	                     \left\{ \begin{array}{ccc} 
                                     j_I & k_3 & g''_2 \\
                                     h_3 & j_I &  1
                                     \end{array} \right\}}^{D_1}
	     \overbrace{(-1)^{h_3+k_3+1}A(h_4,k_4)
	                     \left\{ \begin{array}{ccc} 
                                     j_3 & h_3 & h_4 \\
                                      1  & k_4 & k_3
                                     \end{array} \right\}}^{A_2}
	     \overbrace{(-1)^{h_3-k_3}A(h_3,g''_3)
	                     \left\{ \begin{array}{ccc} 
                                     j_I & g''_2 & h_3 \\
                                     j_3 & h_4 & g''_3
                                     \end{array} \right\}}^{B_2} = }\nonumber\\\nonumber\\
             &\hspace{0cm} =&
	          (-1)^{g''_2-j_1-j_2+k_3}A(k_3,g''_3)A(h_4,k_4) 
		          \underbrace{ \sum_{h_3}(-1)^{h_3}(2h_3+1)
	                     \left\{ \begin{array}{ccc} 
                                     j_I & k_3 & g''_2 \\
                                     h_3 & j_I &  1
				     \end{array} \right\}
	                     \left\{ \begin{array}{ccc} 
                                     j_3 & h_3 & h_4 \\
                                      1  & k_4 & k_3
				     \end{array} \right\}
	                     \left\{ \begin{array}{ccc} 
                                     j_I & g''_2 & h_3 \\
                                     j_3 & h_4   & g''_3 
                                     \end{array} \right\} } 
	                   _{\begin{array}{l}
			       \stackrel{\ref{symmetry1},\ref{symmetry2}}{=}
			        \sum_{h_3}(-1)^{h_3}(2h_3+1)
	                	\left\{ \begin{array}{ccc} 
                                	g''_2 & k_3 & j_I \\
                                	  1   & j_I & h_3
					\end{array} \right\}
	                	\left\{ \begin{array}{ccc} 
                                	k_3 & k_4 & j_3 \\
                                	h_4 & h_3 &  1
					\end{array} \right\}
	                	\left\{ \begin{array}{ccc} 
                                	g''_2 & j_3 & g''_3 \\
                                	h_4   & j_I & h_3 
                                	\end{array} \right\} \\	       
			       \stackrel{\ref{EBI}}{=} 
				(-1)^{-(\overbrace{k_4+j_I+1+g''_3}^{integer}+k_3+g''_2+h_4+j_I)}
	                        \left\{ \begin{array}{ccc} 
                                         k_4   & k_3   & j_3 \\
                                         g''_2 & g''_3 & j_I
                                        \end{array} \right\}
	                        \left\{ \begin{array}{ccc} 
                                         j_I & k_4 & g''_3 \\
                                         h_4 & j_I &  1
                                     \end{array} \right\} \\ 
			       \stackrel{\ref{integer conditions}}{=}
			        (-1)^{k_4+j_I+1+g''_3-k_3-g''_2-h_4-j_3-j_I}
	                        \left\{ \begin{array}{ccc} 
                                         k_4   & k_3   & j_3 \\
                                         g''_2 & g''_3 & j_I
                                        \end{array} \right\}
	                        \left\{ \begin{array}{ccc} 
                                         j_I & k_4 & g''_3 \\
                                         h_4 & j_I &  1
                                     \end{array} \right\}  
		             \end{array}} \nonumber\\ \nonumber\\
             &\hspace{0cm} = &
	                     \underbrace{
                             (-1)^{g''_3+1+k_4-j_1-j_2-j_3-h_4}A(h_4,k_4) 
	                     \left\{ \begin{array}{ccc} 
                                     j_I & k_4 & g''_3 \\
                                     h_4 & j_I &  1
                                     \end{array} \right\} }_{D_2}
			     \underbrace{	     
	                     A(k_3,g''_3)
			     \left\{ \begin{array}{ccc} 
                                     k_4   & k_3   & j_3 \\
                                     g''_2 & g''_3 & j_I
                                     \end{array} \right\} }_{E_2} \nonumber
 	\ea
        \end{footnotesize} \be \ee
      \subitem summation over $k_3$: 
        \begin{footnotesize}
	\ba 
	  \lefteqn{
	  \sum_{k_3}	      
            \overbrace{A(k_3,g''_3)
			     \left\{ \begin{array}{ccc} 
                                     k_4   & k_3   & j_3 \\
                                     g''_2 & g''_3 & j_I
                                     \end{array} \right\} }^{E_2}
	     \overbrace{A(k_3,a_3)       
	                     \left\{ \begin{array}{ccc} 
                                     j_I & a_2 & k_3 \\
                                     j_3 & k_4 & a_3
                                     \end{array} \right\}}^{C_2} = } \nonumber\\ \nonumber \\
             &\hspace{3cm} \stackrel{\ref{summation k2},\ref{symmetry1},\ref{symmetry2}}{=}&
	                     \frac{A(a_3,g''_3)}{(2a_3+1)} 
			     \underbrace{\sum_{k_3}(2k_3+1)(2a_3+1)
	                        \left\{ \begin{array}{ccc} 
                                        a_2 & j_3 & g''_3 \\
                                        k_4 & j_I & k_3
                                        \end{array} \right\}
	                	\left\{ \begin{array}{ccc} 
                                	a_2 & j_3 & a_3 \\
                                	k_4 & j_I & k_3
                                	\end{array} \right\} }
			      _{\stackrel{\ref{orthogonality relation}}{=}~\delta_{g''_3a_3}} \nonumber\\
             &\hspace{3cm} =& \delta_{g''_3a_3} \nonumber
 	\ea
        \end{footnotesize} \be \label{summation k3} \ee

   \item third step
      \subitem summation over $h_4$: 
        \begin{footnotesize}
	\ba
	  \lefteqn{\sum_{h_4}	      
	     \overbrace{(-1)^{g''_3+1+k_4-j_1-j_2-j_3-h_4}A(h_4,k_4)       
	                     \left\{ \begin{array}{ccc} 
                                     j_I & k_4 & g''_3 \\
                                     h_4 & j_I &  1
                                     \end{array} \right\}}^{D_2}
	     \overbrace{(-1)^{h_4+k_4+1}A(h_5,k_5)
	                     \left\{ \begin{array}{ccc} 
                                     j_4 & h_4 & h_5 \\
                                      1  & k_5 & k_4
                                     \end{array} \right\}}^{A_3}
	     \overbrace{(-1)^{h_4-k_4}A(h_4,g''_4)
	                     \left\{ \begin{array}{ccc} 
                                     j_I & g''_3 & h_4 \\
                                     j_4 & h_5 & g''_4
                                     \end{array} \right\}}^{B_3} = }\nonumber\\\nonumber\\
             &\hspace{0cm} =&
	          (-1)^{g''_3-j_1-j_2-j_3+k_4}A(k_4,g''_4)A(h_5,k_5) 
		          \underbrace{ \sum_{h_4}(-1)^{h_4}(2h_4+1)
	                     \left\{ \begin{array}{ccc} 
                                     j_I & k_4 & g''_3 \\
                                     h_4 & j_I &  1
				     \end{array} \right\}
	                     \left\{ \begin{array}{ccc} 
                                     j_4 & h_4 & h_5 \\
                                      1  & k_5 & k_4
				     \end{array} \right\}
	                     \left\{ \begin{array}{ccc} 
                                     j_I & g''_3 & h_4 \\
                                     j_4 & h_5   & g''_4 
                                     \end{array} \right\} } 
	                   _{\begin{array}{l}
			       \stackrel{\ref{symmetry1},\ref{symmetry2}}{=}
			        \sum_{h_4}(-1)^{h_4}(2h_4+1)
	                	\left\{ \begin{array}{ccc} 
                                	g''_3 & k_4 & j_I \\
                                	  1   & j_I & h_4
					\end{array} \right\}
	                	\left\{ \begin{array}{ccc} 
                                	k_4 & k_5 & j_4 \\
                                	h_5 & h_4 &  1
					\end{array} \right\}
	                	\left\{ \begin{array}{ccc} 
                                	g''_3 & j_4 & g''_4 \\
                                	h_5   & j_I & h_4 
                                	\end{array} \right\} \\	       
			       \stackrel{\ref{EBI}}{=} 
				(-1)^{-(\overbrace{k_5+j_I+1+g''_4}^{integer}+k_4+g''_3+h_5+j_4+j_I)}
	                        \left\{ \begin{array}{ccc} 
                                         k_5   & k_4   & j_4 \\
                                         g''_3 & g''_4 & j_I
                                        \end{array} \right\}
	                        \left\{ \begin{array}{ccc} 
                                         j_I & k_5 & g''_4 \\
                                         h_5 & j_I &  1
                                     \end{array} \right\} \\ 
			       \stackrel{\ref{integer conditions}}{=}
			        (-1)^{k_5+j_I+1+g''_4-k_4-g''_3-h_5-j_4-j_I}
	                        \left\{ \begin{array}{ccc} 
                                         k_5   & k_4   & j_4 \\
                                         g''_3 & g''_4 & j_I
                                        \end{array} \right\}
	                        \left\{ \begin{array}{ccc} 
                                         j_I & k_5 & g''_4 \\
                                         h_5 & j_I &  1
                                     \end{array} \right\} \\ 
		             \end{array}} \nonumber\\ \nonumber\\
             &\hspace{0cm} = &
	                     \underbrace{
                             (-1)^{g''_4+1+k_5-j_1-j_2-j_3-j_4-h_5}A(h_5,k_5) 
	                     \left\{ \begin{array}{ccc} 
                                     j_I & k_5 & g''_4 \\
                                     h_5 & j_I &  1
                                     \end{array} \right\} }_{D_3}
			     \underbrace{	     
	                     A(k_4,g''_4)
			     \left\{ \begin{array}{ccc} 
                                     k_5   & k_4   & j_4 \\
                                     g''_3 & g''_4 & j_I
                                     \end{array} \right\} }_{E_3} \nonumber
 	\ea
        \end{footnotesize} \be \ee \\[1cm]
      \subitem summation over $k_4$: 
        \begin{footnotesize}
	\ba
	  \lefteqn{
	  \sum_{k_4}	      
            \overbrace{A(k_4,g''_4)
			     \left\{ \begin{array}{ccc} 
                                     k_5   & k_4   & j_4 \\
                                     g''_3 & g''_4 & j_I
                                     \end{array} \right\} }^{E_3}
	     \overbrace{A(k_4,a_4)       
	                     \left\{ \begin{array}{ccc} 
                                     j_I & a_3 & k_4 \\
                                     j_4 & k_5 & a_4
                                     \end{array} \right\}}^{C_3} = } \nonumber\\ \nonumber \\
             &\hspace{3cm} \stackrel{\ref{summation k3},\ref{symmetry1},\ref{symmetry2}}{=}&
	                     \frac{A(a_4,g''_4)}{(2a_4+1)} 
			     \underbrace{\sum_{k_4}(2k_4+1)(2a_4+1)
	                        \left\{ \begin{array}{ccc} 
                                        a_3 & j_4 & g''_4 \\
                                        k_5 & j_I & k_4
                                        \end{array} \right\}
	                	\left\{ \begin{array}{ccc} 
                                	a_3 & j_4 & a_4 \\
                                	k_5 & j_I & k_4
                                	\end{array} \right\} }
			      _{\stackrel{\ref{orthogonality relation}}{=}~\delta_{g''_4a_4}} \nonumber\\
             &\hspace{3cm} =& \delta_{g''_4a_4} \nonumber
 	\ea
        \end{footnotesize} \be \ee
   \end{itemize}

   In this way we successively carry out all the summations until the last 
step:
   \begin{itemize}
   \item last step
      \subitem summation over $h_{I-1}$: 
        \begin{footnotesize}
	\ba
	  \lefteqn{\sum_{h_{I-1}}	      
	     \overbrace{(-1)^{g''_{I-1}+1+k_{I-1}-\sum_{n=1}^{I-2}j_n-h_{I-1}}A(h_{I-1},k_{I-1})       
	                     \left\{ \begin{array}{ccc} 
                                     j_I & k_{I-1} & g''_{I-2} \\
                                     h_{I-1} & j_I &  1
                                     \end{array} \right\}}^{D_{I-3}} \times }\nonumber\\
           \lefteqn{ ~~~\times
	     \overbrace{(-1)^{h_{I-1}+k_{I-1}+1}A(h_I,k_I)
	                     \left\{ \begin{array}{ccc} 
                                     j_{I-1} & h_{I-1} & h_I \\
                                        1    & k_I     & k_{I-1}
                                     \end{array} \right\}}^{A_{I-2}}
	     \overbrace{(-1)^{h_{I-1}-k_{I-1}}A(h_{I-1},g''_{I-1})
	                     \left\{ \begin{array}{ccc} 
                                     j_I     & g''_{I-2} & h_{I-1} \\
                                     j_{I-1} & g''_I     & g''_{I-1}
                                     \end{array} \right\}}^{B_{I-2}} = }\nonumber\\\nonumber\\\nonumber\\ 
             &\hspace{0cm} =&
	          (-1)^{g''_{I-2}-\sum_{n=1}^{I-2}j_n+k_{I-1}}A(k_{I-1},g''_{I-1})A(h_I,k_I)~\times\nonumber\\
             &~&\times ~~~~\underbrace{ \sum_{h_{I-1}}(-1)^{h_{I-1}}(2h_{I-1}+1)
	                     \left\{ \begin{array}{ccc} 
                                     j_I     & k_{I-1} & g''_{I-2} \\
                                     h_{I-1} & j_I     &    1
				     \end{array} \right\}
	                     \left\{ \begin{array}{ccc} 
                                     j_{I-1} & h_{I-1} & h_I \\
                                        1    & k_I     & k_{I-1}
				     \end{array} \right\}
	                     \left\{ \begin{array}{ccc} 
                                     j_I     & g''_{I-2} & h_{I-1} \\
                                     j_{I-1} & g''_I     & g''_{I-1} 
                                     \end{array} \right\} } 
	                   _{\begin{array}{l}
			       \stackrel{\ref{symmetry1},\ref{symmetry2}}{=}
			        \sum_{h_{I-1}}(-1)^{h_{I-1}}(2h_{I-1}+1)
	                	\left\{ \begin{array}{ccc} 
                                	g''_{I-2} & k_{I-1} & j_I \\
                                	    1     & j_I     & h_{I-1}
					\end{array} \right\}
	                	\left\{ \begin{array}{ccc} 
                                	k_{I-1} & k_I     & j_{I-1} \\
                                	h_I     & h_{I-1} &   1
					\end{array} \right\}
	                	\left\{ \begin{array}{ccc} 
                                	g''_{I-2} & j_{I-1} & g''_{I-1} \\
                                	h_I       & j_I     & h_{I-1} 
                                	\end{array} \right\} \\	       
			       \stackrel{\ref{EBI}}{=} 
                                  (-1)^{-(\overbrace{k_I+j_I+1+g''_{I-1}}^{integer}
                                           +k_{I-1}+g''_{I-2}+h_I+j_{I-1}+j_I)}
	                        \left\{ \begin{array}{ccc} 
                                         k_I       & k_{I-1}   & j_{I-1} \\
                                         g''_{I-2} & g''_{I-1} & j_I
                                        \end{array} \right\}
	                        \left\{ \begin{array}{ccc} 
                                         j_I & k_I & g''_{I-1} \\
                                         h_I & j_I &    1
                                     \end{array} \right\} \\ 
			       \stackrel{\ref{integer conditions}}{=}
			        (-1)^{k_I+j_I+1+g''_{I-1}-k_{I-1}-g''_{I-2}-h_I-j_{I-1}-j_I}
	                        \left\{ \begin{array}{ccc} 
                                         k_I       & k_{I-1}   & j_{I-1} \\
                                         g''_{I-2} & g''_{I-1} & j_I
                                        \end{array} \right\}
	                        \left\{ \begin{array}{ccc} 
                                         j_I & k_I & g''_{I-1} \\
                                         h_I & j_I &    1
                                     \end{array} \right\} \\ 
		             \end{array}} \nonumber\\ \nonumber\\
             &\hspace{0cm} = &
	                     \underbrace{
                             (-1)^{g''_{I-1}+1+k_I\sum_{n=1}^{I-1}j_n-h_I}A(h_I,k_I) 
	                     \left\{ \begin{array}{ccc} 
                                     j_I & k_I & g''_{I-1} \\
                                     h_I & j_I &    1
                                     \end{array} \right\} }_{D_{I-2}}
			     \underbrace{	     
	                     A(k_{I-1},g''_{I-1})
			     \left\{ \begin{array}{ccc} 
                                     k_I       & k_{I-1}   & j_{I-1} \\
                                     g''_{I-2} & g''_{I-1} & j_I
                                     \end{array} \right\} }_{E_{I-2}} \nonumber
 	\ea
        \end{footnotesize} \be \label{summation h_(I-1)}\ee
      \subitem summation over $k_{I-1}$: 
        \begin{footnotesize}
	\ba
	  \lefteqn{
	  \sum_{k_{I-1}}	      
            \overbrace{A(k_{I-1},g''_{I-1})
			     \left\{ \begin{array}{ccc} 
                                     k_I       & k_{I-1} & j_{I-1} \\
                                     g''_{I-2} & g''_{I-1} & j_I
                                     \end{array} \right\} }^{E_{I-2}}
	     \overbrace{A(k_{I-1},a_{I-1})       
	                     \left\{ \begin{array}{ccc} 
                                     j_I     & a_{I-2} & k_{I-1} \\
                                     j_{I-1} & k_I     & a_{I-1}
                                     \end{array} \right\}}^{C_{I-2}} = } \nonumber\\ \nonumber \\
             &\hspace{3cm} \stackrel{\ref{symmetry1},\ref{symmetry2}}{=}&
	                     \frac{A(a_{I-1},g''_{I-1})}{(2a_{I-1}+1)} 
			     \underbrace{\sum_{k_{I-1}}(2k_{I-1}+1)(2a_{I-1}+1)
	                        \left\{ \begin{array}{ccc} 
                                        a_{I-2} & j_{I-1} & g''_{I-1} \\
                                        a_I     & j_I     & k_{I-1}
                                        \end{array} \right\}
	                	\left\{ \begin{array}{ccc} 
                                	a_{I-2} & j_{I-1} & a_{I-1} \\
                                	a_I     & j_I     & k_{I-1}
                                	\end{array} \right\} }
			      _{\stackrel{\ref{orthogonality relation}}{=}~\delta_{g''_{I-1}a_{I-1}}} \nonumber\\
             &\hspace{3cm} =& \delta_{g''_{I-1}a_{I-1}} \nonumber
 	\ea
        \end{footnotesize} \be \label{summation k_(I-1)} \ee
	Here we have used that $g''_{I-2}=a_{I-2}$ resulting from the summation over $k_{I-2}$
	and additionally the fact that $a_I=k_I$ from (\ref{kopplungsschemata}).
   \end{itemize}

   At the end of the $\bf part~I$-summation over $h_2 \ldots 
h_{I-1}$,$k_2 \ldots k_{I-1}$ we can 
   now summarize the remaining  terms  of (\ref{zwischenschritt})
resulting in the term $D_{I-2}$ from the
   summation over $h_{I-1}$ in (\ref{summation h_(I-1)})\\
   With $g''_{I-1}=a_{I-1}$ from (\ref{summation k_(I-1)}) , $h_I=g''_I$, $k_I=a_I$ from
   (\ref{kopplungsschemata}) we obtain the drastically simplified formula:
    \ba  
       \lefteqn{\sum_{h_2 \ldots h_{I-1} \atop k_2 \ldots k_{I-1}} 
        \Biggm[\prod_{n=2}^{I}M_n \prod_{m=I+1}^{J-1}\tilde{M}_m 
	\times  (-1)^{a_J} \left\{ \begin{array}{ccc} 
                                  a_J & j_J     & g''_{J-1}\\
                                    1 & a_{J-1} & j_J
                            \end{array} \right\} \Biggm] \times ~remaining~terms~=} \nonumber\\
       &\hspace{0cm}=&
                (-1)^{a_{I-1}+1}(-1)^{a_I-g''_I)}(-1)^{-\sum_{n=1}^{I-1}j_n} A(g''_I,a_I)
            	\left\{ \begin{array}{ccc} 
                      	j_I   & a_I & a_{I-1} \\
                      	g''_I & j_I &  1
                      	\end{array} \right\} \times
		\prod_{m=I+1}^{J-1}\tilde{M}_m 
	             \times  (-1)^{a_J} \left\{ \begin{array}{ccc} 
                                  a_J & j_J     & g''_{J-1}\\
                                    1 & a_{J-1} & j_J
                            \end{array} \right\}  \nonumber 
     \ea\nonumber
    \be \label{part I} \ee
    Now the summation for \fbox{$\bf part~II$ in (\ref{zwischenschritt})} is the last task in order to
    complete our calculation:
    Let us write down this expression (again suppressing the $(-1)^{g''_i-a_i}$-signs) with the product $\prod_{n=2}^{J-1}\delta_{h_nk_n}$ evaluated (this
    cancels the summation over $k_2$ and all the exponents in $(-1)^{h_i-k_i}$):
    
    \begin{footnotesize}
    \ba
     \lefteqn{\sum_{h_2 \ldots h_{I-1} \atop k_2 \ldots k_{I-1}} 
     \Biggm[N \prod_{n=2}^{J-1}\delta_{h_n k_n}\Biggm]\times~remaining~terms=} \nonumber\\ \nonumber\\
             & \begin{array}{lcll}
               ~~~=\big[ j_I(j_I+1)+j_J(j_J+1)\big]   
	       \displaystyle\sum\limits_{h_2 \ldots h_{I-1}} 
  	      &\times 
		 &\underbrace{ A(h_2,g''_2)
		 \left\{ \begin{array}{ccc} 
                	  j_I & j_1 & h_2\\
                	  j_2 & h_3 & g''_2
        	 \end{array} \right\}}_{B_1} 
		 &\underbrace{ A(h_2,a_2)
		 \left\{ \begin{array}{ccc} 
                	  j_I & j_1 & h_2\\
                	  j_2 & k_3 & a_2
        	 \end{array} \right\}}_{C_1} \\
        	& \times
        	 &\underbrace{A(h_3,g''_3)
		 \left\{ \begin{array}{ccc} 
                	  j_I & g''_2 & h_3\\
                	  j_3 & h_4 & g''_3
        	 \end{array} \right\}}_{B_2}
		 &\underbrace{A(h_3,a_3)
		 \left\{ \begin{array}{ccc} 
                	  j_I & a_2 & h_3\\
                	  j_3 & k_4 & a_3
        	 \end{array} \right\}}_{C_2} \\
        	 &\times
		 &\underbrace{A(h_4,g''_4)
		 \left\{ \begin{array}{ccc} 
                	  j_I & g''_3 & h_4\\
                	  j_4 & h_5 & g''_4
        	 \end{array} \right\}}_{B_3} 
		 &\underbrace{A(h_4,a_4)
		 \left\{ \begin{array}{ccc} 
                	  j_I & a_3 & h_4\\
                	  j_4 & k_5 & a_4
        	 \end{array} \right\}}_{C_3} \\

        	&\vdots &~~~~~~~~~~~~~~\vdots & ~~~~~~~~~~~~~~\vdots\\
		&\vdots &~~~~~~~~~~~~~~\vdots & ~~~~~~~~~~~~~~\vdots\\
        	&\times
		 &\underbrace{A(h_{I-1},2g'_{I-1})
		 \left\{ \begin{array}{ccc} 
                	  j_I & g''_{I-2} & h_{I-1}\\
                	  j_{I-1} & g''_I & g'_{I-1}
        	 \end{array} \right\}}_{B_{I-2}}
		 &\underbrace{A(h_{I-1},a_{I-1})
		 \left\{ \begin{array}{ccc} 
                	  j_I & a_{I-2} & h_{I-1}\\
                	  j_{I-1} & a_I & a_{I-1}
        	 \end{array} \right\}}_{C_{I-2}} \nonumber
            \end{array} 	 		    
   \ea 
   \end{footnotesize}\be \label{symmetrischer Teil}\ee 
     
   Looking at (\ref{symmetrischer Teil}) we can see, that every summation gives rise to an orthogonality
   relation between the $6j$-symbols as follows: \\
   Every sum in (\ref{symmetrischer Teil}) has the following form (again using the conventions in
   (\ref{kopplungsschemata})~):
    \begin{footnotesize}
   \ba
     \lefteqn{\sum_{h_n} A(h_n,g''_n)\left\{ \begin{array}{ccc} 
                	                      j_I & g''_{n-1} & h_n\\
                	                      j_n & h_{n+1}   & g''_n
			                     \end{array} \right\}
                           A(h_n,a_n)\left\{ \begin{array}{ccc} 
                	              j_I & a_{n-1} & h_n\\
                	              j_n & h_{n+1} & a_n
			             \end{array} \right\} =} \nonumber\\
     & \hspace{4cm}= & \frac{A(g''_i,a_i)}{(2a_i+1)}\sum_{h_n} (2h_i+1)(2a_i+1) 
                          \left\{ \begin{array}{ccc} 
                	              j_I & g''_{n-1} & h_n\\
                	              j_n & h_{n+1}   & g''_n
			  \end{array} \right\}
                          \left\{ \begin{array}{ccc} 
                	              j_I & a_{n-1} & h_n\\
                	              j_n & h_{n+1} & a_n
			  \end{array} \right\}  \nonumber\\
     & \hspace{4cm}\stackrel{\ref{orthogonality relation}}{=} & \delta_{g''_ia_i} \nonumber
   \ea
   \end{footnotesize}\be \ee
   One has to start with the summation over $h_2$. This gives $\delta_{g''_2a_2}$.\\
   Secondly the summation over $h_3$ is carried out by using the result $\delta_{g''_2a_2}$ from the summation
   before. In this way one can step by step sum over all $h_i$ up to $h_{I-1}$.\\
   The final result for $part~II$ in (\ref{zwischenschritt}) is:
       
   \ba \label{part II}
     \sum_{h_2 \ldots h_{I-1} \atop k_2 \ldots k_{I-1}} 
     \Biggm[N \prod_{n=2}^{J-1}\delta_{h_n k_n}\Biggm]\times~remaining~terms
     &=& \big[ j_I(j_I+1)+j_J(j_J+1)\big]\prod_{n=2}^{J-1}\delta_{g''_na_n}  
   \ea
   \pagebreak
   
   Now we have solved the problem posed in (\ref{basic structure in 6j symbols}) and can write down the
   remarkably simplified expression by using the results of (\ref{part I}), (\ref{part II}). Note, that we
   have inserted $\delta$-expressions coming from lemma \ref{lemma51}.\vspace{2mm}
   \fbox{\parbox{15.5cm}{
   \ba \label{precalculation}
     \lefteqn{\sum_{\vec{g}(IJ)}g_2(IJ)\bigm(g_2(IJ)+1\bigm) <\vec{g}(IJ)|\vec{g}''(12)>
	            <\vec{g}(IJ)|\vec{a}(12)>~=} \nonumber\\ \nonumber\\
     &\hspace{2cm}=&\frac{1}{2}(-1)^{j_J-j_I-\sum_{n=1}^{I-1}j_n-\sum_{m=I+1}^{J-1}j_m}X(j_I,j_J)^{\frac{1}{2}}
     \times\nonumber\\
     & &~~~~\times~ (-1)^{a_{I-1}}(-1)^{a_I-g''_I}(-1)^{-\sum_{n=1}^{I-1}j_n}(-1)^{+1}\sqrt{(2g''_I+1)(2a_I+1)}		    
               \left\{ \begin{array}{ccc} 
       	               a_{I-1} & a_I & j_I\\
                          1    & j_I & g''_I
		      \end{array} \right\} \nonumber\\
     & &~~~~\times~ \prod_{n=I+1}^{J-1}\sqrt{(2g''_n)(2a_n+1)} (-1)^{g''_{n-1}+a_{n-1}+1}		    
               \left\{ \begin{array}{ccc} 
       	               j_n  & g''_{n-1} & g''_n\\
                          1    & a_n & a_{n-1}
		      \end{array} \right\} \nonumber\\
     & &~~~~\times~ (-1)^{a_J}		    
               \left\{ \begin{array}{ccc} 
       	               a_J  & j_J     & g''_{J-1}\\
                        1   & a_{J-1} & j_J
 	      \end{array} \right\} 
  	  \prod_{n=2}^{I-1}\delta_{g''_na_n} \prod_{n=J}^{N}\delta_{g''_na_n} \nonumber\\
     & & +\big[ j_I(j_I+1)+j_J(j_J+1)\big]\prod_{n=2}^{N}\delta_{g''_na_n} 	       
   \ea }}\\[2mm]

For configurations $(I,J)$ one has to take all terms of (\ref{precalculation}), which are in suitable
limits, e.g. if $J=I+1$ then the product $\prod\limits_{n=I+1}^{J-1}\ldots$ is not to be taken into account.
Notice that for special configurations $I<J$ certain terms drop out, e.g. if $J=I+1$ then $\prod\limits_{n=I+1}^{J-1}$ is not taken into account. 
\pagebreak

Let us for clarity explicitly discuss four special cases of the edge-labelling $I,J$, namely 
\big[$I=1$, $J$ arbitrary\big], \big[ $I=1$, $J=2$\big], \big[$I=2$, $J$ arbitrary\big] and
\big[ $I=2$, $J=3$\big].

We will display for every case the parts remaining from (\ref{basic structure in 6j symbols}).\\  
Note, that again $A(x,y)=\sqrt{(2x+1)(2y+1)}$, and the conventions of (\ref{kopplungsschemata}) 
are kept in mind~:

\subsubsection{\fbox{$I=1$, $J$ arbitrary}}
    \begin{samepage}
    
    \begin{footnotesize} 
      \[ \begin{array}{ccccrcr} 
	 \lefteqn{\sum_{\vec{g}(IJ)}g_2(IJ)\bigm(g_2(IJ)+1\bigm) <\vec{g}(IJ)|\vec{g}''(12)>
	               <\vec{g}(IJ)|\vec{a}(12)>~=} 
	 \nonumber\\
	 \nonumber\\
	 \nonumber\\

           \lefteqn{=\sum_{\vec{g}(IJ)}g_2(IJ)\bigm(g_2(IJ)+1\bigm)~\times}
	   \\
	   \\

	  \\& & &\times 
	  &(-1)^{g''_{I+1}} A(g_{I+1},g''_{I+1})
	   \left\{ \begin{array}{ccc} 
                    j_J & g''_I & g_{I+1}\\
                    j_{I+1} & g_{I+2} & g''_{I+1}
           \end{array} \right\} 
	  & &(-1)^{-a_{I+1}} A(g_{I+1},a_{I+1})
	   \left\{ \begin{array}{ccc} 
                    j_J & a_I & g_{I+1}\\
                    j_{I+1} & g_{I+2} & a_{I+1}
           \end{array} \right\} \\

	  \\& & &\times 
	  &(-1)^{g''_{I+2}} A(g_{I+2},g''_{I+2})
	   \left\{ \begin{array}{ccc} 
                    j_J & g''_{I+1} & g_{I+2}\\
                    j_{I+2} & g_{I+3} & g''_{I+2}
           \end{array} \right\} 
	  & & (-1)^{-a_{I+2}} A(g_{I+2},a_{I+2})
	   \left\{ \begin{array}{ccc} 
                    j_J & a_{I+1} & g_{I+2}\\
                    j_{I+2} & g_{I+3} & a_{I+2}
           \end{array} \right\} \\

	  & & &\vdots &\vdots~~~~~~~~~~~~& &\vdots~~~~~~~~~~~~\\
	  & & &\vdots &\vdots~~~~~~~~~~~~& &\vdots~~~~~~~~~~~~\\

	  & & &\times 
	  &(-1)^{g''_{J-1}} A(g_{J-1},g''_{J-1})
	   \left\{ \begin{array}{ccc} 
                    j_J & g''_{J-2} & g_{J-1}\\
                    j_{J-1} & g_J & g''_{J-1}
           \end{array} \right\} 
	  & &(-1)^{-a_{J-1}} A(g_{J-1},a_{J-1})
	   \left\{ \begin{array}{ccc} 
                    j_J & a_{J-2} & g_{J-1}\\
                    j_{J-1} & g_J & a_{J-1}
           \end{array} \right\} \\
          \\
	  \\
          & & =&\multicolumn{4}{l}{\frac{1}{2}(-1)^{J_J+j_1-\sum_{n=2}^{J-1}j_n} X(j_1,j_J)^{\frac{1}{2}} 
	      \displaystyle\prod_{n=2}^{J-1} A(g''_n,a_n)(-1)^{g''_{n-1}+a_{n-1}+1}
	      
	  \left\{ \begin{array}{ccc} 
                    j_n & g''_{n-1} & g_{n}\\
                      1 & a_n & a_{n-1}
           \end{array} \right\} \times }\\
	  \\
	  &&&~~~~\times&\multicolumn{1}{l}{ (-1)^{a_J}
	   \left\{ \begin{array}{ccc} 
                    j_J & a_{J-2} & g_{J-1}\\
                    j_{J-1} & g_J & a_{J-1}
           \end{array} \right\} \displaystyle\prod_{n=J}^{N}\delta_{g''_na_n}}\\
	  \\ 
	  &&&+&\multicolumn{1}{l}{\big[j_1(j_1+1)+j_J(j_J+1) \big]
	       \displaystyle\prod_{n=2}^N \delta_{g''_na_n}} 
	      
       \end{array} \]  
    \end{footnotesize} 
    \\
    \be \label{I=1 J arbitrary}\ee
    \end{samepage}

\subsubsection{\fbox{$I=1$, $J=2$}}
    \begin{samepage}
    
    \begin{footnotesize} 
      \[ \begin{array}{ccccrcr} 
	 \lefteqn{\sum_{\vec{g}(IJ)}g_2(IJ)\bigm(g_2(IJ)+1\bigm) <\vec{g}(IJ)|\vec{g}''(12)>
	               <\vec{g}(IJ)|\vec{a}(12)>~} 
	 \nonumber\\
	 \nonumber\\
	 \nonumber\\
%
%
%
          & =& &\multicolumn{4}{l}{a_2(a_2+1)\displaystyle\prod_{n=2}^{N}\delta{g''_na_n}{}}

       \end{array} \]  
    \end{footnotesize} 
    \be \label{I=1 J=2}\ee
    \end{samepage}

\subsubsection{\fbox{$I=2$, $J$ arbitrary}}
    \begin{footnotesize} 
      \[ \begin{array}{ccccrcr} 
	 \lefteqn{\sum_{\vec{g}(IJ)}g_2(IJ)\bigm(g_2(IJ)+1\bigm) <\vec{g}(IJ)|\vec{g}''(12)>
	               <\vec{g}(IJ)|\vec{a}(12)>~=} 
	 \nonumber\\
	 \nonumber\\
	 \nonumber\\

           \lefteqn{=\sum_{\vec{g}(IJ)}g_2(IJ)\bigm(g_2(IJ)+1\bigm)~\times}
	   \\
	   \\

	  \\& & &\times 
	  &(-1)^{g''_I} A(g_I,g''_I)
	   \left\{ \begin{array}{ccc} 
                    j_J & h_{I-1} & g_I\\
                    j_{I-1} & g_{I+1} & g''_I
           \end{array} \right\} 
	  & &(-1)^{-a_I} A(g_I,a_I)
	   \left\{ \begin{array}{ccc} 
                    j_J & k_{I-1} & g_I\\
                    j_{I-1} & g_{I+1} & a_I
           \end{array} \right\} \\

	  \\& & &\times 
	  &(-1)^{g''_{I+1}} A(g_{I+1},g''_{I+1})
	   \left\{ \begin{array}{ccc} 
                    j_J & g''_I & g_{I+1}\\
                    j_{I+1} & g_{I+2} & g''_{I+1}
           \end{array} \right\} 
	  & &(-1)^{-a_{I+1}} A(g_{I+1},a_{I+1})
	   \left\{ \begin{array}{ccc} 
                    j_J & a_I & g_{I+1}\\
                    j_{I+1} & g_{I+2} & a_{I+1}
           \end{array} \right\} \\

	  \\& & &\times 
	  &(-1)^{g''_{I+2}} A(g_{I+2},g''_{I+2})
	   \left\{ \begin{array}{ccc} 
                    j_J & g''_{I+1} & g_{I+2}\\
                    j_{I+2} & g_{I+3} & g''_{I+2}
           \end{array} \right\} 
	  & & (-1)^{-a_{I+2}} A(g_{I+2},a_{I+2})
	   \left\{ \begin{array}{ccc} 
                    j_J & a_{I+1} & g_{I+2}\\
                    j_{I+2} & g_{I+3} & a_{I+2}
           \end{array} \right\} \\

	  & & &\vdots &\vdots~~~~~~~~~~~~& &\vdots~~~~~~~~~~~~\\
	  & & &\vdots &\vdots~~~~~~~~~~~~& &\vdots~~~~~~~~~~~~\\

	  & & &\times 
	  &(-1)^{g''_{J-1}} A(g_{J-1},g''_{J-1})
	   \left\{ \begin{array}{ccc} 
                    j_J & g''_{J-2} & g_{J-1}\\
                    j_{J-1} & g_J & g''_{J-1}
           \end{array} \right\} 
	  & &(-1)^{-a_{J-1}} A(g_{J-1},a_{J-1})
	   \left\{ \begin{array}{ccc} 
                    j_J & a_{J-2} & g_{J-1}\\
                    j_{J-1} & g_J & a_{J-1}
           \end{array} \right\} \\
	   
          \\
	  \\
          & & =&\multicolumn{4}{l}{\frac{1}{2}(-1)^{J_J-j_2-j_1+1-\sum_{n=3}^{J-1}j_n} X(j_2,j_J)^{\frac{1}{2}}
	        A(a_2,g''_2)
	  \left\{ \begin{array}{ccc} 
                    j_1 & j_2 & g''_2\\
                      1 & a_2 & j_2
           \end{array} \right\} 
	   
	  \displaystyle\prod_{n=3}^{J-1} A(g''_n,a_n)(-1)^{g''_{n-1}+a_{n-1}+1}   
	  \left\{ \begin{array}{ccc} 
                    j_n & g''_{n-1} & g_{n}\\
                      1 & a_n & a_{n-1}
           \end{array} \right\} \times }\\
	  \\
	  &&&~~~~\times&\multicolumn{1}{l}{ (-1)^{a_J}
	   \left\{ \begin{array}{ccc} 
                    j_J & a_{J-2} & g_{J-1}\\
                    j_{J-1} & g_J & a_{J-1}
           \end{array} \right\} \displaystyle\prod_{n=J}^{N}\delta_{g''_na_n}}\\
	  \\ 
	  &&&+&\multicolumn{1}{l}{\big[j_2(j_2+1)+j_J(j_J+1) \big]
	       \displaystyle\prod_{n=2}^N \delta_{g''_na_n}} 
	
       \end{array} \]  
    \end{footnotesize} 
    \be \label{I=2 J arbitrary}\ee

\subsubsection{\fbox{$I=2$, $J=3$}}

    \begin{samepage}
    
    \begin{footnotesize} 
      \[ \begin{array}{ccccrcr} 
	 \lefteqn{\sum_{\vec{g}(IJ)}g_2(IJ)\bigm(g_2(IJ)+1\bigm) <\vec{g}(IJ)|\vec{g}''(12)>
	               <\vec{g}(IJ)|\vec{a}(12)>~=} 
	 \nonumber\\
	 \nonumber\\
	 \nonumber\\

           \lefteqn{=\sum_{\vec{g}(IJ)}g_2(IJ)\bigm(g_2(IJ)+1\bigm)~\times}
	   \\
	   \\
	  \\& & &\times 
	  &(-1)^{g''_I} A(g_I,g''_I)
	   \left\{ \begin{array}{ccc} 
                    j_J & h_{I-1} & g_I\\
                    j_{I-1} & g_{I+1} & g''_I
           \end{array} \right\} 
	  & &(-1)^{-a_I} A(g_I,a_I)
	   \left\{ \begin{array}{ccc} 
                    j_J & k_{I-1} & g_I\\
                    j_{I-1} & g_{I+1} & a_I
           \end{array} \right\} \\
	   \\
	   \\
          & & =&\multicolumn{4}{l}{\frac{1}{2}(-1)^{j_3-j_2-j_1} X(j_2,j_J)^{\frac{1}{2}}
	        A(a_2,g''_2)
	  \left\{ \begin{array}{ccc} 
                    j_1 & j_2 & g''_2\\
                      1 & a_2 & j_2
           \end{array} \right\} 
	   
	   (-1)^{a_3}
	   \left\{ \begin{array}{ccc} 
                    g_3 & j_3 & g''_2\\
                      1 & a_2 & j_3
           \end{array} \right\} } 
	  \\
	  \\ 
	  &&&+&\multicolumn{1}{l}{\big[j_2(j_2+1)+j_3(j_3+1) \big]
	       \displaystyle\prod_{n=2}^N \delta_{g''_na_n}} 
	
       \end{array} \]  
    \end{footnotesize} 
    \\
    \be \label{I=2 J=3}\ee
    \end{samepage}

  \pagebreak

\subsection{Explicit Formula for the Matrix Elements}

   After having finished the precalculations in the last section we are now in the position to evaluate the whole
   matrix element in (\ref{vorformel}) by using (\ref{precalculation}).
   \\
   We have now (again we abbreviate $A(x,y)=\sqrt{(2x+1)(2y+1)}$):
   
   \begin{footnotesize}
   \[ \begin{array}{rcl}
      \lefteqn{<\vec{a}(12)|\hat{q}_{IJK}|\vec{a}'(12)>=}\hspace{1cm}\nonumber\\\nonumber\\
      &=&\displaystyle\sum_{\vec{g}''(12)}
	\displaystyle\sum_{\vec{g}(IJ)} 
          g_2(IJ)\big(g_2(IJ)+1\big)<\vec{g}(IJ)|\vec{g}''(12)><\vec{g}(IJ)|\vec{a}(12)>\times \nonumber\\
      &&~~~~~~~~\times\displaystyle\sum_{\vec{g}(JK)} 
	 g_2(JK)\big(g_2(JK)+1\big)<\vec{g}(JK)|\vec{g}''(12)><\vec{g}(JK)|\vec{a}(12)> \nonumber\\\nonumber\\
      &&~~~~~~~~- \left[ \vec{a}(12) \Longleftrightarrow \vec{a}'(12) \right] 	\nonumber\\\nonumber\\ 
      &\stackrel{\ref{precalculation}}{=}&
      \displaystyle\sum_{\vec{g}''(12)}
      \Bigg[\Big[\frac{1}{2}(-1)^{j_J-j_I-\sum_{n=1}^{I-1}j_n-\sum_{m=I+1}^{J-1}j_m}(-1)^{a_{I-1}}(-1)^{a_I-g''_I}
            (-1)^{-\sum_{n=1}^{I-1}j_n}(-1)^{+1} \times \nonumber\\
      &&~~~~\times~ X(j_I,j_J)^{\frac{1}{2}}A(g''_I,a_I)		    
        	      \left\{ \begin{array}{ccc} 
       	        	       a_{I-1} & a_I & j_I\\
                                  1    & j_I & g''_I
		       \end{array} \right\} 
            \displaystyle\prod_{n=I+1}^{J-1}A(g'',a_n)(-1)^{g''_{n-1}+a_{n-1}+1}		    
        	      \left\{ \begin{array}{ccc} 
       	        	       j_n  & g''_{n-1} & g''_n\\
                               1    & a_n       & a_{n-1}
		      \end{array} \right\} \times \nonumber\\
      &&~~~~\times~ (-1)^{a_J}		    
        	      \left\{ \begin{array}{ccc} 
       	        	       a_J  & j_J     & g''_{J-1}\\
                	        1   & a_{J-1} & j_J
 	              \end{array} \right\} 
  	   \displaystyle\prod_{n=2}^{I-1}\delta_{g''_na_n} \displaystyle\prod_{n=J}^{N}\delta_{g''_na_n} \Big]
        +\Big[\big[ j_I(j_I+1)+j_J(j_J+1)\big]\displaystyle\prod_{n=2}^{N}\delta_{g''_na_n}\Big] \Bigg] \times
              \nonumber\\\nonumber\\	       
      &&\times
      \Bigg[\Big[\frac{1}{2}(-1)^{j_K-j_J-\sum_{n=1}^{J-1}j_n-\sum_{m=J+1}^{K-1}j_m}(-1)^{a_{J-1}}(-1)^{a'_I-g''_I}
            (-1)^{-\sum_{n=1}^{J-1}j_n}(-1)^{+1} \times \nonumber\\
      &&~~~~\times~ X(j_J,j_K)^{\frac{1}{2}}A(g''_J,a'_J)		    
        	      \left\{ \begin{array}{ccc} 
       	        	       a'_{J-1} & a'_J& j_J\\
                                  1     & j_J  & g''_J
		       \end{array} \right\} 
            \displaystyle\prod_{n=J+1}^{K-1}A(g'',a'_n)(-1)^{g''_{n-1}+a'_{n-1}+1}		    
        	      \left\{ \begin{array}{ccc} 
       	        	       j_n  & g''_{n-1} & g''_n\\
                               1    & a'_n      & a'_{n-1}
		      \end{array} \right\} \times \nonumber\\
      &&~~~~\times~ (-1)^{a'_K}		    
        	      \left\{ \begin{array}{ccc} 
       	        	       a'_K & j_K      & g''_{K-1}\\
                	        1   & a'_{K-1} & j_K
 	              \end{array} \right\} 
  	   \displaystyle\prod_{n=2}^{J-1}\delta_{g''_na'_n} \displaystyle\prod_{n=K}^{N}\delta_{g''_na'_n} \Big]
        +\Big[\big[ j_J(j_J+1)+j_K(j_K+1)\big]\displaystyle\prod_{n=2}^{N}\delta_{g''_na'_n}\Big] 
	\Bigg] \times \nonumber \\\nonumber\\ 
      && - \left[ \vec{a}(12) \Longleftrightarrow \vec{a}'(12) \right] \nonumber\\\nonumber\\

     
      &=&\Bigg[\Big[\frac{1}{2}(-1)^{j_J-j_I-\sum_{n=1}^{I-1}j_n-\sum_{m=I+1}^{J-1}j_m}(-1)^{a_{I-1}}
                (-1)^{a_I-a'_I} (-1)^{-\sum_{n=1}^{I-1}j_n}(-1)^{+1} \times \nonumber\\
      &&~~~~\times~ X(j_I,j_J)^{\frac{1}{2}}A(a'_I,a_I)		    
        	      \left\{ \begin{array}{ccc} 
       	        	       a_{I-1} & a_I & j_I\\
                                  1    & j_I & a'_I
		       \end{array} \right\} 
            \displaystyle\prod_{n=I+1}^{J-1}A(a'_n,a_n)(-1)^{a'_{n-1}+a_{n-1}+1}		    
        	      \left\{ \begin{array}{ccc} 
       	        	       j_n  & a'_{n-1} & a'_n\\
                               1    & a_n      & a_{n-1}
		      \end{array} \right\} \times \nonumber\\
      &&~~~~\times~ (-1)^{a_J}		    
        	      \left\{ \begin{array}{ccc} 
       	        	       a_J  & j_J     & a'_{J-1}\\
                	        1   & a_{J-1} & j_J
 	              \end{array} \right\} \Big]\times \nonumber\\
      &&
      \times\Big[\frac{1}{2}(-1)^{j_K-j_J-\sum_{n=1}^{J-1}j_n-\sum_{m=J+1}^{K-1}j_m}(-1)^{a_{J-1}}(-1)^{a'_I-a_I}
            (-1)^{-\sum_{n=1}^{J-1}j_n}(-1)^{+1} \times \nonumber\\
      &&~~~~\times~ X(j_J,j_K)^{\frac{1}{2}}A(a'_J,a'_J)		    
        	      \left\{ \begin{array}{ccc} 
       	        	       a'_{J-1} & a'_J& j_J\\
                                  1     & j_J  & a'_J
		       \end{array} \right\} 
            \displaystyle\prod_{n=J+1}^{K-1}A(a_n,a'_n)(-1)^{a_{n-1}+a'_{n-1}+1}		    
        	      \left\{ \begin{array}{ccc} 
       	        	       j_n  & a_{n-1} & a_n\\
                               1    & a'_n    & a'_{n-1}
		      \end{array} \right\} \times \nonumber\\
      &&~~~~\times~ (-1)^{a'_K}		    
        	      \left\{ \begin{array}{ccc} 
       	        	       a'_K & j_K      & a_{K-1}\\
                	        1   & a'_{K-1} & j_K
 	              \end{array} \right\} \Big]\Bigg]
	 \displaystyle\prod_{n=2}^{I-1}\delta_{a_na'_n}
	 \displaystyle\prod_{n=K}^{N}\delta_{a_na'_n}  \nonumber\\\nonumber\\
	 
     & & +\Bigg[\Big[\frac{1}{2}(-1)^{j_J-j_I-\sum_{n=1}^{I-1}j_n-\sum_{m=I+1}^{J-1}j_m}(-1)^{a_{I-1}}(-1)^{a_I-a'_I}
            (-1)^{-\sum_{n=1}^{I-1}j_n}(-1)^{+1} \times \nonumber\\
      &&~~~~\times~ X(j_I,j_J)^{\frac{1}{2}}A(a'_I,a_I)		    
        	      \left\{ \begin{array}{ccc} 
       	        	       a_{I-1} & a_I & j_I\\
                                  1    & j_I & a'_I
		       \end{array} \right\} 
            \displaystyle\prod_{n=I+1}^{J-1}A(a'_n,a_n)(-1)^{a'_{n-1}+a_{n-1}+1}		    
        	      \left\{ \begin{array}{ccc} 
       	        	       j_n  & a'_{n-1} & a'_n\\
                               1    & a_n      & a_{n-1}
		      \end{array} \right\} \times \nonumber\\
      &&~~~~\times~ (-1)^{a_J}		    
        	      \left\{ \begin{array}{ccc} 
       	        	       a_J  & j_J     & a'_{J-1}\\
                	        1   & a_{J-1} & j_J
 	              \end{array} \right\} \Big]
            \Big[ j_J(j_J+1)+j_K(j_K+1)\Big]  \Bigg]
	    \displaystyle\prod_{n=2}^{I-1}\delta_{a_na'_n}		      
	    \displaystyle\prod_{n=J}^{N}\delta_{a_na'_n}   \nonumber\\\nonumber\\		      
\end{array} \]
\[ \begin{array}{rcl}     
&&+\Bigg[\Big[\frac{1}{2}(-1)^{j_K-j_J-\sum_{n=1}^{J-1}j_n-\sum_{m=J+1}^{K-1}j_m}(-1)^{a_{J-1}}(-1)^{a'_J-a_J}
            (-1)^{-\sum_{n=1}^{J-1}j_n}(-1)^{+1} \times \nonumber\\
      &&~~~~\times~ X(j_J,j_K)^{\frac{1}{2}}A(a_J,a'_J)		    
        	      \left\{ \begin{array}{ccc} 
       	        	       a'_{J-1} & a'_J& j_J\\
                                  1     & j_J  & a_J
		       \end{array} \right\} 
            \displaystyle\prod_{n=J+1}^{K-1}A(a_n,a'_n)(-1)^{a_{n-1}+a'_{n-1}+1}		    
        	      \left\{ \begin{array}{ccc} 
       	        	       j_n  & a_{n-1} & a_n\\
                               1    & a'_n    & a'_{n-1}
		      \end{array} \right\} \times \nonumber\\
      &&~~~~\times~ (-1)^{a'_K}		    
        	      \left\{ \begin{array}{ccc} 
       	        	       a'_K & j_K      & a_{K-1}\\
                	        1   & a'_{K-1} & j_K
 	              \end{array} \right\} \Big] 
  	    \Big[ j_I(j_I+1)+j_J(j_J+1)\Big]  \Bigg]
	    \displaystyle\prod_{n=2}^{J-1}\delta_{a_na'_n}		      
	    \displaystyle\prod_{n=K}^{N}\delta_{a_na'_n} ~~~~+    \nonumber\\\nonumber\\
    &&+\Bigg[\Big[ j_J(j_J+1)+j_K(j_K+1)\Big]\Big[j_I(j_I+1)+j_J(j_J+1)\Big]\Bigg]
         \displaystyle\prod_{n=2}^{N}\delta_{a_na'_n} \nonumber\\\nonumber\\\nonumber\\ 
      && - \Bigg[ \vec{a}(12) \Longleftrightarrow \vec{a}'(12) \Bigg] \nonumber
   \end{array} \] 
   \end{footnotesize}\be \label{Endformel fuer 3nj-Ausdruecke}\ee
   
   Here we have in the last step carried out the summation over $\vec{g}''_{12}$ by evaluating all the $\delta$-expressions.
   Finally we take a closer look at the
   symmetry properties which the four terms of the sum in (\ref{Endformel fuer 3nj-Ausdruecke}) have 
   with respect to
   the interchange $[\vec{a}(12) \Longleftrightarrow \vec{a}'(12)]$, that is the simultaneous replacement
   $a_n \to a'_n, a'_n \to a_n$ for all $n=1 \ldots N$.
   \begin{itemize}
       \item{$\bf Fourth~term:$}
          Due to the product of $\delta_{a_na'_n}$-expressions this term is obviously symmetric under\\
	  $[\vec{a}(12) \Longleftrightarrow \vec{a}'(12)]$
       
       \item{$\bf Third~term:$} The symmetry is not obvious, we will show it part by part:
       \begin{itemize}  
	 \item[1)] In the $(-1)$-exponents $a_{J-1}=a'_{J-1}$, $a'_K=a_K$ by the 
	 $\delta$-expressions at  the end of the term. 
	 
	 \item[2)]The term $(-1)^{a'_J-a_J}$ is not changed by interchanging $a'_J\leftrightarrow a_J$, since 
	 $a'_J-a_J$ is an integer number and therefore the first formula of (\ref{integer exponents})
	 holds. The integer-statement is verified by the fact, that $\vec{a}_{12},\vec{a}'_{12}$ are
	 $Standard~Recoupling~Schemes$, defined as in (\ref{Def Standard basis}). 
	 Therefore they recouple all $j_1,\ldots,j_N$ successively together according to
	 theorem (\ref{Clebsh Gordan theorem}). Since $\vec{a}_{12},\vec{a}'_{12}$ contain temporarily 
	 recoupled angular momenta, namely
	 $a_k=a_k(a_{k-1},j_K), a'_k=a'_k(a'_{k-1},j_k)$ for which 
	 $|a_{k-1}-j_k| \le a_k \le a_{k-1}+j_k, |a'_{k-1}-j_k|\le a'_k \le a'_{k-1}+j_k$   
	 the integer- or half-integer- property of each component $a_k,a'_K$ is only caused by the order
	 the involved spins $j_1,\ldots,j_N$ are coupled together. Since this order is the same in  
	 $\vec{a}_{12},\vec{a}'_{12}$, the components $a_k,a'_K$ are simultaneously (for every 
	 $k=1\ldots N$) either half-integer or integer and therefore every sum or difference $a_k\pm a'_k$
	 is integer. 

	 \item[3)]The same statement as in $1)$ holds for $a'_{J-1},a'_K$ as the entries in the upper
	 left corner of the two $6j$-symbols before and after the product in the middle of them.
	 \item[4)]
	 In the product of the $6j$-symbols the exponent of the $(-1)$ contains only a sum $a_{n-1}+a'_{n-1}$ and is therefore 
	 symmetric. 
	 \item[5)]All prefactors $A(a_k,a'_k)$, $k=J \ldots K-1$ are symmetric, too. 
	 \item[6)]Finally all
	 $6j$-symbols in the third term turn out beeing symmetric, if we recall, that they are invariant
	 under an interchange of their last two columns (\ref{symmetry1}) 
	 followed by an interchange of the upper and lower arguments of their last two columns. 
	 (\ref{symmetry2}).    
       \end{itemize}

      \item{$\bf Second~term:$} The symmetry is again not obvious, so we will show it part by part:
      \begin{itemize}  
	 \item[1)] In the $(-1)$-exponents $a_{I-1}=a'_{I-1}$, $a'_J=a_J$ by the 
	 $\delta$-expressions at  the end of the term. 
	 
	 \item[2)]The term $(-1)^{a_I-a'_I}$ is again not changed by interchanging 
	 $a_I\leftrightarrow a'_I$, by the same integer statement as under point $2)$ in the Third term
	 discussion.

	 \item[3)]The same statement as in $1)$ holds for $a_{I-1},a_J$ as the entries in the upper
	 left corner of the two $6j$-symbols before and after the product in the middle of them.
	 \item[4)]
	 In the product of the $6j$-symbols the exponent of the $(-1)$ contains only a sum 
	 $a'_{n-1}+a_{n-1}$ and is therefore  symmetric. 
	 \item[5)]All prefactors $A(a'_k,a_k)$, $k=I \ldots J-1$ are symmetric, too. 
	 \item[6)]Finally (again) all
	 $6j$-symbols in the third term turn out being symmetric, if we recall, that they are invariant
	 under an interchange of their last two columns (\ref{symmetry1}) 
	 followed by an interchange of the upper and lower arguments of their last two columns. 
	 (\ref{symmetry2}).    
       \end{itemize}\pagebreak

       \item{$\bf First~term$}: This term is $\bf not$ symmetric under 
       $[\vec{a}(12) \Longleftrightarrow \vec{a}'(12)]$. Arguments similar 
to the ones we gave
       in the previous points now let us conclude that:
       \begin{itemize}
         \item[1)]The prefactors $(-1)^{a_J},(-1)^{a'_{J-1}}$ are not symmetric with respect to the
	 interchange of \linebreak  $a_J \to a'_J, a'_{J-1} \to a_{J-1}$.
	 \item[2)]The two $6j$-symbols containing $a_J, a'_{J-1}$ in the upper left corner are not
	 symmetric either.
       \end{itemize}
   \end{itemize}

The analysis has revealed that the last three terms in the sum of 
   (\ref{Endformel fuer 3nj-Ausdruecke}) are symmetric in $\vec{a}(12)$ 
and $\vec{a}'(12)$, hence after an 
antisymmetrization with respect to the interchange of
   $\vec{a}(12)$ and $\vec{a}'(12)$ only the first term in (\ref{Endformel 
fuer 3nj-Ausdruecke}) 
survives.\\ 
\\
Summarizing, we get by explicitly writing down all the terms occurring 
through the antisymmetrization:
    
   \begin{footnotesize}
   \[ \begin{array}{rcl}
      \lefteqn{<\vec{a}(12)|\hat{q}_{IJK}|\vec{a}'(12)>=}\hspace{1cm}\nonumber\\\nonumber\\
    
      &=&\Bigg[\Big[\frac{1}{2}(-1)^{j_J-j_I-\sum_{n=1}^{I-1}j_n-\sum_{m=I+1}^{J-1}j_m}(-1)^{a_{I-1}}
                (-1)^{a_I-a'_I} (-1)^{-\sum_{n=1}^{I-1}j_n}(-1)^{+1} \times \nonumber\\
      &&~~~~\times~ X(j_I,j_J)^{\frac{1}{2}}A(a'_I,a_I)		    
        	      \left\{ \begin{array}{ccc} 
       	        	       a_{I-1} & a_I & j_I\\
                                  1    & j_I & a'_I
		       \end{array} \right\} 
            \displaystyle\prod_{n=I+1}^{J-1}A(a'_n,a_n)(-1)^{a'_{n-1}+a_{n-1}+1}		    
        	      \left\{ \begin{array}{ccc} 
       	        	       j_n  & a'_{n-1} & a'_n\\
                               1    & a_n      & a_{n-1}
		      \end{array} \right\} \times \nonumber\\
      &&~~~~\times~ \underline{(-1)^{a_J}		    
                     \left\{ \begin{array}{ccc} 
       	        	       a_J  & j_J     & a'_{J-1}\\
                	        1   & a_{J-1} & j_J
 	              \end{array} \right\} } \Big]\times \nonumber\\
      &&
      \times\Big[\frac{1}{2}(-1)^{j_K-j_J-\sum_{n=1}^{J-1}j_n-\sum_{m=J+1}^{K-1}j_m}
            \underline{(-1)^{a'_{J-1}}}(-1)^{a'_I-a_I}
            (-1)^{-\sum_{n=1}^{J-1}j_n}(-1)^{+1} \times \nonumber\\
      &&~~~~\times~ X(j_J,j_K)^{\frac{1}{2}}A(a'_J,a_J)		    
            \underline{\left\{ \begin{array}{ccc} 
       	        	       a'_{J-1} & a'_J& j_J\\
                                  1     & j_J  & a'_J
		       \end{array} \right\} } 
            \displaystyle\prod_{n=J+1}^{K-1}A(a_n,a'_n)(-1)^{a_{n-1}+a'_{n-1}+1}		    
        	      \left\{ \begin{array}{ccc} 
       	        	       j_n  & a_{n-1} & a_n\\
                               1    & a'_n    & a'_{n-1}
		      \end{array} \right\} \times \nonumber\\
      &&~~~~\times~ (-1)^{a'_K}		    
        	      \left\{ \begin{array}{ccc} 
       	        	       a'_K & j_K      & a_{K-1}\\
                	        1   & a'_{K-1} & j_K
 	              \end{array} \right\} \Big]\Bigg]
	 \displaystyle\prod_{n=2}^{I-1}\delta_{a_na'_n}
	 \displaystyle\prod_{n=K}^{N}\delta_{a_na'_n}  \nonumber\\\nonumber\\

      &&-\Bigg[\Big[\frac{1}{2}(-1)^{j_J-j_I-\sum_{n=1}^{I-1}j_n-\sum_{m=I+1}^{J-1}j_m}(-1)^{a'_{I-1}}
                (-1)^{a'_I-a_I} (-1)^{-\sum_{n=1}^{I-1}j_n}(-1)^{+1} \times \nonumber\\
      &&~~~~\times~ X(j_I,j_J)^{\frac{1}{2}}A(a_I,a'_I)		    
        	      \left\{ \begin{array}{ccc} 
       	        	       a'_{I-1} & a'_I & j_I\\
                                  1    & j_I & a_I
		       \end{array} \right\} 
            \displaystyle\prod_{n=I+1}^{J-1}A(a_n,a'_n)(-1)^{a_{n-1}+a'_{n-1}+1}		    
        	      \left\{ \begin{array}{ccc} 
       	        	       j_n  & a_{n-1} & a_n\\
                               1    & a'_n      & a'_{n-1}
		      \end{array} \right\} \times \nonumber\\
      &&~~~~\times~ \underline{(-1)^{a_J}		    
        	    \left\{ \begin{array}{ccc} 
       	        	       a'_J  & j_J     & a_{J-1}\\
                	        1   & a'_{J-1} & j_J
 	              \end{array} \right\} } \Big]\times \nonumber\\
      &&
      \times\Big[\frac{1}{2}(-1)^{j_K-j_J-\sum_{n=1}^{J-1}j_n-\sum_{m=J+1}^{K-1}j_m}
            \underline{(-1)^{a_{J-1}}}(-1)^{a'_I-a_I}
            (-1)^{-\sum_{n=1}^{J-1}j_n}(-1)^{+1} \times \nonumber\\
      &&~~~~\times~ X(j_J,j_K)^{\frac{1}{2}}A(a_J,a'_J)		    
        	      \underline{\left\{ \begin{array}{ccc} 
       	        	       a_{J-1} & a_J & j_J\\
                                  1    & j_J & a'_J
		       \end{array} \right\} } 
            \displaystyle\prod_{n=J+1}^{K-1}A(a'_n,a_n)(-1)^{a'_{n-1}+a_{n-1}+1}		    
        	      \left\{ \begin{array}{ccc} 
       	        	       j_n  & a'_{n-1} & a'_n\\
                               1    & a_n    & a_{n-1}
		      \end{array} \right\} \times \nonumber\\
      &&~~~~\times~ (-1)^{a_K}		    
        	      \left\{ \begin{array}{ccc} 
       	        	       a_K & j_K      & a'_{K-1}\\
                	        1   & a_{K-1} & j_K
 	              \end{array} \right\} \Big]\Bigg]
	 \displaystyle\prod_{n=2}^{I-1}\delta_{a_na'_n}
	 \displaystyle\prod_{n=K}^{N}\delta_{a_na'_n}  \nonumber\\\nonumber\\
         
   \end{array} \] 
   \end{footnotesize}\be \label{fast am Ende}\ee
   
   Here we have underlined the terms, which are different with respect to the antisymmetrization
   (recall, that $(-1)^{a_I-a'_I}=(-1)^{-(a_I-a'_I)}=(-1)^{a'_I-a_I}$ because the exponent is an 
   integer number.
   All the other terms are symmetric under the interchange $\vec{a}(12)\Longleftrightarrow\vec{a}'(12)$, again
   because of the symmetry properties (\ref{symmetry1}), (\ref{symmetry2}) and the symmetrization by the
   $\delta$-expressions.
   \\
   Before we write down the final result we can simplify the exponents in (\ref{fast am Ende}):
   \begin{footnotesize}
   \ba   
      \lefteqn{(-1)^{-\sum_{n=1}^{I-1}j_n-\sum_{m=I+1}^{J-1}j_m}(-1)^{-\sum_{n=1}^{I-1}j_n}
      (-1)^{-\sum_{n=1}^{J-1}j_n-\sum_{m=J+1}^{K-1}j_m}(-1)^{-\sum_{n=1}^{J-1}j_n}~=} \nonumber\\\nonumber\\
      & \hspace{6cm} = & (-1)^{-2\sum_{n=1}^{I-1}j_n}(-1)^{-\sum_{n=I+1}^{J-1}j_n}
          (-1)^{-2\sum_{n=1}^{J-1}}(-1)^{-\sum_{n=J+1}^{K-1}j_n} \nonumber\\       
      & \hspace{6cm} = & (-1)^{-\sum_{n=I+1}^{J-1}j_n}
          (-1)^{-2\sum_{n=I}^{J-1}j_n}(-1)^{-\sum_{n=J+1}^{K-1}j_n} \nonumber\\
      & \hspace{6cm} = & (-1)^{2j_I-3\sum_{n=I+1}^{J-1}j_n-\sum_{n=J+1}^{K-1}j_n} \nonumber\\
      & \hspace{6cm} = & (-1)^{2j_I+\sum_{n=I+1}^{J-1}j_n-\sum_{n=J+1}^{K-1}j_n} \nonumber
   \ea
   \end{footnotesize}

\pagebreak   
   
Now we are able to give a closed expression of the matrix elements of 
$\hat{q}_{IJK}$ in terms of
standard recoupling scheme basis (\ref{vorformel}). In order to avoid 
confusion we assume that $I>1,\;J>I+1$, the remaining cases will be 
discussed below:\\
\\
{\bf THEOREM}
\hfill\vspace{5mm}\linebreak
\fbox{\parbox{16.0cm}{
\ba \label{Endgueltige Formel fuer das Matrixelement}
  \lefteqn{<\vec{a}|\hat{q}_{IJK}|\vec{a}'>~=}\hspace{0cm}\nonumber\\\nonumber\\
   &=&\frac{1}{4}(-1)^{j_K+j_I+a_{I-1}+a_K}(-1)^{a_I-a'_I}(-1)^{\sum_{n=I+1}^{J-1}j_n}
           (-1)^{-\sum_{p=J+1}^{K-1}j_p} \times \nonumber\\
   & &\times X(j_I,j_J)^{\frac{1}{2}}X(j_J,j_K)^{\frac{1}{2}} 
           \sqrt{(2a_I+1)(2a'_I+1)}\sqrt{(2a_J+1)(2a'_J+1)} \times \nonumber\\
   & &\times \left\{ \begin{array}{ccc} 
       	        	       a_{I-1} & j_I  & a_I\\
                	          1    & a'_I & j_I
 	              \end{array} \right\}	  
	\Bigg[ \prod_{n=I+1}^{J-1} \sqrt{(2a'_n+1)(2a_n+1)}(-1)^{a'_{n-1}+a_{n-1}+1}
	     \left\{ \begin{array}{ccc} 
       	        	       j_n & a'_{n-1}  & a'_n\\
                	        1  & a_n & a_{n-1}
 	              \end{array} \right\}
	\Bigg]	 \times \nonumber\\      	      
    & &	\times\Bigg[ \prod_{n=J+1}^{K-1} \sqrt{(2a'_n+1)(2a_n+1)}(-1)^{a'_{n-1}+a_{n-1}+1}
	     \left\{ \begin{array}{ccc} 
       	        	       j_n & a'_{n-1}  & a'_n\\
                	        1  & a_n & a_{n-1}
 	              \end{array} \right\}
	\Bigg]
	\left\{ \begin{array}{ccc} 
       	        	       a_{K} & j_K  & a_{K-1}\\
                	          1    & a'_{K-1} & j_K
 	              \end{array} \right\}   \times \nonumber\\
    & & \times\Bigg[(-1)^{a'_J+a'_{J-1}}
	\left\{ \begin{array}{ccc} 
       	        	       a_J & j_J     & a'_{J-1}\\
                	        1  & a_{J-1} & j_J
 	              \end{array} \right\}   
	\left\{ \begin{array}{ccc} 
       	        	       a'_{J-1} & j_J  & a'_J\\
                	          1     & a_J  & j_J
 	              \end{array} \right\}   
    -(-1)^{a_J+a_{J-1}}
	\left\{ \begin{array}{ccc} 
       	        	       a'_J & j_J     & a'_{J-1}\\
                	        1  & a_{J-1} & j_J
 	              \end{array} \right\}   
	\left\{ \begin{array}{ccc} 
       	        	       a_{J-1} & j_J  & a'_J\\
                	          1     & a_J  & j_J
 	              \end{array} \right\}\Bigg] \times \nonumber\\
     & &\times \prod_{n=2}^{I-1}\delta_{a_na'_n}\prod_{n=K}^{N}\delta_{a_na'_n}		      		      		      
\ea }}
\\\\[5mm]
 with $X(j_1,j_2)=2j_1(2j_1+1)(2j_1+2)2j_2(2j_2+1)(2j_2+2)$. Notice that 
 all still appearing $6j$-symbols are just abbreviations for the following 
 simple expressions in which no summations or producs (factorials) need to 
be carried out any longer as 
compared to (\ref{A1}) e.g. (using $s=a+b+c$):

\ba    \label{special 6j I}
	\left\{ \begin{array}{ccc} 
       	        	       a & b & c \\
                	       1 & c & b
 	              \end{array} \right\} 
    &=&(-1)^{s+1} \frac{2[b(b+1)c(c+1)-a(a+1)]}{[2b(2b+1)(2b+2)2c(2c+1)(2c+2)]^\frac{1}{2}}    
    \\ \label{special 6j II}
	\left\{ \begin{array}{ccc} 
       	        	       a & b   & c \\
                	       1 & c-1 & b
 	              \end{array} \right\} 
    &=& (-1)^s \Bigg[\frac{2(s+1)(s-2a)(s-2b)(s-2c+1)}{2b(2b+1)(2b+2)(2c-1)2c(2c+1)}\Bigg]^\frac{1}{2}   
    \\ \label{special 6j III}
	\left\{ \begin{array}{ccc} 
       	        	       a & b   & c \\
                	       1 & c-1 & b-1
 	              \end{array} \right\} 
    &=& (-1)^s \Bigg[\frac{s(s+1)(s-2a-1)(s-2a)}{(2b-1)2b(2b+1)(2c-1)2c(2c+1)}\Bigg]^\frac{1}{2}   
    \\ \label{special 6j IV}
	\left\{ \begin{array}{ccc} 
       	        	       a & b   & c \\
                	       1 & c-1 & b+1
 	              \end{array} \right\} 
    &=& (-1)^s \Bigg[\frac{(s-2b-1)(s-2b)(s-2c+1)(s-2c+2)}{(2b+1)(2b+2)(2b+3)(2c-1)2c(2c+1)}\Bigg]^\frac{1}{2}   
\ea
\\
{\bf Remark: Gauge invariance}\\
Recall that a vertex is said to be gauge invariant if the angular momenta coming from the edges $e_1,\ldots,e_N$ meeting in
the vertex $v$ are coupled to a resulting angular momentum $j=g_N=0$. Using the notation from 
(\ref{introduce recoupling scheme}) this means:
\be
  J~:=~\sum_{i=1}^NJ_i~=:~G_N~=~G_{N-1}+J_N~ {\stackrel{!}{=}~0}
\ee
This implies:
\ba
  G_{N-1}&=&G_{N-2}+J_{N-1}~\stackrel{!}{=}~-J_N\\
  \leadsto G_{N-2}&=&G_{N-3}+J_{N-2}~=~-J_N-J_{N-1}
\ea
But that gives $g_{N-1}=j_N$ and a certain restriction on which values
$g_{N-2}$ can take due to the Clebsh-Gordon-theorem (\ref{Clebsh Gordan theorem}) :
\be \label{restriction to g_{N-2}}
  \max{\left( \large|j_{N-2}-g_{N-3}\large|,\large|j_N-j_{N-1}\large| \right)} ~\le g_{N-2}~
  \le~ \min{\left( j_{N-2}+g_{N-3}~,~j_N+j_{N-1}\right)}
\ee
This relation will become useful when we consider gauge invariance later.

\pagebreak
As promised we now display the remaining cases of (\ref{Endgueltige 
Formel fuer das Matrixelement})
 explicitly. They are obtained, if some of the special
cases (\ref{I=1 J arbitrary}) $\ldots$ (\ref{I=2 J=3}) are involved.

\subsubsection{\fbox{I=1~~J,K arbitrary}}

\ba
  <\vec{a}|\hat{q}_{1JK}|\vec{a}'>
   &=&\frac{1}{4}(-1)^{j_K-j_1+a_K+1}(-1)^{\sum_{n=2}^{J-1}j_n}
           (-1)^{-\sum_{p=J+1}^{K-1}j_p} \times \nonumber\\
   & &\times X(j_1,j_J)^{\frac{1}{2}}X(j_J,j_K)^{\frac{1}{2}} 
           \sqrt{(2a_J+1)(2a'_J+1)} \times \nonumber\\
   & &\times 	  
	\Bigg[ \prod_{n=2}^{J-1} \sqrt{(2a'_n+1)(2a_n+1)}(-1)^{a'_{n-1}+a_{n-1}+1}
	     \left\{ \begin{array}{ccc} 
       	        	       j_n & a'_{n-1}  & a'_n\\
                	        1  & a_n & a_{n-1}
 	              \end{array} \right\}
	\Bigg]	 \times \nonumber\\      	      
    & &	\times\Bigg[ \prod_{n=J+1}^{K-1} \sqrt{(2a'_n+1)(2a_n+1)}(-1)^{a'_{n-1}+a_{n-1}+1}
	     \left\{ \begin{array}{ccc} 
       	        	       j_n & a'_{n-1}  & a'_n\\
                	        1  & a_n & a_{n-1}
 	              \end{array} \right\}
	\Bigg]
	\left\{ \begin{array}{ccc} 
       	        	       a_{K} & j_K  & a_{K-1}\\
                	          1    & a'_{K-1} & j_K
 	              \end{array} \right\}   \times \nonumber\\
    & & \times\Bigg[(-1)^{a'_J+a'_{J-1}}
	\left\{ \begin{array}{ccc} 
       	        	       a_J & j_J     & a'_{J-1}\\
                	        1  & a_{J-1} & j_J
 	              \end{array} \right\}   
	\left\{ \begin{array}{ccc} 
       	        	       a'_{J-1} & j_J  & a'_J\\
                	          1     & a_J  & j_J
 	              \end{array} \right\}   
    -(-1)^{a_J+a_{J-1}}
	\left\{ \begin{array}{ccc} 
       	        	       a'_J & j_J     & a'_{J-1}\\
                	        1  & a_{J-1} & j_J
 	              \end{array} \right\}   
	\left\{ \begin{array}{ccc} 
       	        	       a_{J-1} & j_J  & a'_J\\
                	          1     & a_J  & j_J
 	              \end{array} \right\}\Bigg] \times \nonumber\\
     & &\times \prod_{n=K}^{N}\delta_{a_na'_n}		      		      		      
\ea
\subsubsection{\fbox{I=1~J=2~~K arbitrary}}
\ba
  <\vec{a}|\hat{q}_{12K}|\vec{a}'>
   &=&\frac{1}{4}(-1)^{j_K-j_1-j_2+a_K+1}(-1)^{\sum_{n=2}^{J-1}j_n}
           (-1)^{-\sum_{p=J+1}^{K-1}j_p} \times \nonumber\\
   & &\times X(j_2,j_K)^{\frac{1}{2}} 
           \sqrt{(2a_2+1)(2a'_2+1)} 
	     \left\{ \begin{array}{ccc} 
       	        	       j_1 & j_2  & a_2\\
                	        1  & a'_2 & j_2
 	              \end{array} \right\}
		 \times \nonumber\\      	      
    & &	\times\Bigg[ \prod_{n=3}^{K-1} \sqrt{(2a'_n+1)(2a_n+1)}(-1)^{a'_{n-1}+a_{n-1}+1}
	     \left\{ \begin{array}{ccc} 
       	        	       j_n & a'_{n-1}  & a'_n\\
                	        1  & a_n & a_{n-1}
 	              \end{array} \right\}
	\Bigg]
	\left\{ \begin{array}{ccc} 
       	        	       a_{K} & j_K  & a_{K-1}\\
                	          1    & a'_{K-1} & j_K
 	              \end{array} \right\}   \times \nonumber\\
    & & \times\Bigg[ a_2(a_2+1)-a'_2(a'_2+1)\Bigg] \prod_{n=K}^{N}\delta_{a_na'_n}		      		      		      
\ea

\subsubsection{\fbox{I=1~J=2~K=3}}
This case is actually the easiest. We will use it in the next section. 
Therefore we will write down the
calculation explicitly. We start with (\ref{I=1 J=2}), (\ref{I=2 J=3}) to obtain
\ba \label{I=1 J=2 K=3 I}
  <\vec{a}|\hat{q}_{123}|\vec{a}'>
   &=& a_2(a_2+1)\prod_{n=2}^N \delta_{g''_na_n} \Bigg[ \frac{1}{2}(-1)^{j_2-j_1+j_3}
      X(j_2,j_3)^{\frac{1}{2}} \sqrt{(2g_2''+1)(2a_2'+1)}
	\left\{ \begin{array}{ccc} 
    	            j_1 & j_2  & g''_2\\
              	     1  & a'_2 & j_2
 	        \end{array} \right\} \times \nonumber\\
   & &\hspace{3cm}\times~ (-1)^{a_3}
	\left\{ \begin{array}{ccc} 
    	            a_3 & j_3  & g''_2\\
              	     1  & a'_2 & j_3
        \end{array} \right\} \prod_{n=3}^N \delta_{a_na'_n}
	~+~ \big[j_2(j_2+1)+j_3(j_3+1)\big]\prod_{n=2}^N \delta_{a_na'_n} \Bigg]\nonumber \\
   & &-~\Big( \vec{a} \longleftrightarrow \vec{a'}\Big) \nonumber\\\nonumber\\
   &=&\bigg(a_2(a_2+1)-a'_2(a'_2+1) \bigg)
      \Bigg[ \frac{1}{2}(-1)^{j_2-j_1+j_3}
	    X(j_2,j_3)^{\frac{1}{2}} \sqrt{(2a_2'+1)(2a_2+1)}
	      \left\{ \begin{array}{ccc} 
    	        	  j_1 & j_2  & a_2\\
              		   1  & a'_2 & j_2
 	              \end{array} \right\} \times \nonumber\\
	 & &\hspace{3cm}\times~ (-1)^{a_3}
	      \left\{ \begin{array}{ccc} 
    	        	  a_3 & j_3  & a_2\\
              		   1  & a'_2 & j_3
              \end{array} \right\} \Bigg]\prod_{n=3}^{N}\delta_{a_na'_n}
\ea

Now the conditions for the arguments in the definition of the $6j$-symbols (\ref{A1})
gives certain relations for $a_2$, $a'_2$ occurring in (\ref{I=1 J=2 K=3 I}), namely:
\[
  a'_2= \left\{ \begin{array}{l}
                   a_2-1 \\
		   a_2   \\
		   a_2+1 
                \end{array} \right.
\]
We can choose either the first or the third case to obtain a non-vanishing matrix element. We choose
\fbox{$a'_2=a_2-1$} (the other choice would give us a sign, due to the antisymmetry of $\hat{q}$). So we
continue (considering only the nontrivial information $a_2$, $a'_2$ contained in the recoupling-schemes
 $\vec{a},\vec{a}'$ ):

\ba \label{I=1 J=2 K=3 II}
  <a_2|\hat{q}_{123}|a_2-1>
   &=&\bigg(a_2(a_2+1)-(a_2-1)a_2 \bigg)
      \Bigg[ \frac{1}{2}(-1)^{j_2-j_1+j_3}
	    X(j_2,j_3)^{\frac{1}{2}} \sqrt{(2(a_2-1)+1)(2a_2+1)}
	      \left\{ \begin{array}{ccc} 
    	        	  j_1 & j_2  & a_2\\
              		   1  & a_2-1 & j_2
 	              \end{array} \right\} \times \nonumber\\
	 & &\hspace{3cm}\times~ (-1)^{a_3}
	      \left\{ \begin{array}{ccc} 
    	        	  a_3 & j_3  & a_2\\
              		   1  & a_2-1 & j_3
              \end{array} \right\} \Bigg]
\ea
Now we can rewrite the second $6j$-symbol in (\ref{I=1 J=2 K=3 II}) by using the identity 
\footnote{This follows directly from (\ref{special 6j II}) }:

\[	      \left\{ \begin{array}{ccc} 
    	        	  a & b   & c\\
              		  1 & c-1 & b
              \end{array} \right\} 
	 =\sqrt{\frac{2a(2a+1)(2a+2)}{2b(2b+1)(2b+2)}}
               \left\{ \begin{array}{ccc} 
    	        	  b & a   & c\\
              		  1 & c-1 & a
               \end{array} \right\} 	      	      
\]

\ba \label{I=1 J=2 K=3 III}
  <a_2|\hat{q}_{123}|a_2-1>
   &=&2a_2 \frac{1}{2}(-1)^{j_2-j_1+j_3}
	    X(j_2,j_3)^{\frac{1}{2}} \sqrt{(2a_2-1)(2a_2+1)}
	      \left\{ \begin{array}{ccc} 
    	        	  j_1 & j_2  & a_2\\
              		   1  & a_2-1 & j_2
 	              \end{array} \right\} \times \hspace{4cm}\nonumber\\
	 & &\hspace{3cm}\times~ (-1)^{a_3}
             \sqrt{\frac{2a_3(2a_3+1)(2a_3+2)}{2j_3(2j_3+1)(2j_3+2)}}
	     \left\{ \begin{array}{ccc} 
    	        	  j_3 & a_3  &  a_2\\
              		   1  & a_2-1 & a_3
             \end{array} \right\} 
\ea

At the last step we once more use (\ref{special 6j II}) to express all $6j$-symbols in 
(\ref{I=1 J=2 K=3 III}) explicitly. 
Furthermore we use the fact, that $j_3+a_2+a_3$ is an integer number and therefore
$(-1)^{2(j_3+a_2+a_3)}=1$ (one can see this by
applying the integer conditions (\ref{integer conditions}) to the second $6j$-symbol in (\ref{I=1 J=2 K=3
III})~). Additionally remember the shortcut introduced earlier: 
$X(j_2,j_3)=2j_2(2j_2+1)(2j_2+2)2j_3(2j_3+1)(2j_3+2)$. After carefully expanding all the terms and
cancelling all identical terms in the nominator and the denominator the explicit result is:

\ba 
  <a_2|\hat{q}_{123}|a_2-1>
   &=& \frac{1}{\sqrt{(2a_2-1)(2a_2+1)}}\bigg[ (j_1+j_2+a_2+1)(-j_1+j_2+a_2)(j_1-j_2+a_2)(j_1+j_2-a_2+1)
     \nonumber\\
   & & \hspace{3.4cm}                            (j_3+a_3+a_2+1)(-j_3+a_3+a_2)(j_3-a_3+a_2)(j_3+a_3-a_2+1)
					\bigg]^{\frac{1}{2}} \nonumber\\
  &=&~-~<a_2-1|\hat{q}_{123}|a_2>                               \nonumber
\ea
\be \label{derivation de Pietri}\ee

The analytical result for this special case coincides with the result already obtained by graphical methods in \cite{dePietri}, if one
considers the gauge invariant 4-vertex (see next section!) by putting $a_2 \rightarrow j_{12}$ and 
$a_3 \rightarrow j_4$ (as a consequence of applying the definition of the standard recoupling schemes
(definition \ref{Def Standard basis}) to a 4-vertex).


\subsubsection{Comparison: Computational Effort}
At the end of this section we want to compare the computational effort one has to invest for calculating the
matrix element (\ref{vorformel}) using the full definition in terms of $6j$-symbols 
(\ref{basic structure in 6j symbols})
or the derived formula (\ref{Endgueltige Formel fuer das Matrixelement}) instead.
We will give here only a rough estimate, since for the full definition (\ref{basic structure in 6j symbols})
the calculation can hardly be done for all possible combinations of arguments.

Consider an $N$-valent monochromatic vertex\footnote{Of course this 
special case is not that expensive, because it is the most symmetric 
case. But it illustrates
the estimates for the general case with different spins.}
 $v$ with $N$ outgoing edges $e_1,\ldots,e_N$,
each carrying the spin $j_1=\ldots =j_N=j_{max}$. Assume, we had to calculate the matrix element 
(\ref{vorformel}) for a certain combination of the triple $I<J<K$, namely 
$I\approx (J-I) \approx (K-J):=L\sim \frac{N}{3}\gg 1$.

Let us first discuss the full definition using (\ref{vorformel}) with (\ref{basic structure in 6j symbols})
inserted.\\
Consider first the definition (\ref{A1}) of the $6j$-symbols: we will only pay attention to the $w$-coefficient,
since the number of $\Delta$-coefficients is constant. By the requirement for the summation variable $n$ 
($\max[j_1+j_2+j_{12},j_1+j+j_{23},j_3+j_2+j_{23},j_3+j+j_{12}] \le n \le
       \min[j_1+j_2+j_3+j,j_2+j_{12}+j+j_{23},j_{12}+j_1+j_{23}+j_3]$)  
we can (if we put $j_1=j_2=j_3=j=j_{max}$, $j_{23}=j_{12}=2j$) extract $0 \le n \le 4j$. That is we have
approximately $4j_{max}$ summations and therefore $7\cdot 4j_{max}$ factorials to calculate for every 
$6j$-symbol.
\\\\
Now look at the definition of the $3nj$-symbol in terms of $6j$-symbols (\ref{6j-hauptformel}): We have
approximately $2I=2L$ $6j$-symbols, due to summation over the intermediate recoupling steps  $h_k$ and
additionally $J-I=L$ $6j$-symbols not involved into that summation, a constant number which we can drop. 
Now in the worst case:\\
$0\le h_1 \le 2j_{max},~j_{max}\le h_2 \le 2j_{max}+j_{max}, \ldots, (I-1)j_{max} \le h_{I-1}\le (I+1)j_{max}$. 
Therefore each $h_k$ ($2\le k \le (I-1)$) can take $2j_{max}$ different values and we have thus about
$(2j_{max})^I$ possible combinations for the $h_k$.
So we had to calculate at all $2I(2j_{max})^I=2L(2j_{max})^L$ $6j$-symbols.
\\\\
Every term in the sum (\ref{vorformel}) contains a product of 4 $3nj$-symbols. Now the summation
over $\vec{g}(IJ),\vec{g}(JK),\vec{g}''(12)$ again gives (under the assumption, that each intermediate
angular momentum $g(IJ)_k,g(JK)_k,g(12)''_k$ can take $2j_{max}$ different values, but only
$(J-I)\approx (K-J):=L$ intermediate steps of each summation contribute to calculations, 
due to the $\delta$-terms in (\ref{hauptformel})~),
for each matrix element approximately $(2\cdot 2j_{max})^{3L}$ $3nj$-symbols to calculate.
\\\\
Summarizing these 3 steps we end up with a computational effort of approximately:
\[  
7\cdot 4j_{max}\cdot 2L(2j_{max})^L \cdot (2\cdot 2j_{max})^{3L} \sim (j_{max})^{4L} 
                        \sim (j_{max})^{\frac{4}{3}N}
\]
calculations of \bf factorials \rm occurring in (\ref{A1}).
\\\\ 
It is much easier to discuss the effort one has in case of using the derived equation 
(\ref{Endgueltige Formel fuer das Matrixelement}). We have only a product of special $6j$-symbols
containing \bf no summation and factorials at all\rm. So one only has to carry out the product consisting of 
only \fbox{$K-I=2L\sim \frac{2}{3}N$ factors}, {\bf independent} of 
$j_{max}$. 

Be aware that this estimate given is only rough, one could introduce the symmetry properties
(\ref{symmetry1}), (\ref{symmetry2}) and additionally keep in memory  previously calculated $6j$- or 
$3nj$-symbols to
save calculation time. Nevertheless the computational effort for the calculation of the matrix element 
(\ref{vorformel}) depends on $j_{max}$ if one uses the original formulas  
(\ref{vorformel}) with (\ref{basic structure in 6j symbols}). This is no longer the case if one uses  
(\ref{Endgueltige Formel fuer das Matrixelement}). It is clear, that if one wants to numerically compute all the matrix elements then one cannot get very much over $j_{max}=2$ with (\ref{vorformel}).

\subsubsection{Conclusion}

We have shown in the last section, that it is possible to explicitly
evaluate the matrix elements of the volume operator in (\ref{vorformel}).
Here by 'explicitly' we mean that there are no more $6j$-symbols in the final expression.
The derived formula is a simple algebraic function of the spin quantum numbers, no factorials appear any
longer and no conditional summations, implicit in Racah's formula for the $6j$-symbol, have to be carried
out anymore. Thus the computational effort in order to evaluate the matrix elements has decreased by a
huge order of magnitude, which grows with growing maximal spin $j_{max}$.
This simplification has been achieved by the discovery of a nontrivial fact, namely that the highly 
involved formula (\ref{vorformel}) or (\ref{basic structure in 6j symbols}) is like a telescopic sum of the form $\sum_{n=1}^{N}(a_n-a_{n-1})=a_N-a_0 $ once
one takes the orthogonality relations of the $6j$-symbols and the Elliot-Biedenharn identity into account.

 A first observation is that the matrices defined by 
(\ref{Endgueltige Formel fuer das Matrixelement}), show a banded structure, that is a rich selection rule 
structure.
Non-vanishing entries are
only on certain parallels to the main diagonal, because of the restrictions of the presence of an entry 1 in
every $6j$-symbol contained in (\ref{Endgueltige Formel fuer das Matrixelement}).
\\[15.5cm]
   

\section{Gauge Invariant 4-Vertex}
In this section we will examine in more detail the gauge invariant 4-vertex, that is the following
configuration of edges:

\begin{figure}[!hbt]
\begin{center}
  \begin{minipage}[t]{8cm} 
      \includegraphics[height=4cm]{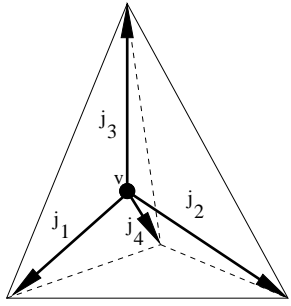}
      \caption{The configuration at the 4-vertex: 4 outgoing edges each carrying a representation of
      $SU(2)$ with a weight according to $j_1,j_2,j_3,j_4$ }
      \label{4vertex setup}
  \end{minipage} 
\end{center}
\end{figure}

We have 4 edges $e_1,\ldots,e_4$ outgoing at the vertex $v$ carrying the spins $j_1,\ldots,j_4$ and the
according representations of $SU(2)$, $\pi_{j_1},\ldots,\pi_{j_4}$.

The square $\hat{Q}_v=\hat{V}^2_v$ of the volume operator represented in term of the standard-recoupling
scheme basis was
(note that in what follows we will
stick to the squared version of the volume operator, therefore of all eigenvalues we write down the square
root has to be taken in order to obtain the spectral behaviour of the 
volume operator itself):

\ba \label{Q 4-vertex}
  \hat{Q}_v&:=&Z\cdot 
                \sum_{I<J<K} \epsilon(I,J,K)~\big[(J_{IJ})^2,(J_{JK})^2 \big]  \nonumber\\
           & =&Z\cdot 
                \sum_{I<J<K} \epsilon(I,J,K)~\hat{q}_{IJK} 
\ea
Since we have 4 edges the summation in (\ref{Q 4-vertex}) has to be extended over the
combinations:\linebreak
$(I<J<K)=(1,2,3), (1,2,4), (1,3,4), (2,3,4)$.

The point is now that  due to gauge invariance the 4 angular momenta $j_1,\ldots,j_4$
should couple to a resulting zero angular momentum $j=0$ at the vertex $v$. For this reason 
for the angular momentum operators $J_1,\ldots,J_4$ holds due to (\ref{restriction to g_{N-2}}):
\be
  J_1+J_2+J_3+J_4 \stackrel{}{=}0
\ee
which implies
\be
  J_4=-(J_1+J_2+J_3) 
\ee
It follows that:
\ba
  \hat{q}_{124}=-\big(\hat{q}_{121}+\hat{q}_{122}+\hat{q}_{123} \big)=-\hat{q}_{123} \nonumber\\
  \hat{q}_{134}=-\big(\hat{q}_{131}+\hat{q}_{132}+\hat{q}_{133} \big)=-\hat{q}_{132} \nonumber\\
  \hat{q}_{234}=-\big(\hat{q}_{231}+\hat{q}_{232}+\hat{q}_{233} \big)=-\hat{q}_{231} 
\ea
Here we have used the fact that $\hat{q}_{IJJ}+\hat{q}_{IJI}=0~~~\forall I,J$
\footnote{To see this, just take the definition $\hat{q}_{IJK}=\big[(J_{IJ})^2,(J_{JK})^2 \big]$ and 
expand the resulting commutators. Alternatively use the antisymmetry of 
$\hat{q}_{IJJ}\sim \epsilon_{ijk}J^j_IJ^j_JJ^k_J$ to see that 
$\hat{q}_{IJJ}+\hat{q}_{IJI}=\hat{q}_{IJJ}-\hat{q}_{JII}=0$ .}.\\

Thus (\ref{Q 4-vertex}) reduces to:
\ba
  \sum_{I<J<K} \epsilon(I,J,K)\hat{q}_{IJK} 
  &=& \epsilon(1,2,3)\hat{q}_{123}+\epsilon(1,2,4)\hat{q}_{124}
     +\epsilon(1,3,4)\hat{q}_{134}+\epsilon(2,3,4)\hat{q}_{234}\nonumber\\
  &=& \big[\epsilon(1,2,3)-\epsilon(1,2,4)+\epsilon(1,3,4)-\epsilon(2,3,4)\big]\hat{q}_{123} \nonumber\\
  \nonumber\\
  &=& 2\cdot\hat{q}_{123}  
\ea
where we have used in the last line the configuration of the 4 edges outgoing from $v$
described above to obtain the
orientation ($\pm 1$) of every triple of tangent vectors corresponding to 3 distinct edges
(Note that for different orientations of $e_1,\ldots,e_4$ just the prefactor 2 changes.).
This brings us precisely into the situation of the last example in the previous section. 
We can now explicitly write down the
matrix-elements of $\hat{Q}_v$ represented in a basis of standard gauge invariant recoupling schemes
$|\vec{a}(12)~j=0~M=0>=|a_2>$, $|\vec{a}'(12)~j=0~M=0>=|a'_2>$.
For simplicity we relabel $a_2 \rightarrow j_{12}$, $a_3=a'_3 \rightarrow j_4$, $a'_2 \rightarrow j'_{12}$,
$a_3=a'_3 \rightarrow j_4$.

Now the non vanishing matrix-elements in (\ref{derivation de Pietri}) are:
\ba 
  \lefteqn{<j_{12}|\hat{q}_{123}|j_{12}-1>~=}\hspace{1cm} \nonumber\\\nonumber\\
   &=& \frac{1}{\sqrt{(2j_{12}-1)(2j_{12}+1)}}
         \bigg[       (j_1+j_2+j_{12}+1)(-j_1+j_2+j_{12})(j_1-j_2+j_{12})(j_1+j_2-j_{12}+1) \nonumber\\
   & & \hspace{3.4cm} (j_3+j_4+j_{12}+1)(-j_3+j_4+j_{12})(j_3-j_4+j_{12})(j_3+j_4-j_{12}+1)
					\bigg]^{\frac{1}{2}} \nonumber\\
  &=&~-~<j_{12}-1|\hat{q}_{123}|j_{12}>  \nonumber                             
\ea
\be \label{matrix element 4 vertex}\ee

By (\ref{restriction to g_{N-2}}) we get certain restrictions for the values that $j_{12}$ may
take \footnote{We may label w.l.g. edges in such a way that 
$0<j_1\le j_2 \le j_3 \le j_4 \le (j_1+j_2+j_3)$ 
and $j_1+j_2+j_3+j_4=$integral}:
\be \label{allowed values for j12 at the 4 vertex 1}
  \max{(|j_1-j_2|,|j_3-j_4|)}\le j_{12} \le \min{(j_1+j_2,j_3+j_4)}
\ee
Therefore the dimension $n$ of the matrix-representation $A$ of $\hat{q}_{123}$ in the standard basis
is given by.
\ba \label{allowed values for j12 at the 4 vertex}
  n:=\dim{A}&=&\min{(j_1+j_2,j_3+j_4)} - \max{(|j_1-j_2|,|j_3-j_4|)} +1 \nonumber\\
            &=& j_{12}^{max}-j_{12}^{min}+1
\ea

We find for the matrix $A$ (labelling the rows by $j_{12}$ and the
columns by $j'_{12}$, where the first row/column equals $j_{12}=j'_{12}=j_{12}^{min}$ increasing down to 
the last row/column with $j_{12}=j'_{12}=j_{12}^{max}$ and using the abbreviation for the matrix element
$a_k:=i\cdot<j_{12}^{min}+k|\hat{q}_{123}|j_{12}^{min}+k-1>$ (where $i$ is the imaginary unit
\footnote{This only changes the spectrum of $A$ from being antisymmetric to hermitian and therefore
rotates its spectrum from purely imaginary to purely real.}),
$k=1,\ldots,n-1$ and $a_0=a_n=0$ 
\footnote{just insert $j_{12}=j_{12}^{min}$ or $j_{12}=j_{12}^{min}+n=j_{12}^{max}$ into the matrix
element (\ref{matrix element 4 vertex})}):

\setlength{\arraycolsep}{1.2mm}
\be \label{general matrix 4 vertex}
  A=\left( \begin{array}{cccccc}
             0      & -a_1    & 0      & \cdots  & \cdots  & 0         \\
	     a_1    &  0      & -a_2   &         &         & \vdots    \\
	     0      & a_2     &  0     & \ddots  &         & \vdots    \\
	     \vdots &         & \ddots & \ddots  & \ddots  & \vdots    \\
	     \vdots &         &        & \ddots  & \ddots  & -a_{n-1}  \\
	     0      & \cdots  & \cdots & \cdots  & a_{n-1} & 0 
           \end{array} \right)
\ee
\setlength{\arraycolsep}{0.7mm}

That is: The matrix $A$ possesses a banded matrix structure which is called a $Jacobi$-matrix. 
Note, that the $a_k$ are purely imaginary, because $A$ is hermitian and its
eigenvalues are real.
We will discuss the spectral theory of $A$ by analytical and numerical methods. Note the following
advantages of the gauge invariant case over the gauge variant:

\begin{itemize}
   \item{The dimension of $A$ scales only linearly with the spins outgoing at the vertex $v$ (this advantage
          will be useful for the numerical studies).}
   \item{There is no sum over matrices left any longer, as it would be the case for the gauge variant     4-vertex.}
   \item{Due to the formulation in a recoupling scheme basis we have automatically implemented gauge
         invariance.}
\end{itemize}

\subsection{Analytical Investigations}
\subsubsection{Eigenvalues}
As pointed out before, all eigenvalues $\lambda$ of $A$ are real and come in pairs $\pm\lambda$. The
special case of zero-eigenvalues will be discussed below.
One can find upper bounds for the eigenvalues by applying the theorem of $Ger\check{s}gorin$ (see
\cite{Gantmacher},p.465):

\begin{Theorem}{$Ger\check{s}gorin$} \label{Gersgorin discs}
   
   Every characteristic root $\lambda$ of a ($n\times n$)-matrix $A$ lies at least in one of the discs\\
   \[
     |a_{ii}-\lambda| \le \displaystyle\sum_{{j=1}\atop{j\ne i}}^n |a_{ij}| \hspace{1cm} i=1,\ldots,n
   \] 
\end{Theorem}

That is, every eigenvalue lies in a disc centered at the diagonal element $a_{ii}$ with radius of the sum of 
moduli of the off-diagonal-elements $a_{ij},~i\ne j$ of the $i$-th row or column (called the $i$-th
row- or column-sum).
In case of the gauge invariant 4-vertex this theorem simplifies due to the banded matrix structure of (\ref{general matrix 4 vertex}) and the fact that $a_{ii}=0$ to
\be
  |\lambda|\le \sum\limits_{j \ne i}|a_{ij}| = |a_{i~i-1}| + a_{i~i+1}         
\ee

We will give an upper and a lower bound for the eigenvalue-spectrum in terms of the leading polynomial order of the largest angular momentum $j_{max}=\max(j_1,\ldots,j_4)$.
\\
 By inspection of (\ref{general matrix 4 vertex}) one can see that the row 
 sum introduced in theorem (\ref{Gersgorin discs}) is dependent of the 
 modulus of each of the matrix elements 
 $a_i(j_{12}):=<j_{12}~|~\hat{q}_{123}~|~j_{12}-1>,~i=1\ldots n-1$. This 
 observation will be useful for obtaining an upper bound for the modulus 
 of the eigenvalues of $A$. On the other hand we could also use theorem 
 (\ref{Gersgorin discs}) for giving a lower bound of the eigenvalues if we 
 could guarantee the existence of the inverse $A^{-1}$ that is the absence 
 of the eigenvalue 0, which will be discussed explicitly in appendix 
 \ref{diskussion 4-vertex}. Then the upper bound of the eigenvalues of 
 $A^{-1}$ would give us a lower bound of the non-zero eigenvalues of $A$. 
However, due to the general formula for the matrix element of the inverse 
matrix
\be
  (A^{-1})_{ij}=\frac{\det{M_{ij}}}{\det{A}}
\ee 

(where $M_{ij}$ denotes the sub-determinant of A with row $i$ and column $j$ deleted) all the entries of $A^{-1}$ will be of the order $\frac{1}{a_i(j_{12})}$.
We can therefore try to find the extrema of the matrixelement 
(\ref{matrix element 4 vertex}) in terms of the spins $j_1,\ldots,j_4$ by 
partial differentation \footnote{We will list here only these solutions 
which allow positive values for $j_1,\ldots,j_4$ and mutually 
different $j_1,j_2,j_3,j_4$~!}. Note, that we have 
the freedom to choose $j_1\le j_2\le j_3\le j_4=j_{max}$, since the matrix element (\ref{matrix element 4 vertex}) is symmetric under permutations of  $j_1,\ldots,j_4$ :

\ba \label{Abschaetzung}
  \frac{\partial{~a(j_{12})}}{\partial j_1} ~\stackrel{!}{=}0&\Leftrightarrow&
   j_1^{(0)}=-\frac{1}{2}+\frac{1}{2}\sqrt{1+4j_{12}^2+4j_2+4j_2^2}\nonumber\\
  \frac{\partial{~a(j_{12})}}{\partial j_2} ~\stackrel{!}{=}0&\Leftrightarrow&
   j_2^{(0)}=-\frac{1}{2}+\frac{1}{2}\sqrt{1+4j_{12}^2+4j_1+4j_1^2}\nonumber\\
  \frac{\partial{~a(j_{12})}}{\partial j_3} ~\stackrel{!}{=}0&\Leftrightarrow&
   j_3^{(0)}=-\frac{1}{2}+\frac{1}{2}\sqrt{1+4j_{12}^2+4j_4+4j_4^2}\nonumber\\
  \frac{\partial{~a(j_{12})}}{\partial j_4} ~\stackrel{!}{=}0&\Leftrightarrow&
   j_4^{(0)}=-\frac{1}{2}+\frac{1}{2}\sqrt{1+4j_{12}^2+4j_3+4j_3^2}
\ea

 If we want all relations of (\ref{Abschaetzung}) to be fulfilled at the 
 same time with strictly positive values of $j_1,\ldots,j_4$ then we have 
to demand 
\ba
  j_1\stackrel{!}{=}j_2:=m&~~\mbox{and}~~&j_3\stackrel{!}{=}j_4:=j_{max}
\ea

Therefore (\ref{allowed values for j12 at the 4 vertex}) reads due to the ordering $j_1\le j_2\le j_3\le j_4=j_{max}$ as
\be
  0 \le j_{12} \le 2m 
\ee
And the matrix element (\ref{matrix element 4 vertex}) simplifies to
\be \label{extremal me}
  a(j_{12})=\frac{j_{12}^2}{\sqrt{4j_{12}^2-1}}
   \Big[\big[(2a+1)^2-j_{12}^2\big]\big[(2j_{max}+1)^2-j_{12}^2\big] \Big]^{\frac{1}{2}}
\ee
\begin{enumerate}
   \item{\bf Largest Eigenvalue\\}{
         By inspection of (\ref{extremal me}) we can maximize the order of $j_{max}$ contained in the              matrix element by putting 
	 \ba
	   m \sim j_{max}  &~~~\leadsto~~~& a(j_{12}) \sim j_{max}^3
	 \ea
	 The result is an upper bound on the growth of the maximum eigenvalues of the matrix $A$ with the          maximum angular momentum $j_{max}$ \footnote{This result coincides with that already obtained in          \cite{Seifert}. Note, that the number of terms in each row/column - sum is equal to 2 due to the          special structure (\ref{general matrix 4 vertex}) of the matrix $A$ and therefore independent of          $j_{max}$.}:
	 \ba \label{estimate largest eigenvalue}
	   | \lambda_{max}(j_{max}) | \sim j_{max}^3 &~~\Rightarrow~~&
	   | V_{max}(j_{max}) | \sim j_{max}^{\frac{3}{2}}
	 \ea}

   \item{\bf Smallest non-zero Eigenvalue\\}{
         By assuming the existence of the inverse $A^{-1}$ with a sparse population of entries of the order $\frac{1}{a_k}$ one can minimize the order of $j_{max}$ contained in the              matrix element (that is maximizing the matrix elements of $A^{-1}$) by putting 
	 \ba
	   m \sim 1 \ll j_{max}  &~~~\leadsto~~~& a(j_{12}) \sim j_{max} 
	 \ea
 	 The result is an upper bound on the growth of the maximum eigenvalues of the matrix $A^{-1}$              with the maximum angular momentum $j_{max}$ and therefore for the smallest non-zero eigenvalue            of $A$ ~\footnote{Note, that one should really use $(A^{-1})_{ij}$ which can be a complicated polynomial in the $A_{ij}$. This is indeed the case as shown in appendix \ref{diskussion 4-vertex} and therefore the result presented here is at best a rough estimate.} :
	 \ba \label{estimate smallest eigenvalue}
	    | \lambda_{min}(j_{max}) | \sim j_{max}&~~\Rightarrow~~&
	   | V_{min}(j_{max}) | \sim j_{max}^{\frac{1}{2}}
	 \ea
	 }

\end{enumerate}



These are first estimates, we will come back to this, 
when we discuss the numerical investigations, since 
the theorem (\ref{Gersgorin discs}) and our approximations tell us nothing 
about the numerical coefficints in front of the leading powers of
$j_{max}$. Nevertheless this estimate will give 
us a certain criterion for completeness of numerically calculated eigenvalues: Since the smallest eigenvalue $\lambda_{min}$ grows with $j_{max}$ we can at a certain value of $j_{max}$ be sure having calculated the complete volume-spectrum for all $V \le V_{min}$ as we shall see in the numerical section below.  

\subsubsection{Eigenvectors for $\lambda=0$ \label{Eigenvektor fuer 0}}

Posing the eigenvalue problem $A\Psi=\lambda \Psi$ for the matrix $A$ we obtain a
three term recursion relation every eigenvector $\Psi$ of $A$ has to fulfill:
\be \label{recursion I}
  a_{k-1}\Psi_{k-1}-a_k\Psi_{k+1}=\lambda\Psi_k     \hspace{1cm} \mbox{with } a_0=a_n=0,~
                                                                               \lambda \in \mathbbm{R},~ 
									       a_k(0<k<n)~\mbox{purely imaginary}
\ee

We can now check, whether the eigenvalue $\lambda=0$ belongs to the spectrum. This decouples the recursion
relation (\ref{recursion I}) to give
\be \label{recursion III}
  a_{k-1}\Psi_{k-1}-a_k\Psi_{k+1}=0
\ee

Now for consistency of (\ref{recursion III}):
\be \label{consistency on Psi}
   \begin{array}{crrcc}
      \fbox{k=n:}& a_{n-1}\Psi_{n-1}=0 &\hspace{1cm}\mbox{and therefore}&\Psi_{n-1}=\Psi_{n-3}&=\ldots=0
      \\
      \fbox{k=1:}& -a_1\Psi_2=0        &\hspace{1cm}\mbox{and therefore}&\Psi_2=\Psi_4&=\ldots=0
   \end{array}
\ee

But this means that the matrix $A$ can only have an eigenvector $\Psi \ne 0$ if the dimension $n$
of $A$ is odd, because if $n$ would be even, then all components of $\Psi$ would be forced to vanish by
(\ref{consistency on Psi}).

That means, that we will only obtain $\lambda=0$ as a eigenvalue in configurations with odd dimension of
$A$.
We can now construct explicitly the eigenvector $\Psi$ for odd $n$:
First we choose $\Psi_1=x$ with $x=const \in \mathbbm{C}$. Then from
\be \label{recursion IV}
  \frac{a_{k-1}}{a_k}=\frac{\Psi_{k+1}}{\Psi_{k-1}}
\ee
We find the general expression:
\be \label{kernel I}
  \Psi_{2r+1}=x\cdot \frac{\prod\limits_{s=1}^r a_{2s-1}}{\prod\limits_{s=1}^r a_{2s}} 
                 \hspace{1cm}\begin{array}{ll} 
		               \\
			       \\
		                 \mbox{where}& r=1,2,\ldots,\frac{n-1}{2} \\
				             & n=\dim A \\
					     & x=\Psi_1
                             \end{array}
\ee
Since the $a$'s were chosen to be purely imaginary, all odd components $\Psi_3,\ldots,\Psi_n$ of $\Psi$
are real, all even components $\Psi_2,\ldots,\Psi_{n-1}$ are identical zero, because of 
(\ref{consistency on Psi}).
Finally we can fix $\Psi_1=x$  
to be (up to the sign of $x=\Psi_1$):
\be \label{kernel II}
  x=\Psi_1=\pm\left[1+\sum_{r=1}^{\frac{n-1}{2}}  
              \left| \frac{\prod\limits_{s=1}^r a_{2s-1}}{\prod\limits_{s=1}^r a_{2s}}\right|
	                   ^2\right]^{-\frac{1}{2}} 
\ee

Summarizing, the eigenvalue $\lambda=0$ only occurs in the spectra of matrices
$A$, possessing an odd dimension $n$, as a single eigenvalue (since its eigenspace is only one-dimensional due to the
uniqueness of construction of the eigenvector $\Psi$ up to a constant rescaling).
Hence we have shown that $A$ is singular iff $n$ odd and in that case $\lambda=0$ has multiplicity 1.


\subsubsection{Monochromatic 4-Vertex ($j_1=j_2=j_3=j_4=j$)}%

Observing this special case, the matrix elements in (\ref{matrix element 4 vertex}) simplify dramatically
to 
\ba \label{monochromatic 4 vertex}
  <j_{12}|\hat{q}_{123}|j_{12}-1> &=& \frac{1}{\sqrt{4(j_{12})^2-1}} ~ (j_{12})^2 ~\bigg[ n^2-(j_{12})^2
          \bigg]    
\ea
where $0 \le j_{12} \le 2j$ and $\dim A=n=2j+1$

\pagebreak

\subsection{Numerical Investigations}
In this section we will describe numerical calculations done for the gauge invariant 4-vertex. We
will (after describing the setup) sketch the computational effort first. Secondly we will give a conjecture about the volume gap,
that is the smallest non-vanishing eigenvalue as a result of the calculations. As a third step we will take
a look on the accuracy of the upper bound given by theorem \ref{Gersgorin discs}. Finally we will present
some spectral estimates.

\subsubsection{General Setup}
We calculated for the gauge invariant 4-vertex the spectra of all possible edge-spin-configurations
$j_1,j_2,j_3,j_4$ up to a maximal spin of $j_{max}=50$ using the mathematical software $Maple~7$.
Since the matrix element (\ref{matrix element 4 vertex}) is symmetric with respect to interchange 
of the $j$'s we calculated the spectra of all $\hat{Q}_v(j_1,j_2,j_2,j_3)$ for
\ba \label{order of angular momenta}
   0 < j_1\le j_2\le j_3\le j_4 \le \min(j_1+j_2+j_3,j_{max}) &\hspace{0.5 cm} &
   \mbox{and~~~~$j_1+j_2+j_3+j_4$ integral} 
\ea 
Thus we compute less than  $\frac{1}{4!}$ configurations, without losing any information.
The conditions on the right side of (\ref{order of angular momenta}) ensure, that we exclude all trivial 
configurations, since if 
$j_1+j_2+j_3 <j_4$ or $j_1+j_2+j_3+j_4$ not integral it would be impossible to recouple them to resulting 
zero-angular momentum, being the definition for gauge invariance.  If one 
spin would be identical
zero, then the obtained configuration would also be trivial, 
since it would describe then effectively a
gauge invariant $3$-vertex, which vanishes identically, hence we also 
impose $j_1\cdot j_2\cdot j_3\cdot j_4>0$.

The possible values for the intermediate recoupled $j_{12}$ are then due to
(\ref{allowed values for j12 at the 4 vertex 1}) and the above introduced order of the angular momenta
(since we have sorted the $j_1,\ldots,j_4$ by their modulus and each $j \ge 0$, we can leave out the
modulus-notation in (\ref{allowed values for j12 at the 4 vertex 1}) by writing the $j$'s in certain order) :

\be \label{allowed values for j12 in computation}
  \max(j_2-j_1,j_4-j_3) \le  j_{12} \le \min(j_1+j_2,j_3+j_4)=j_1+j_2
\ee
The dimension of the matrix $A$ of such a configuration is then according to 
(\ref{allowed values for j12 at the 4 vertex}) given by:
\be
  \dim A =\dim(j_1,j_2,j_3,j_4)=j_1+j_2-\max(j_2-j_1,j_4-j_3)+1
\ee
  
For every configuration then the matrix elements according to (\ref{matrix element 4 vertex})
are calculated 
and inserted into a numerical matrix. This matrix is then numerically diagonalized, its eigenvalues are
sorted ascending and from this spectrum all eigenvalues $ \ge 0$ are 
taken.
These data are then written into a file linewise, each line starting with the values of $j_1,\ldots,j_4$ and
the total number of saved eigenvalues followed by the sorted list of the eigenvalues itself.
Of course we have to keep in mind the multiplicity 2 of every saved eigenvalue $> 0$. Additionally we have to pay attention on the ordering procedure we applied on $j_1 \ldots j_4$ 
whenever we work with the number of eigenvalues,  since we have suppressed 
certain multiplicities.  
The following table gives the resulting 
multiplicity-factors by which eigenvalue numbers resulting from 
corresponding spin-configurations should be multiplied \footnote{In 
general we have $N!$ possibilities to arrange a list of $N$ elements, but 
having M identical elements each of them with multiplicity $K_1, 
K_2,\ldots,K_M$ in that list we can only have $\frac{N!}{K_1! \cdot K_2! 
\cdot \ldots \cdot K_M!}$ different arrangements of the elements of that 
list.}:\\

%
%

\begin{tabular}{|c||c|c|c|c|c|}\hline\label{multiplicity tabular} 
	~&all spins different& one pair equal & two pairs equal & three 
spins equal& all spins equal\\\hline\hline
	no ordering & $\frac{4!}{1!}=24 $&$\frac{4!}{2!}=12$& 
$\frac{4!}{2!\cdot 2!}=6$ &
	                                                        
$\frac{4!}{3!}=4$&$\frac{4!}{4!}=1$\\\hline 
\end{tabular}

\vfill
\pagebreak

\subsubsection{Computational Effort}
According to the setup above we would expect that:\\

\begin{itemize}
  \item{{\bf{Number of Configurations}}\\First of all we choose integer spins for simplicity, that is $j_J \rightarrow a_J=2\cdot j_J$ and assume all $a$`s to be different (We therefore neglect configurations
 containig equal spins which are dominated by those with differnt spins). 
We choose one of the 4 integer-spins to be maximal and constant, say $a_K=a_{max}:=\max{(a_L)}, L=1\ldots 4$) and then
by condition (\ref{order of angular momenta}) we have  \mbox{$a_K=\le \min{(\sum\limits_{L\ne K}a_L, a_{max})}$}.  Therefore $a_{max}\le \sum\limits_{L\ne K}a_L$. 
Labelling the three remaining $a_L$ by $a_1,a_2,a_3$ we get from that $a_3 \ge a_{max}-a_2-a_1$. But this is only the question of counting the number $N$ of points of a three dimensional cubic lattice fulfilling the last condition which is given by:

\be
  N(a_{max}) = \sum_{a_1=1}^{a_{max}}~\sum_{a_2=1}^{a_{max}}~\sum_{a_3=1}^{a_{max}} 1 -
               \sum_{a_1=1}^{a_{max}-2}~~\sum_{a_2=1}^{a_{max}-a_1}~~\sum_{a_3=1}^{a_{max}-a_1-a_2} 1 
\ee 
 Finally we add $\sum\limits_{a_{max}=1}^{c_{max}} N(a_{max})$. The result 
 is the total number of calculations $N(c_{max})$. Due to the integer 
condition  in (\ref{order of angular momenta}), which becomes an 
even-number condition in the $a$`s, we have to divide that number by 2 and 
to multiply it by 4, since we have chosen one of the 4 integer spins to maximal arbitrarily. If we finally plug in $c_{max}=2\cdot j_{max}$, we get for the number of configurations $N(j_{max}\ge \frac{3}{2})$ with all spins different:
\be
  N(j_{max}) = \frac{20}{3}\cdot j_{max}^4 + 12\cdot j_{max}^3 + \frac{7}{3}\cdot j_{max}^2 - 5\cdot j_{max} + 2
\ee 

This can be compared to the numerically fitted curve of the number of configurations $N_{num}$:
\be
  N_{num}(j_{max})=6.67\cdot j_{max}^4 + 13.33 \cdot j_{max}^3 +11.42\cdot j_{max}^2 
                                       -  8.30 \cdot j_{max}   + 2.22
\ee 

\begin{figure}[!hbt]
\begin{center}
  \begin{minipage}[t]{7cm} 
      \includegraphics[height=7cm]{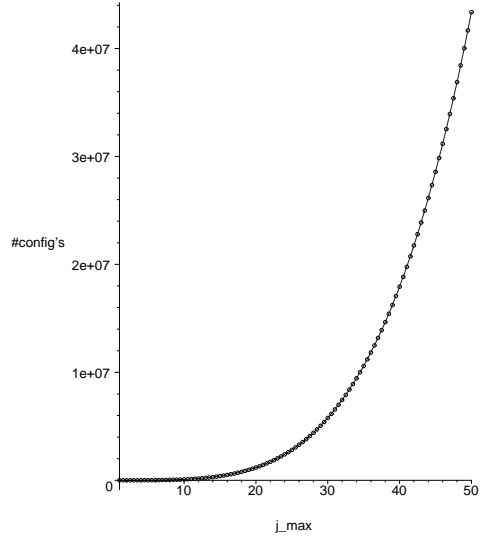}
      \caption{Number of configurations $N(j_{max})$}
      \label{number of configurations}
  \end{minipage} 
\end{center}
\end{figure}}

\item{{\bf{Number of Eigenvalues}}\\ To calculate the expected number of eigenvalues $E(j_{max}$ we have to sum over the dimensions of the individual representation matrix of the Volume Operator which is given by 
\be
  dim=\min{(j_1+j_2,j_3+j_4)}-\max{(|j_2-j_1|,|j_4-j_3|)}
\ee 
One would expect that $E(j_{max})\sim j_{max}^5$. From the numerical calculation we get for the total number of eigenvalues (including 0-eigenvalues and multiplicities):
\be
  E(j_{max})=2.67\cdot j_{max}^5 + 10.00 \cdot j_{max}^4 + 15.52 \cdot j_{max}^3 + 7.92\cdot j_{max}^2
                                  +27.90 \cdot j_{max} - 77.23
\ee
\begin{figure}[!hbt]
\begin{center}
   \begin{minipage}[t]{7cm}
      \includegraphics[height=7cm]{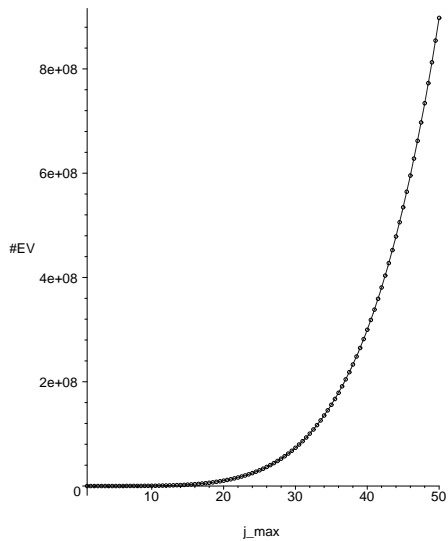}
      \caption{Number of eigenvalues obtained in dependence of $j_{max}$}
      \label{number of eigenvalues}  
   \end{minipage}   
\end{center}
\end{figure}
}

\end{itemize}

\subsubsection{First Impressions}%
We drove the calculations up to a value of $j_{max}=50$ \footnote{This maximal value is limited by to the capacities of the mathematical software $Maple~7$ on the computer used (Intel XEON machine with two  1.7 GHz processors). 
Future calculations will go much further, since $Maple~7$ is only an 
interpreter programming 
language.}. Our primary goal is to obtain some hints on a possibly analytical eigenvalue distribution function for large $j_{max}$. 

Therefore as a start we scanned through all the configurations saved in a file and computed an
eigenvalue-density by distributing the possible eigenvalues $\lambda$ into discrete intervals of 
width $\Delta\lambda=0.5$ where $\lambda$ belongs to the interval $I_n=round(\frac{\lambda}{\Delta\lambda}+0.5)=[\lambda_n - \Delta\lambda,\lambda_n]$ where $\lambda_n=n\cdot \Delta\lambda$, $n=1,2,\ldots$. 

For this we take the square root of the calculated 
eigenvalues, since we computed the spectrum of the square of the eigenvalues of $\hat{V}_v$ and the 
interval width $\Delta\lambda$ does not scale linearly if we would first sort in the eigenvalues of $\hat{Q}_v$ and then took the square root. 
Furthermore, we drop the prefactor $Z$ in what follows.

\begin{figure}[!hbt]
\begin{center}
  \begin{minipage}[!hbt]{14cm} 
      \label{3d impression}
      \includegraphics[height=13cm,width=15cm]{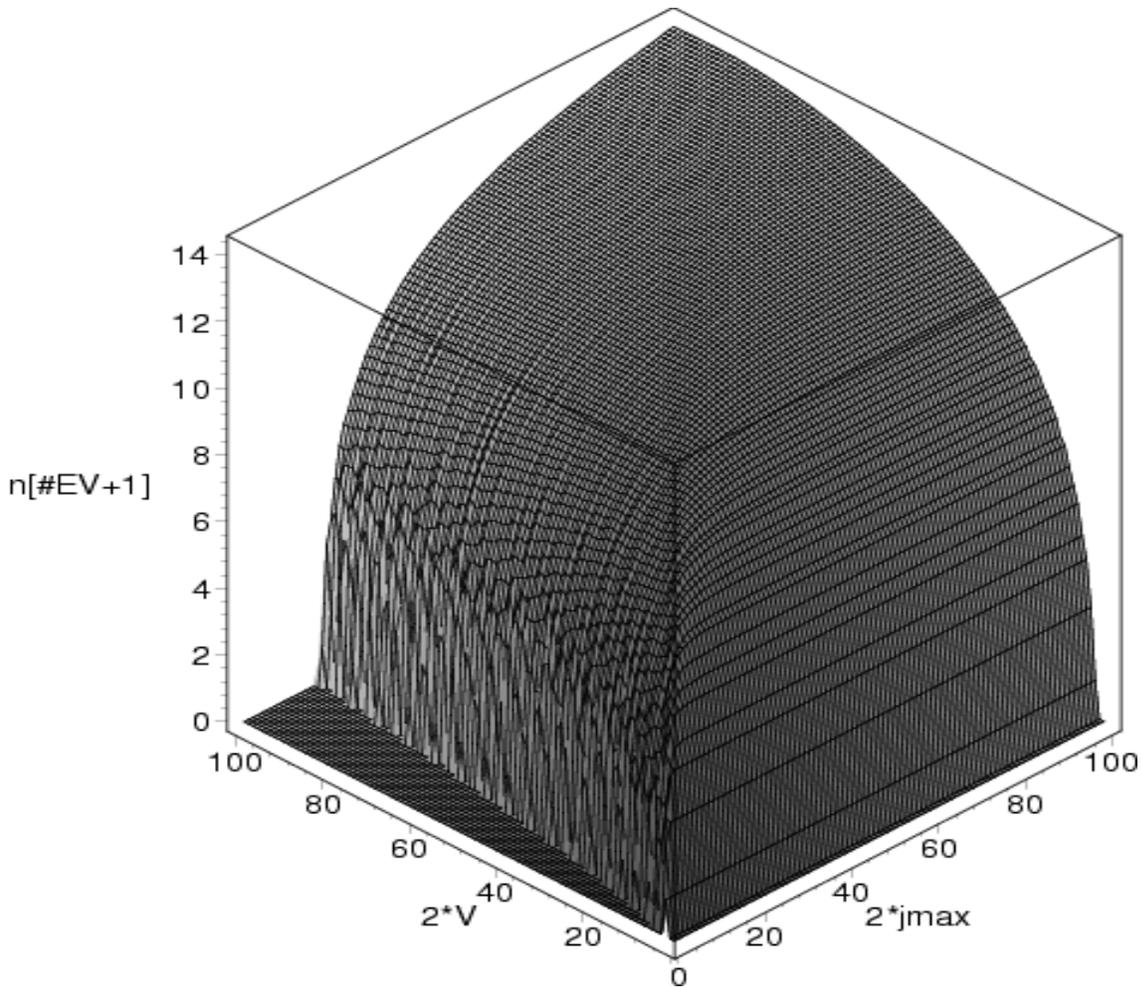}
      \caption{The logarithm of the number of eigenvalues in the intervals                                 
              $I_n=[\lambda_n-\Delta\lambda,\lambda_N]$ $\lambda_n=n\cdot\Delta\lambda$ as a function of
              $ j_{max}$. }    
  \end{minipage} 
\end{center}
\end{figure}

Then we plot the logarithm+1 of the total number of eigenvalues $>0$ 
(including their multiplicity)
according to the multiplicity tabular given before)
in a certain interval $I_n$, against $2\cdot j_{max}$ and the volume-eigenvalues denoted by $2\cdot V$ 
(the form of the axes-labels with prefactors $2$ as well as 
the $+1$ in the logarithm are only for technical reasons). The result is given in figure 
\ref{3d impression}.
    
The diagramme suggests, that for each interval $I_n$ only configurations up to $j_{max}^{(n)}$ matter. Configurations with $j_{max}>j_{max}^{(n)}$ do not increase the number of eigenvalues in the interval $I_n$. 
Therefore we are led to the idea that it would be interesting to look separately at
configurations with fixed $j_{max}$ instead of counting all eigenvalues belonging to an interval
coming from all configurations with $j_4 \le j_{max}$. 

Additionally it would be good to know the
effect that our cutoff $j_{max}=50$ has on the calculated eigenvalue spectrum. 
This is done in what follows.

\subsubsection{Lower Bound for the Spectrum} 
For this purpose
we will split the matrices to be calculated\footnote{due to our setup} into the sets of matrices 
indexed by configurations with fixed $j_4=j=j_{max}$. We then inspect the ordered spectra of every set and
try to find some regularity in the spin-configurations $j_1,j_2,j_3,j_4$ producing the eigenvalues.

As we have calculated all configurations up to $j_{max}=50$ we have 100 
sets of matrices $S_j$ where each set is labelled by $j=\frac{1}{2},1,\ldots,j_{max}-\frac{1}{2},j_{max}$. 

In each set we consider the first 100 positive eigenvalues ordered by values.
Additionally to every eigenvalue we denote the spin configuration of the matrix giving rise to it and the position $k$ the eigenvalue takes in the ordered list of eigenvalues of this set $S_j$. The thus achieved datasets are written
into a file.

Hence we create a function:
\be
 \lambda~:~~~(j,k)\longrightarrow \lambda_k(j) 
\ee
where $\lambda_k(j)$ is the $k$-th eigenvalue in $S_j$.

It turns out that the map $(j,k)\longrightarrow \lambda_k(j)$ indeed 
displays a 
regularity, that is, the eigenvalues seem to produce series being 
separated from each other. Every series can be associated with a certain position in the ordered spectrum of a matrix set $S_j$ with given $j_4 = j \le j_{max}$ . 
The positions are taken from a minimal $j$ on, as series $k$  will not have contributions from $S_j$ with too low $j$. 
It turns out, that the lowest eigenvalues of each matrix set $S_j$ are 
precisely the lowest eigenvalues of the low dimensional matrices with low 
spin configurations $j_1\le j_2\le j_3\le j\le (j_1+j_2+j_3)$.

Remarkably each of these matrices has rank smaller than 9 (that is, the 
nontrivial part of the characteristic polynomial can be reduced to a 
polynomial of degree less than or equal to 4), hence we can find analytic 
expressions for these lowest eigenvalues.
 
We will give here a table containing the first 12 series of eigenvalues.
In the second column we write down the smallest $j$ from which (by inspection of the data) the noted order $k$ is reached. Additionally we note the spin-configuration. The eigenvalues given are always the smallest
ones$\ne 0$ of the according matrix giving rise to $\lambda_k(j)$ with the given spin configuration.      
    
Surprisingly the first smallest eigenvalues are not equally distributed between even and odd
configurations (the latter possessing 0-eigenvalues), but mainly contributed by the even configurations (we
give in the second table the first odd configurations).  Note again that 
these eigenvalues of $\hat{Q}_v$ 
are the square of the eigenvalues of $\hat{V}_v$ and that each eigenvalue has 
a multiplicity accordingly to the multiplicity table given above.
\pagebreak  
\\
\underline{Even Configurations}\\
\\
\renewcommand{\arraystretch}{1.7}
\begin{tabular}{|c|c||c|c|c|c||l|c|}\hline
    $k$ & valid $j\ge $  &  $j_1$        & $j_2$         & $j_3$ & $j_4$ & $\lambda_k(j)$ &$c_k$\\ \hline\hline
 1  & $\frac{1}{2} $ & $\frac{1}{2}$ & $\frac{1}{2}$ & $j$             & $j$ & $2\sqrt{j(j+1)}$&$2$\\ \hline
 2  & $   2        $ & $\frac{1}{2}$ & $    1      $ & $j-\frac{1}{2}$ & $j$ &      
      $2\sqrt{(j+1)(2j-1)}$&$2\sqrt{2}$\\ \hline
 3  & $   3        $ & $\frac{3}{2}$ & $\frac{3}{2}$ & $j$             & $j$ &
    $2\sqrt{17j^2+17j-21-\sqrt{208j^4+416j^3-344j^2-552j+441}}$&$2\sqrt{17-\sqrt{208}}$\\ \hline
 4  & $\frac{13}{2}$ & $\frac{1}{2}$ & $\frac{3}{2}$ & $j-1$           & $j$
    &$2\sqrt{3}\sqrt{(j+1)(j-1)}$ &$2\sqrt{3}$\\ \hline      
 5  & $   4        $ & $\frac{1}{2}$ & $2          $ & $j-\frac{3}{2}$ & $j$ &
    $2\sqrt{2}\sqrt{(j+1)(2j-3)}$ &$4$\\ \hline
 6  & $\frac{7}{2} $ & $1$           & $1$           & $j-1$           & $j$ 
    &$4\sqrt{j(j+1)}$ &$4$\\ \hline
 7  & $   20       $ & $\frac{1}{2}$ & $\frac{3}{2}$ & $j$             & $j$ 
    &$2\sqrt{(2j+3)(2j-1)}$&$4$ \\ \hline
 8  & $\frac{11}{2}$ & $\frac{3}{2}$ & $2$           & $j-\frac{1}{2}$ & $j$ 
    &$\sqrt{108j^2+54j-216-6\sqrt{228j^4+228j^3-903j^2-480j+1152}}$&$\sqrt{108-6\sqrt{228}}$ \\ \hline
 9  & $\frac{65}{2}$ & $\frac{5}{2}$ & $\frac{5}{2}$ & $j$             & $j$ 
    &too long but analytical expression &~\\ \hline
 10 & $10          $ & $\frac{1}{2}$ & $\frac{5}{2}$ & $j-2$           & $j$ 
    &$2\sqrt{5}\sqrt{(j+1)(j-2)}$&$2\sqrt{5}$ \\ \hline
 11 & $11          $ & $\frac{1}{2}$ & $3          $ & $j-\frac{5}{2}$ & $j$ 
    &$2\sqrt{3}\sqrt{(j+1)(2j-5)}$&$2\sqrt{6}$ \\ \hline
 12 & $18          $ & $1$           & $\frac{3}{2}$ & $j-\frac{3}{2}$ & $j$ 
    &$2\sqrt{3}\sqrt{(j+1)(2j-3)}$&$2\sqrt{6}$ \\ \hline
\end{tabular} 
\renewcommand{\arraystretch}{1}
\\\\\\
\underline{Odd Configurations}\\\\
\renewcommand{\arraystretch}{1.7}
\begin{tabular}{|c|c|c|c|c||l|}\hline
    k&     $j_1$        & $j_2$         & $j_3$ & $j_4$ & $\lambda_k(j)$ \\ \hline\hline

    24   & $1$           & $\frac{3}{2}$ & $j-\frac{1}{2}$ & $j$ 
    &$2\sqrt{14j^2+7j-16}$ \\ \hline
    61   & $2$           & $2$           & $j$   & $j$ 
    &$2\sqrt{52j^2+52j-114-18\sqrt{4j^4+8j^3-16j^2-20j+33}}$ \\ \hline
\end{tabular} \\
\renewcommand{\arraystretch}{1}

All of these expressions have to leading order the form 
\be
  \lim_{j\rightarrow \infty}\frac{\lambda_k(j)}{j}=c_k
\ee
where $c_k$ increases with $k$. Thus for sufficiently large $j$ the series 
$j \longrightarrow \lambda_k(j)$ are approximate lines of different inclination.

As an illustration we will give a plot of the first 8 eigenvalue-series, that is we plot for each $j$ the
first 8 eigenvalues of the associated matrix-series $S_j$. Then we connect the first, second, third ,...
eigenvalues with a line. 
Here it becomes obvious, that for small $j$ not all series are present, 
that is $\lambda_k(j)$ is ill defined below the threshold $j$ given in 
the table for each $k$. 

\begin{figure}[!hbt]
\begin{center}
  \begin{minipage}[t]{12cm} 
      \includegraphics[height=13cm]{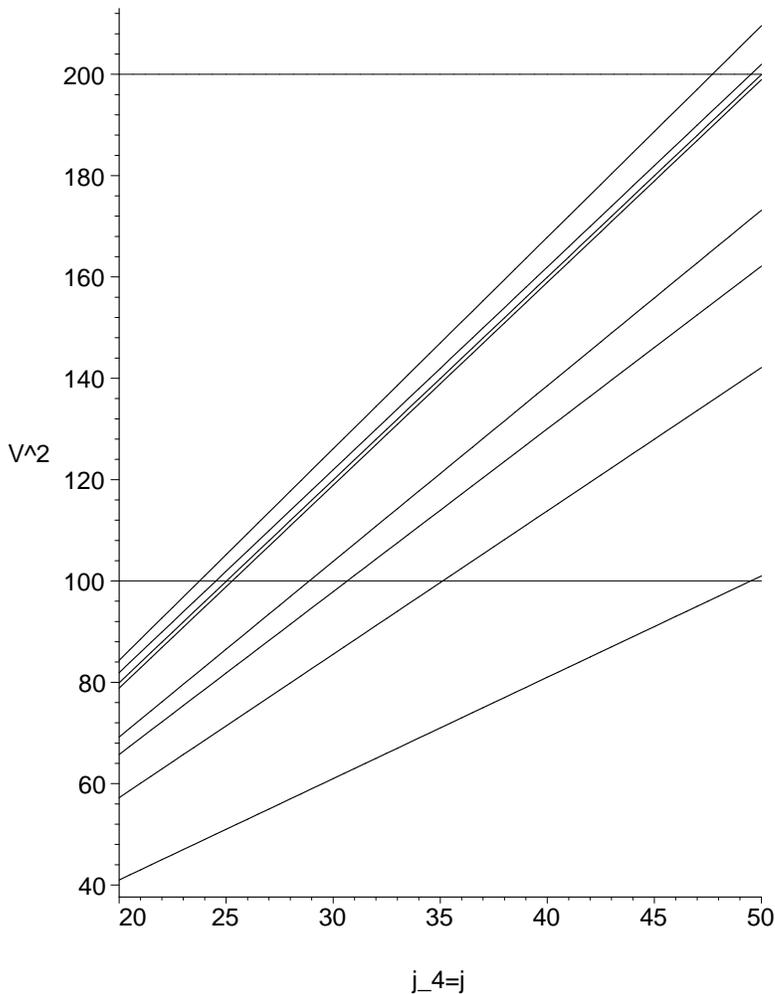}
      \caption{The evolution of the first positive 8 eigenvalues 
$\lambda_k(j):=V^2,\;k=1,..,8$ of $\hat{Q}_v$ in dependence of $j_4=j$. 
      Note that each line represents eigenvalues with multiplicity given 
by the table in section  \ref{multiplicity tabular} 
.}
      \label{first 100 eigenvalues 1}
  \end{minipage} 
\end{center}
\end{figure}

Thus we are given a certain numerical criterion to decide, which part of the spectrum of the
volume operator for the 4-vertex is already entirely calculated for a given cutoff $j_{max}$: 
Given an eigenvalue $\lambda^2$, draw a horizontal line in figure \ref{first 100 eigenvalues 1} and find the
intersection with the first eigenvalue series that is, $k=1$: $\exists~ j~ 
~\mbox{with}~~ \lambda_1(j)^2=\lambda^2 $. 
The value $j(\lambda)$ at which this happens gives the
maximal value $j_{max}(\lambda)$ which we have to consider in order to find configurations giving rise to
eigenvalues $\le \lambda$, because all eigenvalues produced by $j>j(\lambda)$ 
are larger than $\lambda$ because numerically $\lambda_k(j)>\lambda_1(j)
\;\;\forall\;\; k>1$.

According to the table above $\lambda_1(j)^2=2\sqrt{j(j+1)}\stackrel{!}{=} 
\lambda^2$ and therefore 
$j(\lambda)=-\frac{1}{2}+\sqrt{\frac{1}{2}+\frac{\lambda^4}{4}}$. 
 Thus for $j_{max}=50$ we can trust to have computed the complete 
spectrum only for $\lambda\le 
 \lambda_{max}(j_{max})=\sqrt{2}~\sqrt[4]{j_{max}(j_{max}+1)}$, i.e. 
$\lambda_{max}=1.4 \cdot \sqrt{50}\approx 10$.
\\[4cm]

\subsubsection{Upper Bound for the Spectrum}
By observation of the numerical matrices in $S_j$ it turns out, that the maximal eigenvalues 
$\lambda_{max^{(j)}}=(V_{max}^{(j)})^2$ 
of configurations with fixed $j_4=j$ are contributed by matrices of the monochromatic vertex, 
that is $j_1=j_2=j_3=j_4=j$ as we expected from our estimates  
(\ref{estimate largest eigenvalue}).
The matrix elements of this special case we already wrote down in (\ref{monochromatic 4 vertex})~ 
($0 \le j_{12} \le 2j$):
\ba \label{monochromatic 4 vertex II}
 \frac{1}{i}a_k(k=j_{12}):= <j_{12}|\hat{q}_{123}|j_{12}-1> &=& \frac{1}{\sqrt{4(j_{12})^2-1}} ~ (j_{12})^2 ~
                                      \bigg[ (2j+1)^2-(j_{12})^2 \bigg]    
\ea

Now the theorem (\ref{Gersgorin discs}) provides us with upper bounds for the moduli of eigenvalues of a
matrix in terms of its row- or column-sums. It is natural to ask now how the biggest eigenvalue 
$\lambda_{max}^{(j)}$  and the maximal row- or column-sum of the monochromatic matrix $A$ of type 
(\ref{general matrix 4 vertex}) fit together.
It is clear from the structure of $A$ that the inequality given in theorem \ref{Gersgorin discs} for the 
biggest eigenvalue reads:
\[
   |\lambda_{max}^{(j)}| \le \max\big[|a_k|+|a_{k+1}| \big] =:L_{max} \hspace{1cm} k=1,\ldots,n-2
\] 

Therefore we look for the maximal matrix element of $A$ defined by (\ref{monochromatic 4 vertex II}) by
differentiating the given expression with respect to $j_{12}$ and find the value $j_{12}^{max}$ of 
$j_{12}$ maximizing the matrix element. There are several extrema, but the desired one turns out to be:
\be
  j_{12}^{max}=\frac{1}{6}\sqrt{24j^2+24j+12+6\sqrt{16j^4+32j^3+8j^2-8j-2}}
\ee
where $\bf{for~large}$ $j$
\be
  j_{12}^{max}~\stackrel{j \rightarrow \infty}{\longrightarrow}~\frac{2}{\sqrt{3}}
\ee

Since $j_{12}$ can only take positive integer values we then choose the maximal row sum which is given by
\[
  L_{max}=|a_{round(j_{12}^{max})}|+|a_{round(j_{12}^{max})-1}|
\]
Plotting the quotient $\lambda_{max}^{(j)}/L_{max}$ as a function of $j=j_{max}$ we find, that this ratio 
converges to $1$ as $j$ increases:

\begin{figure}[!hbt]
  \begin{center}
    \begin{minipage}[t]{6.5cm} 
      \includegraphics[height=7cm]{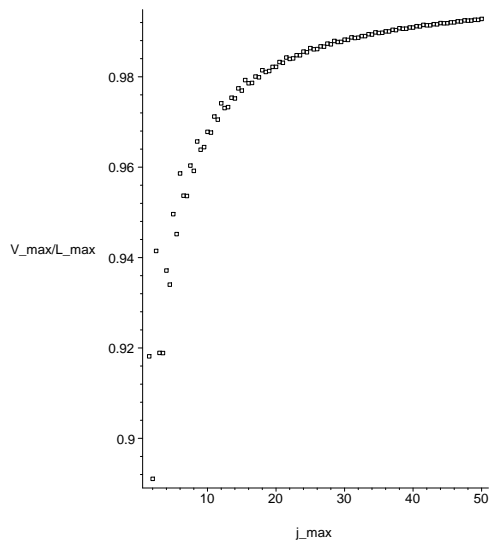}
      \caption{The quotient $\lambda_{max}^{(j)}/L_{max}$ as a function of $j$ }\label{abb.}
    \end{minipage} 
  \end{center}
\end{figure}

Therefore we have numerical evidence for the following large $j$ behavior 
of the biggest eigenvalue in a matrix set $S_j$. 

\be \label{upper bound for eigenvalues}
  \lambda_{max}^{(j)}~\stackrel{j \rightarrow \infty}{\longrightarrow} ~L_{max}(j)
               \approx 2|a_{j_{12}^{max}}(j_{12}^{max}=\frac{1}{\sqrt{3}}~j)| 
	       \approx \sqrt{3}~\frac{11}{9}~j^3
\ee
Here we have inserted $j_{12}^{max}$ in equation (\ref{monochromatic 4 vertex II}) for the
matrix elements to obtain $a_{j_{12}^{max}}$ and approximated the maximal row sum $L_{max}$ by   
$2\cdot|a_{j_{12}^{max}}|$. Finally we keep only the leading order of $j$ in the expression and arrive at
the result (\ref{upper bound for eigenvalues}).  
This coincides with the result obtained in \cite{Seifert}.

\pagebreak

\subsubsection{Spectral Density} 

We now turn to a first investigation of how to get a reliable numerical estimate for the spectral density.

Let us first briefly discuss the behaviour of the 0-eigenvalues. We have proven that they only occur as a
single member of the spectra of matrices with odd dimension. Therefore counting the number of odd
configurations is equal to counting the number of 0-eigenvalues. Since the total number of configurations grows with $j_{max}^4$ we also fit the total number of
0-eigenvalues with a fourth order polynomial, whose coefficient in the leading order should approximately be
half the coefficient we found when we fitted the total number of configurations, since we expect odd and
even configurations to be nearly equally distributed (under restriction of 
(\ref{order of angular momenta}).~).

\begin{figure}[!hbt]
\begin{center}
  \begin{minipage}[t]{7.5cm} 
      \includegraphics[height=9cm]{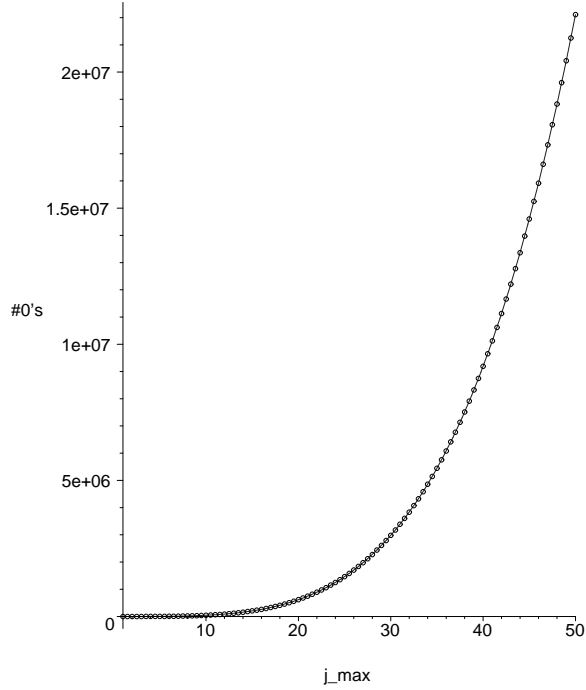}
      \caption{The total number of 0-eigenvalues contained in all configurations allowed by 
       (\ref{order of angular momenta}) depending on $j_{max}$}
  \end{minipage} 
\end{center}
\end{figure}

The fitted polynomial is obtained to be:
\[
   \#0-eigenvalues(j_{max})=3.33\cdot j_{max}^4+10.00\cdot j_{max}^3+10.66\cdot j_{max}^2  -7.10\cdot j_{max} -.85
\]
Indeed the coefficient $3.33$ of $j_{max}^4$ is half the coefficient obtained for the total number of
configurations before. The difference of the other coefficients seems to be caused by the restrictions
given in (\ref{order of angular momenta}). In what follows we will omit the number of 0-eigenvalues, since
their behaviour does not contribute to the spectrum of non-zero-eigenvalues.
Notice that their relative number as compared to the total number of all
all eigenvalues is of the order of $j_{max}^{-1}$. 
\\\\
Let us first recall how we define an eigenvalue density.
We will take the square roots of the eigenvalues $>0$ of $\hat{Q}_v$ obtained in the numerical computation
and split the real axis labelling the eigenvalues $V$ of $\hat{V}$ into identical intervals of the
length $\Delta V$.
Then each eigenvalue $V=n \cdot \Delta V$ ($n=1,2,\ldots$) is assigned to an interval-number $I_n$ defined by 
$I_n=round\big(\frac{V}{\Delta V} +0.5 \big) $. In the third step we add up all eigenvalues
belonging to the same interval $I_n$ to get the number of eigenvalues in the interval 
$[V-\Delta V,V] $.\\

Now we define the Interval density $N_I$ of eigenvalues in the interval $I$ for fixed $j_4=j$ by:
\be
  N_I(j):=\frac{\#eigenvalues(I)}{\Delta V}
\ee

Since we want to have a normalized density $\rho$:
\[
 \int\limits_{V_{min}}^{V_{max}}\rho_I~dV\stackrel{!}{=}1	   
\]
we divide the interval densities $N_I$ by the number of eigenvalues different from 0 to get
the final definition of the eigenvalue density: 
\be \label{definition eigenvalue density}
  \rho_I(j):=\frac{\#eigenvalues(I)}{\Delta V \cdot (\#total~eigenvalues-\#0~eigenvalues)}
\ee

These densities are then represented by a point at $V=n\cdot \Delta V$ for each interval $I_n$. These points are joined then (they can be fitted by polynomials for instance) to display the desired normalized eigenvalue density $\rho$.
This gives us for instance for $j_4=j=50$ the following plot for a fit with a fourth order polynomial 
in the eigenvalues $V$ ($\Delta V = 0.5$):

\begin{figure}[!hbt]
\begin{center}
  \begin{minipage}[t]{7.5cm} 
      \includegraphics[height=9cm]{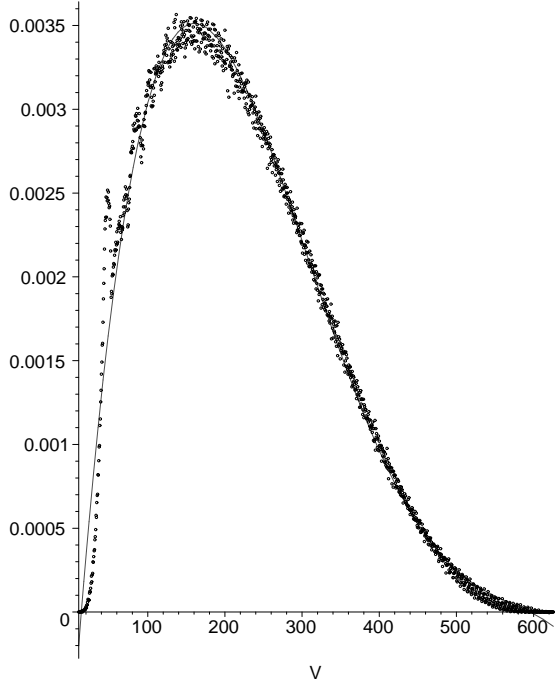}
      \caption{The eigenvalue density for $j_4=50$ (points) fitted by a 4th order polynomial (solid
      line).}
      \label{single spectrum j_4=50}
  \end{minipage} 
\end{center}
\end{figure}

\begin{figure}[!hbt]
\begin{center}
  \begin{minipage}[t]{7.5cm} 
      \includegraphics[height=8.5cm]{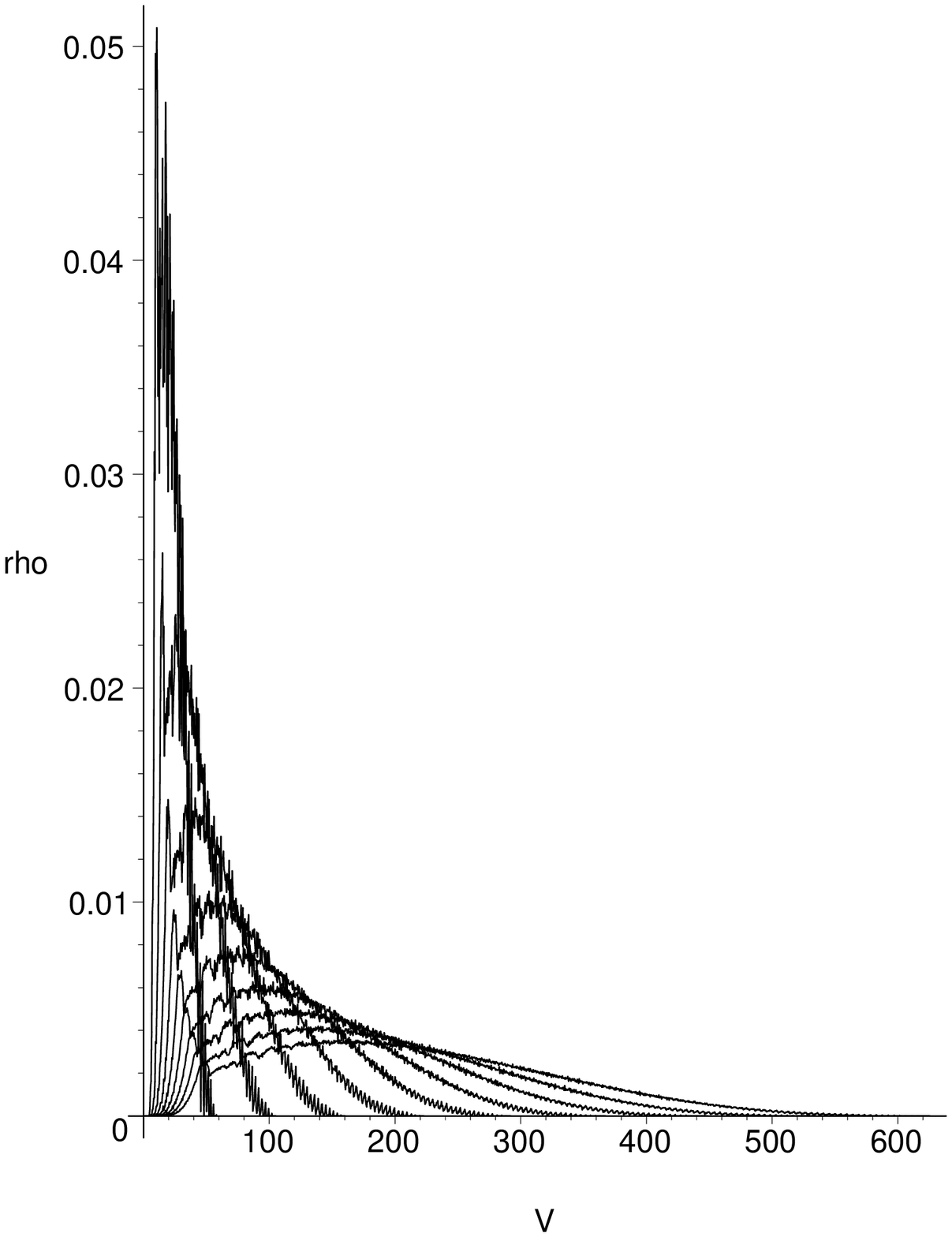}
       \caption{The eigenvalue densities for $j_4=10,15,\ldots,50$ (just 
look at the biggest eigenvalues in order to
      identify, which curve belongs to which $j$). The points representing the eigenvalue
      density, are joined by lines.}
      \label{single spectra}
  \end{minipage} 
\hfill
   \begin{minipage}[t]{7cm}
      \includegraphics[height=8.5cm]{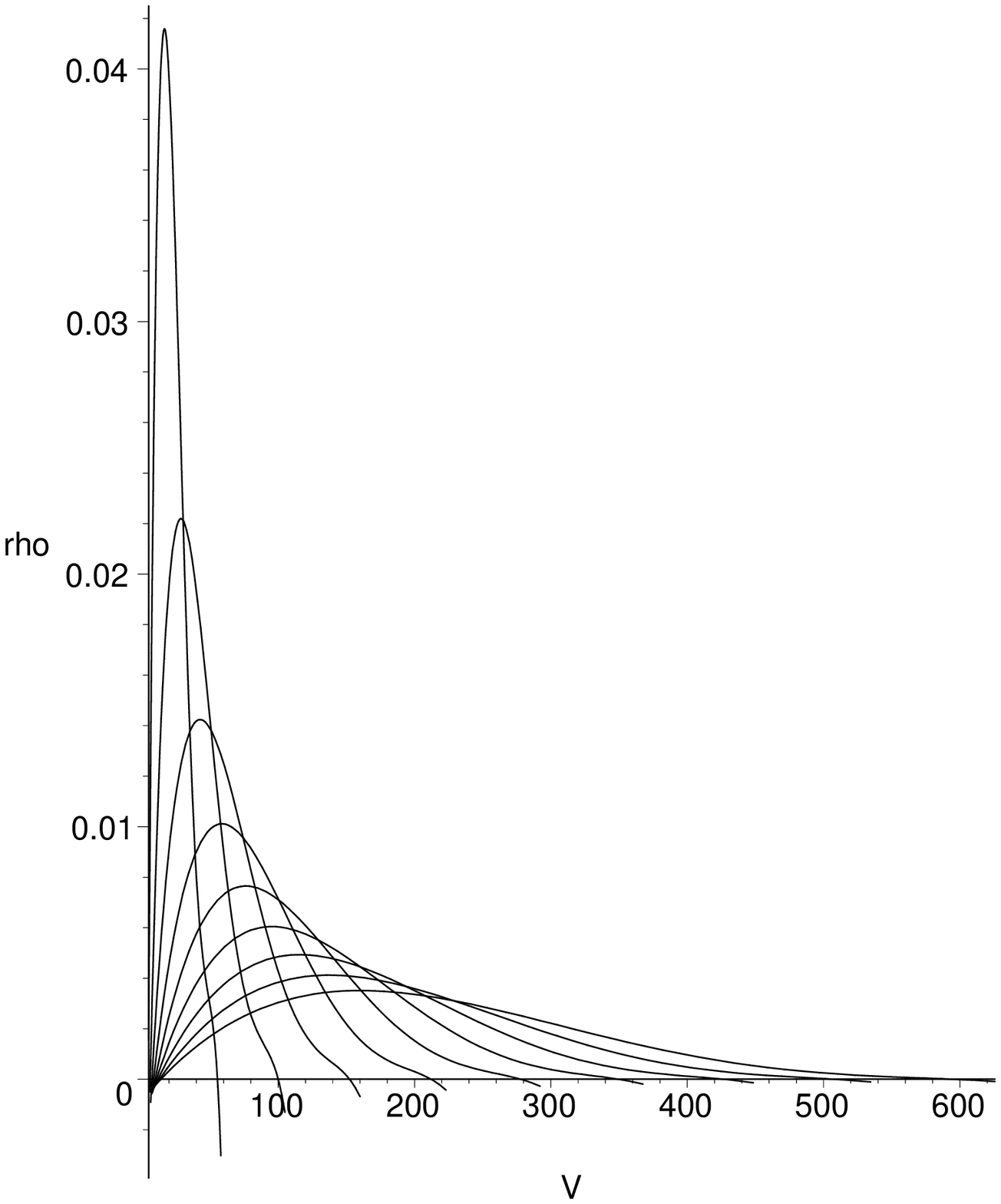}
      \caption{A fit of the spectra described in figure \ref{single 
spectra} by polynomials of 4th order.}
      \label{single spectra fitted}  
   \end{minipage}   
\end{center}
\end{figure}

\begin{figure}[!hbt]
\begin{center}
  \begin{minipage}[t]{7.5cm} 
      \includegraphics[height=8.8cm]{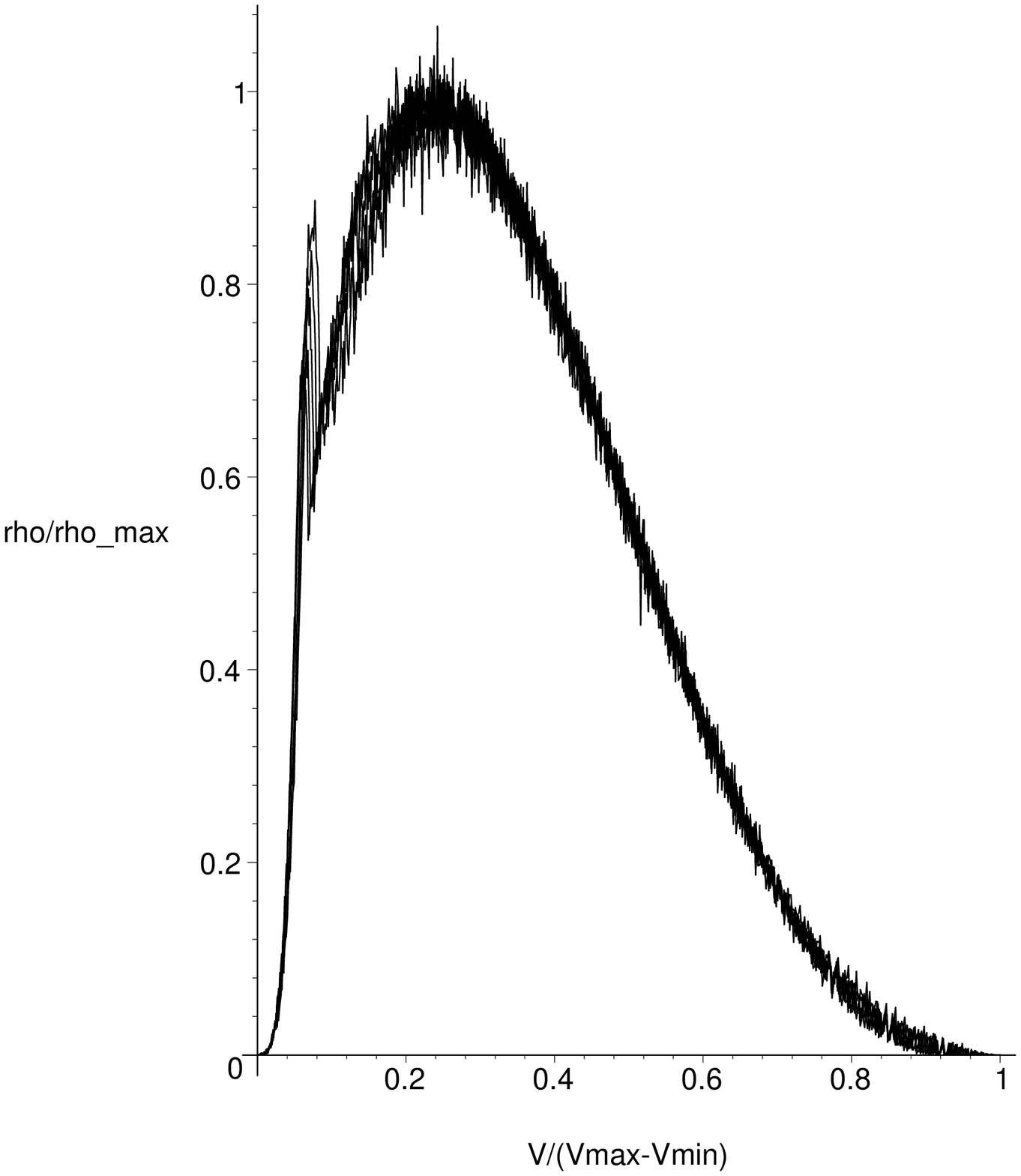}
      \caption{The eigenvalue densities for $j_4=30,35,\ldots,50$ in a 'fully normalized' rescaling, that is
      $V \to \frac{V-V_{min}}{V_{min}-V_{max}}$, $\rho_I \to \frac{\rho_I}{\max(\rho_I)}$}
      \label{fully normalized single spectra}
  \end{minipage} 
\hfill
   \begin{minipage}[t]{7cm}
      \includegraphics[height=8.8cm]{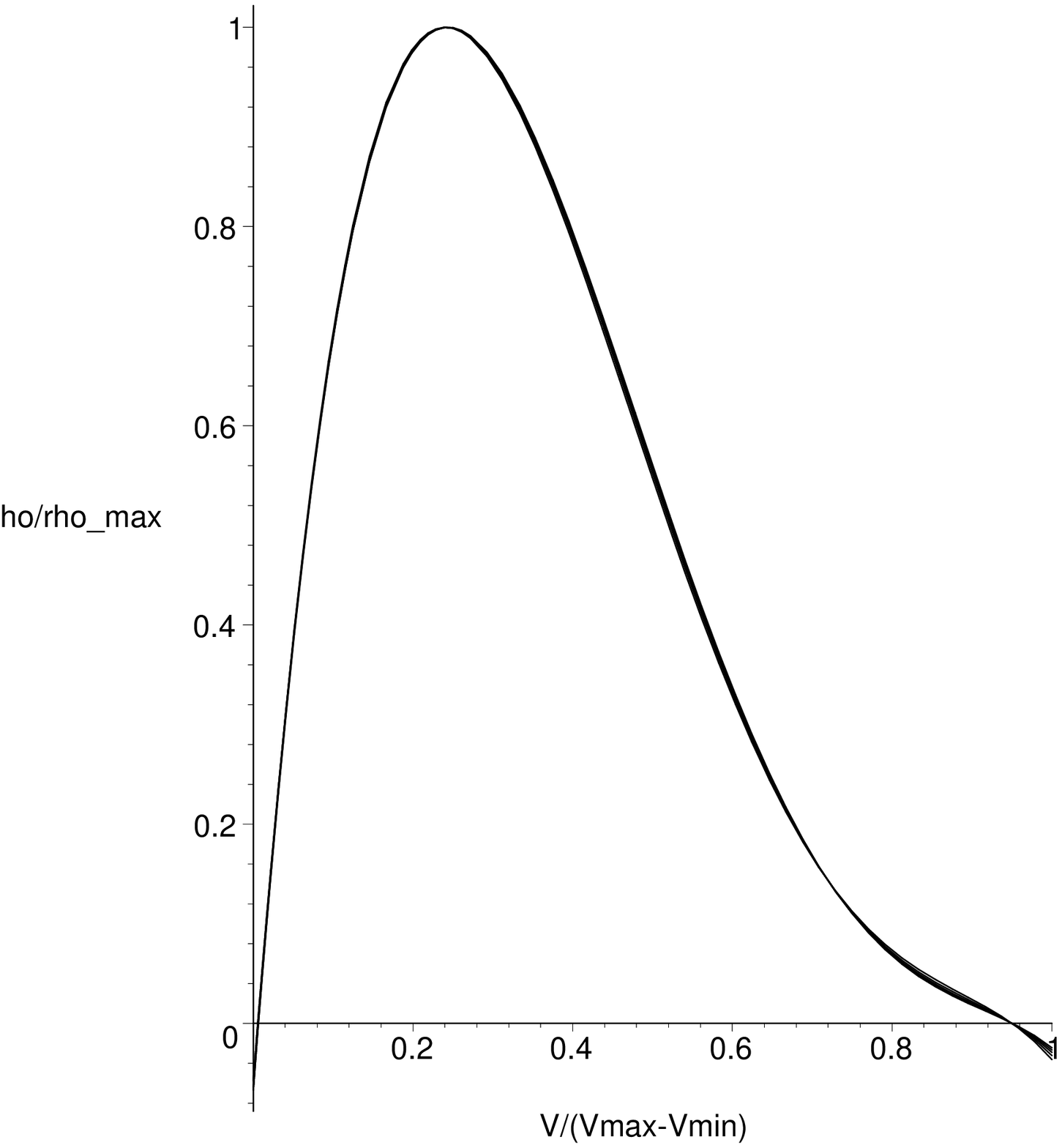}
      \caption{The fit curves of the spectra in 'fully normalized' rescaling }
      \label{fully normalized single spectra fitted}  
   \end{minipage}   
\end{center}
\end{figure}

This can be done for every matrix set $S_j$ with fixed $j=j_4$.
Remarkably it seems to be true, that the eigenvalue densities in fig.\ref{single spectra} are fitted
quite well by 4th-order polynomials. But it is even more surprising that if we rescale the obtained densities,
by putting their width $W:=V_{max}(j)-V_{min}(j) \to 1$ and their height $H(j):=\max(\rho_I(j)) \to 1$ (where
$\max(\rho_I(j))$ is taken from the fit curves) and plotting the resulting rescaled distributions for different values of $j_4=j$ into the interval $[0,1]$ as given in figure 
\ref{fully normalized single spectra} and figure \ref{fully normalized single spectra fitted}, the
distribution seem to possess a similar shape.

Hence, the normalized distrbutions seem to be {\underline{independent of $j$}}, that is, universal.
This discussion suggests to try to define a limit distribution.
By taking into account the behaviour of the ratio of the distance between the begin of the distributions $V_{min}(j)$ to the value $V(H(j))$ at which the maximum $H(j)$ of the distribution is situated and the total length of the distribution $V_{max}(j)-V_{min}(j)$.
\[
   \Delta(j):=\frac{V(H(j))-V_{min}(j)}{V_{max}(j)-V_{min}(j)}
\]
The ratio $\Delta(j)$ should tend to a constant value for $j=j_4 \to \infty$ in the presence of a limit common
shape of all distributions.
Moreover we want to find out the quality of the fits taken by calculating the 
average squared difference between the fitted curves and the real spectra, given by (all quantities at given $j$):
\[
   \chi^2:=\frac{1}{\max(\rho_I)(V_{max}-V_{min})}
         \displaystyle\sum_{I(V_{min})}^{I(V_{max})} \bigg(\rho_I-\rho_I^{(fitted)}\bigg)^2
\]

These quantities seem to behave in a way which is convenient for us (see figures 
\ref{ratio max rho and distribution width}, \ref{chi_2 between spectra and their fits}) :
\begin{figure}[!hbt]
\begin{center}
  \begin{minipage}[t]{7.5cm} 
      \includegraphics[height=8cm]{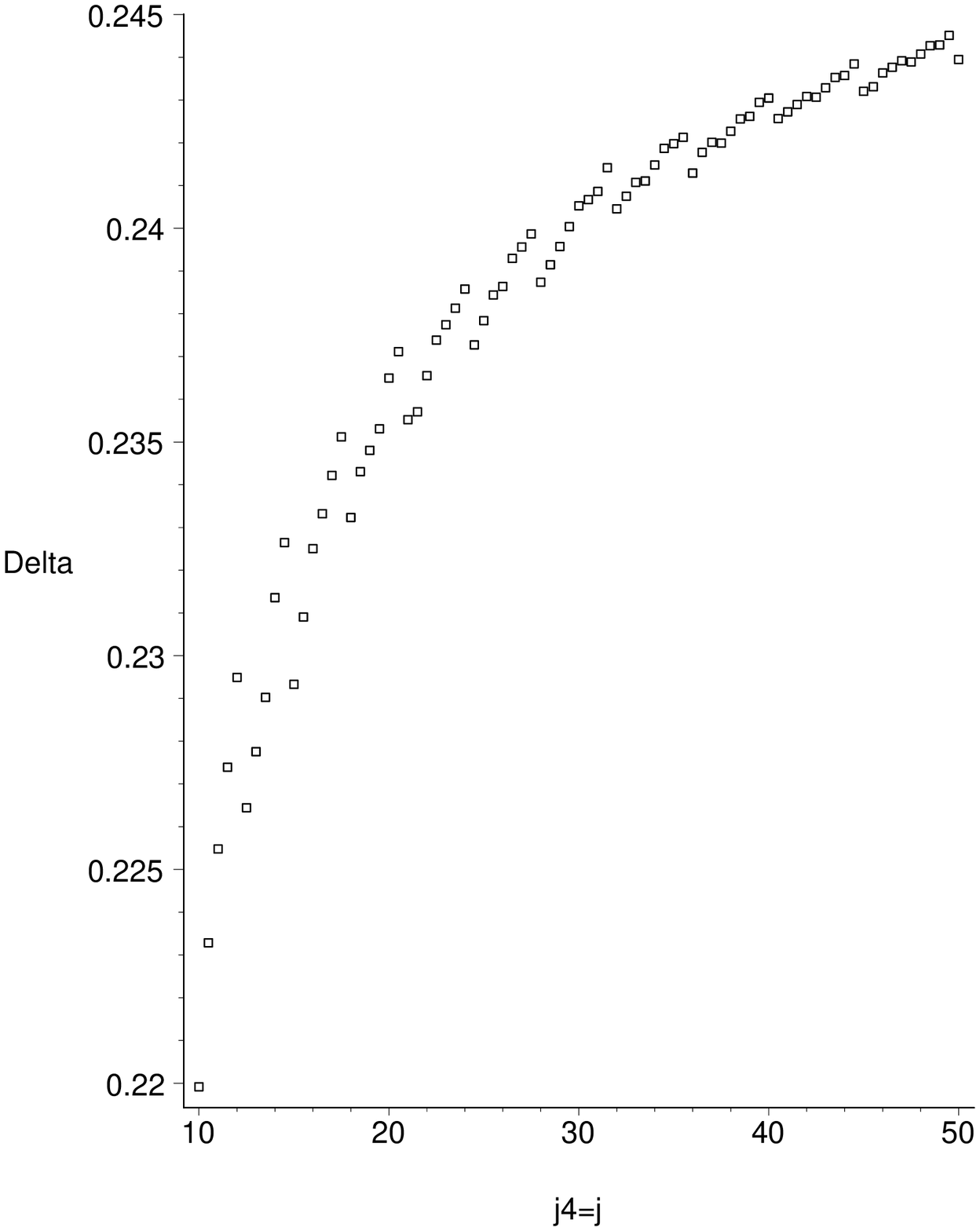}
      \caption{The ratio $\Delta(j)$ dependent on $j_4=j$}
      \label{ratio max rho and distribution width}
  \end{minipage} 
\hfill
   \begin{minipage}[t]{7.5cm}
      \includegraphics[height=8cm]{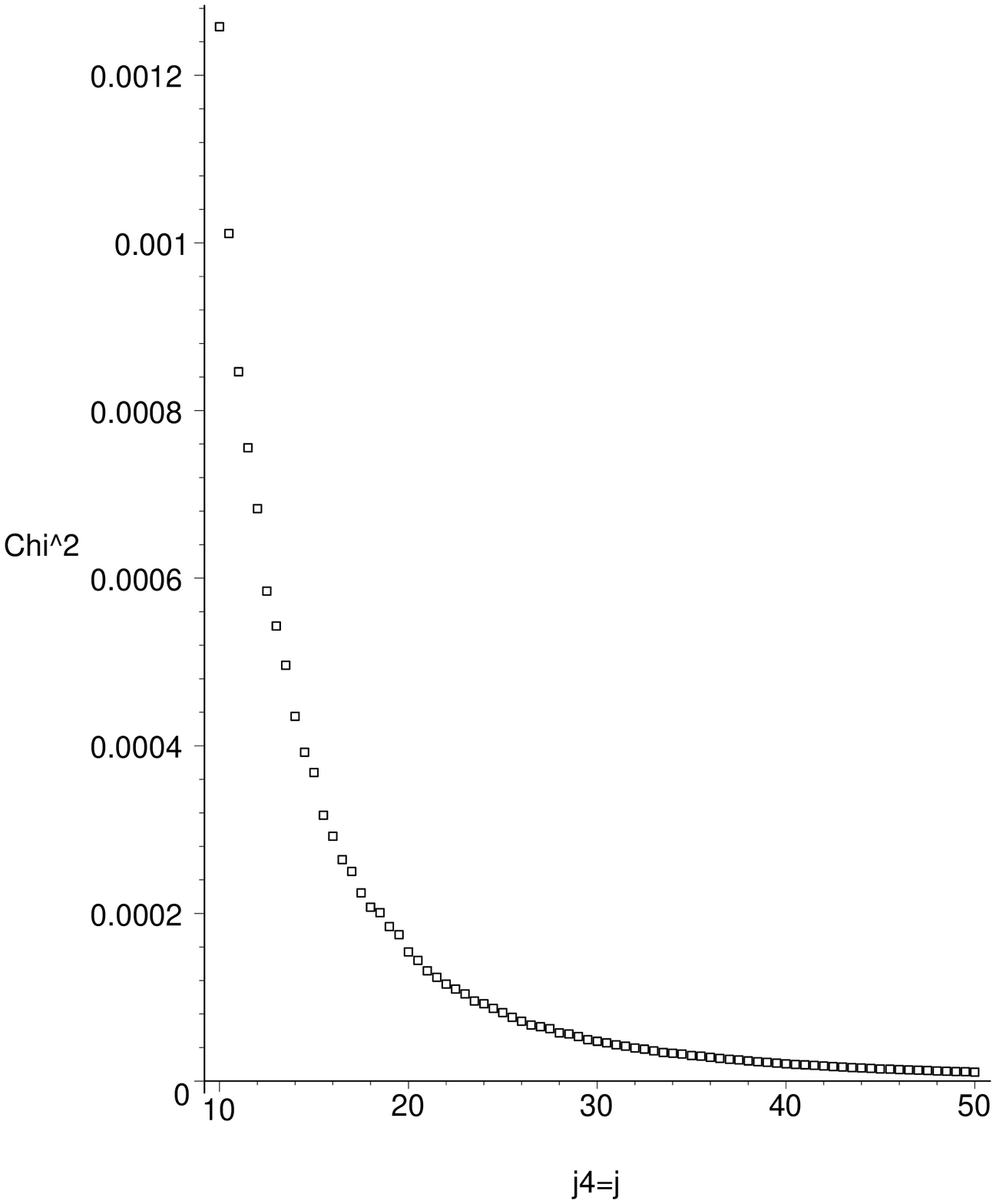}
      \caption{The sum of the squared distances between points of the eigenvalue densities and the fitted
      curves, $\chi^2$, (defined as above) in dependence of $j_4=j$}
      \label{chi_2 between spectra and their fits}  
   \end{minipage}   
\end{center}
\end{figure}

That is, the fit quality improves as $j$ is growing and it seems possible, that the ratio $\Delta(j)$ indeed has a certain  limit of $\approx 0.25$~.


\subsubsection{Density of the Eigenvalues of the Volume Operator - Discussion}
Despite the interesting properties and the occurrence of some systematics in the spectral density in the
matrix sets, we were unable up to this point to give an estimate of the density of eigenvalues of the
whole volume operator so far.
 
The problem is first, that the $4$-th order polynomials were chosen only for
the reason, that they contain the lowest number of parameters ($5$), the achieved spectra can be
satisfactorily fitted by. There has more computational work to be done to ensure the presence of la limiting
eigenvalue distribution. If this turned out to be the case, then we have to look for certain points,
the parameters of the fitted curves of the eigenvalue distribution are fixed by. 

What we know (at least 
our computations up to now encourage us to think that we know)
are 4 parameters: The maximum volume $V_{max} \sim j_{max}^{\frac{3}{2}}$, 
the minimal volume $V_{min} \sim  j_{max}^{\frac{1}{2}}$, the maximum of the distribution, situated
(see fig. \ref{ratio max rho and distribution width}) at $\sim 0.25 \cdot V_{max}$, and we know the
total number $E^{(j_4=j_{max})}$ of eigenvalues , which can be obtained  
from the total number of eigenvalues $E^{(tot)}$ for
each $j_{max}$ by considering $E^{(j_4=j_{max})}=E^{(tot)}(j_{max})- E^{(tot)}(j_{max}-\frac{1}{2})$.
Then we could compute the eigenvalue density in an interval $[V_1,V_2]$, 
by a superposition of all configurations with $V_{min} <V_2$ and $V_{max}>V_1$.

The second problem is then to give a fit-function for the total density of eigenvalues for all
configurations. There is one conjecture, based on a similar behaviour for the spectrum of the area
operator (which is easier to handle), that this eigenvalue density should behave as 
\[
  \rho(V)=\alpha e^{\beta V^{\gamma}} 
\]
We have tried to fit the part of the spectrum we have fully calculated $(V\le V_{max}(j_{max}=50))$ by
this formula. But it seemed to be impossible to give certain values to the three
parameters, especially to $\gamma$. Maybe the calculated part of the spectrum is still to small,
i.e. not sufficient for statistics or one must consider also higher valent vertices. Of course, the
conjecture could also be wrong. So it is
left as an open (but inspired by our done calculations not hopeless) task, to fix the density of
eigenvalues. 
\\\\
Let us conclude by displaying here the complete calculated part of the eigenvalue density of the volume
operator for $j_{max}=50$, according to our numerical criterion, that we can rely on for given $j_{max}$ on
the part of the spectrum with $V\le \sqrt{2\sqrt{j_{max}(j_{max}+1)}}$. That is for $j_{max}=50$ (as we
calculated) we have obtained the full spectrum up to $V \approx 10$. 

By inspection of figure 
\ref{first 100 eigenvalues 1} we can extend this part up to $V \approx \sqrt{200}\sim 14$ if we draw a horizontal line
into figure \ref{first 100 eigenvalues 1} at a given $10 \le V\le 14$ and count the eigenvalue series
situated below that line. If we assume, that these series grow linearly with growing $j_{max}$, as we expect, and there do not occur additional eigenvalue series at higher $j_{max}$ then we
can simply extend the curves $j\longrightarrow \lambda_k(j)$ linearly for $k=1\ldots 8$ and can thereby estimate (approximately) the additional contributions not yet calculated explicitly for $V^2\le 200$ that is for the $k=7$ eigenvalue series.

\begin{figure}[!hbt]
\begin{center}
  \begin{minipage}[t]{12.5cm} 
      \includegraphics[height=9cm,width=10cm]{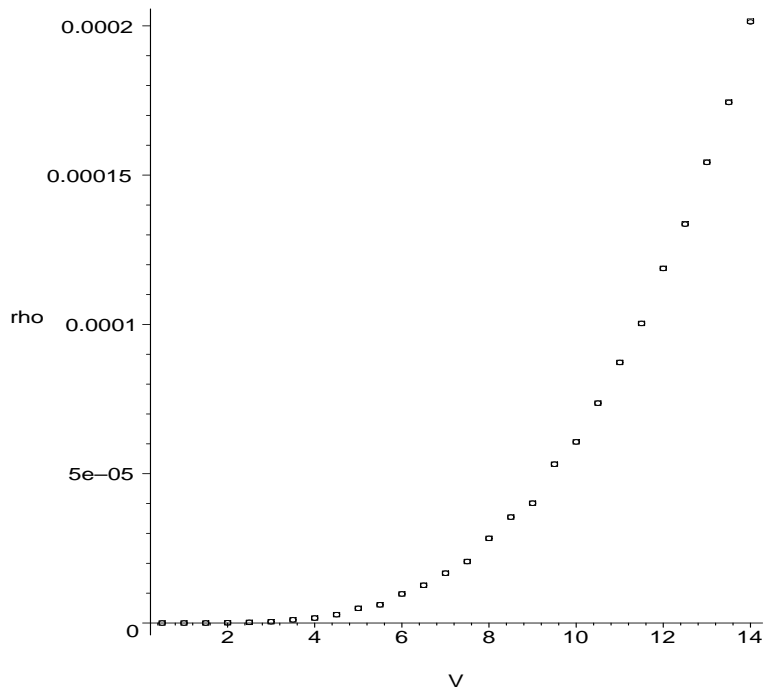}
      \caption{The eigenvalue density of the total volume operator (circles) and the extended eigenvalue
      density (boxes) for $j_{max}=50$~ on a 4-valent vertex}
      \label{complete spectrum j_max=50}
  \end{minipage} 
\end{center}
\end{figure}

Therefore we display the original
spectrum (circles) and the extended spectrum (boxes) normalized with respect to the total number of
non-zero eigenvalues of the original spectrum (we have chosen again an interval width $\Delta V=0.5$, the
eigenvalue density is defined as in (\ref{definition eigenvalue density})~) in figure 
\ref{complete spectrum j_max=50}.

\pagebreak

\section{Summary \& Outlook}
In this article we have analyzed the spectral properties of the Volume Operator defined in Loop Quantum
Gravity.

We discussed the matrix representation of the volume operator with respect to gauge
invariant spin network functions and were able to derive a drastically simplified formula for the matrix elements of the volume operator with respect to the latter.

It turned out that there exist certain selection rules for the matrix elements and  all the matrices are
$i$-times an antisymmetric matrix with the structure of a Jacobi matrix, meaning that non-vanishing matrix
elements are only situated on certain off-diagonals. 
\\\\
We were able to determine the kernel, that is, the eigenstates for the eigenvalue 0,
of the volume operator with respect to the gauge invariant 4-vertex
analytically as given in (\ref{kernel I}), (\ref{kernel II}).
We have done numerical investigations for the gauge invariant 4-vertex.
Our numerical investigations support the analytical estimate, that there exists a smallest eigenvalue 
$V_{min}$ dependent of the the maximal spin $j_{max}$ via $V_{min}\sim \big({j_{max}}\big)^{\frac{1}{2}}$ 
and a maximal eigenvalue $V_{max}\sim \big(j_{max}\big)^{\frac{3}{2}}$.
Therefore we were able to find certain numerical indicators for the completeness of a numerically computed
part of the spectrum. Moreover, we found that the geometrical intuition is 
reflected in the spectrum: At given $j_{max}$ the lowest non -- zero 
eigenvalues come from rather ``distorted'', almost flat tetrahedra with 
large spin on some edges and low spin on the others. On the other hand, 
the largest eigenvalues come from regular tetrahedra with large spin on 
all edges. \\\\
For future analysis one should extend the numerical calculations for the 
gauge invariant 4-vertex and higher n-valent vertices to verify and 
possibliy extend (for higher valence vertices) the regularities of the 
spectra obtained (for the gauge invariant 4-vertex) for single matrix 
sets $S_j$ with $j_4=j$  at higher spins. In particular it would be 
interesing to see whether there exists a volume gap as $n \longrightarrow 
\infty$. This, however, requires more computing power and better 
programming than we have used in this paper.

The formula derived for the matrix  elements with respect to a gauge 
invariant n-valent vertex can be
used to analyze the whole spectrum of the volume operator numerically and  
analytically. Further 
simplifications are conceivable. 
\\
Notice again that the restriction to 4-valued vertices and $j_{max}\le 50$ was only due to the computational capacity of the used mathematical software $MAPLE~7$ and computer. 
By using a compiler-based programming language and optimized numerical 
matrix-diagonalizing routines we expect to be able to go much beyond the computational limits above.
\\\\
Due to the results presented in this paper it seems to the authors, that 
there are good chances for getting sufficient control
about the spectral behaviour of the volume operator in the future, 
especially when it comes to dynamical questions in LQG.\\
\\
As a first qualitative application in that respect, notice the 
following:\\
We have shown analytically that the volume operator of full LQG has zero
eigenvalues at arbitrarily large $j_{max}$ and that their number grows as
$j_{max}^4$ as compared to the total number of eigenvalues which grows as 
$j_{max}^5$, at least for the gauge invariant four vertex which should be 
the most interesting case from a triangulation point of view. Moreover,
the volume gap increases as $j_{max}^{1/2}$. 
It follows that the full spectrum contains many ``flat 
directions'' or ``valleys'' of zero volume and the walls of the valleys 
presumably get steeper as we increase $j_{max}$. Therefore we might find 
arbitrarily large 
eigenvalues as close as we want to zero eigenvalues and hence the 
``derivative'' (rather: difference) of the spectrum around zero volume 
which enters the 
Hamiltonian constraint through the curvature operator, while 
well -- defined 
as shown in \cite{TT:QSD I,TT:QSD II}, could be unbounded from above. 
Therefore,
the full spectrum could not share an important property of the spectrum 
in the cosmological truncation of LQG \cite{4a} where the 
derivative of the spectrum at zero volume is bounded from above. 
As this property has been somewhat important in \cite{4a}, some of
the results of \cite{4a} might have to be revisited in the full theory. 
The 
challenge would be to show that the curvature expectation value remains 
bounded 
when the system is prepared in a semiclassical state for LQG,
see e.g. \cite{6} and references therein. First evidence for 
this and further analysis will be presented soon in \cite{7} thus 
reaffirming the spectacular results of \cite{4a} in the full theory.

\section*{Acknowledgements}

J.B. thanks the Gottlieb Daimler- und Karl Benz Stiftung for financial 
support. This work was supported in part by a grant from NSERC of Canada
to the Perimeter Institute for Theoretical Physics.

\pagebreak


\begin{appendix}

\section{Basics of Recoupling Theory}

\subsection{Angular Momentum in Quantum Mechanics}

For the angular momentum operator $\vec{J}=(J_1,J_2,J_3)$ (where each component has to be seen as an
operator) we have the following commutation relations:
\be \label{CR J}  \big[J_i,J_j\big]=i\cdot \epsilon_{ijk} J_k \hspace{1cm}
                  \big[J^2,J_j\big]=0     
\ee
with $J^2=\vec{J}~^2=J_1^2+J_2^2+J_3^2$.%

Additionally we can define 
\be J_+=J_1+iJ_2 \hspace{1cm} J_-=J_1-iJ_2 \ee\ 
with (using (\ref{CR J})~):
\be   
  \big[ J^2,J_{\pm} \big]=0    \hspace{5mm} \big[J_3,J_{+} \big]=J_+ \hspace{5mm}
  \big[ J^2,J_{-} \big]=-J_{-} \hspace{5mm} \big[J_{+},J_{-} \big]=2J_3 
\ee%

Since every angular momentum state is completely determined
\footnote{In the sense that we have a maximal set of simulaneously measureable observables.} 
by its total angular momentum quantum number $j$ (where $J^2=j(j+1)$) and one component say 
$J_3$ (where $J_3=-j,-j+1,\ldots,j-1,j$ ) we then associate for certain $j$ a $2j+1$ dimensional  
\footnote{for fixed $j$ there are $2j+1$ values which $J_3$ can take.} 
Hilbert space $\mathcal H$ equipped with an orthonormal basis $|j~m>$, $m=J_3$ where:

\be \label{conditions in single HS} 
  <j~m|j~m'>=\delta_{mm'} 
\ee

The $|~j~m>$ simultaneously diagonalize the 2 operators of the squared angular momentum $J^2$ and the 
magnetic quantum number $J_3$~~\cite{Edmonds}:
\be \label{operators in single HS 1}
      J^2|j~m> = j(j+1) |j~m> \hspace{1cm} J_3|j~m>=m|j~m> 
\ee 
That is, $|j~m>$ is a maximal set of simultanous eigenvectors of $J^2$ and $J_3$.

On these eigenvectors the other operators act as
\be \label{operators in single HS 2}
    J_+|j~m>=\sqrt{j(j+1)-m(m+1)}~|j~m+1>\hspace{3mm} J_-|j~m>=\sqrt{j(j+1)-m(m-1)}~|j~m-1>
\ee

\subsection{Fundamental Recoupling}

Equipped with a small part of representation theory we can easily understand what happens if we couple
several angular momenta. For that we first repeat the well known theorem of Clebsh \& Gordan on
tensor products of representations of $SU(2)$:
\begin{Theorem}{Clebsh \& Gordan} \label{Clebsh Gordan theorem}\\
   Having two irreducible representations $\pi_{j_1}$, $\pi_{j_2}$ of $SU(2)$
   with weights $j_1,j_2$ 
   their tensor product space splits into a direct sum of irreducible representations $\pi_{j_{12}}$ with 
   $|j_1-j_2|\le j_{12}\le j_1+j_2$ such that
   
   \[ \pi_{j_1} \otimes \pi_{j_2}=\pi_{j_1+j_2}\oplus \pi_{j_1+j_2-1} \oplus \ldots 
                                \oplus \pi_{|j_1-j_2+1|}\oplus \pi_{|j_1-j_2|} \]
\end{Theorem}

Equivalently we can write for the resulting representation space 
$\mathcal{H}^{(D)}=\mathcal{H}^{(D_1)} \otimes \mathcal{H}^{(D_2)}$ 
(where $D_1=2j_1+1,D_2=2j_2+1,D=D_1\cdot D_2$ denote the dimensions of the Hilbert spaces):
\be \label{Hilbert space decay}
  \mathcal{H}^{(D)} =\mathcal{H}^{(D_1)} \otimes \mathcal{H}^{(D_2)}
   =\bigoplus\limits_{j_{12}=|j_1-j_2|}^{j_1+j_2}\mathcal{H}^{(2j_{12} +1)} 
\ee
 
Or in other words: 
If we couple two angular momenta $j_1,j_2$ we can get resulting angular momenta $j_{12}$
varying in the range $|j_1-j_2| \le j_{12} \le j_1+j_2$.

The tensor product space of two representations of $SU(2)$ decomposes into a direct sum of
representation spaces, that is one space for every possible value of recoupling  $j_{12}$ with the
according dimension $2j_{12}+1$.

\subsection{Recoupling of $Two$ Angular Momenta} \label{couple TWO}

According to (\ref{Hilbert space decay}) we can expand for each value of $j_{12}$ the elements
$|j_1~j_2;j_{12}(j_1,j_2)~M> ~\in \mathcal{H}^{(2j_{12}+1)}$ called
`coupled states' into the tensor basis $|j_1~m_1> \otimes~|j_2~m_2> $ of $\mathcal{H}^{(D)}$:
\ba 
   |j_1~j_2;j_{12}(j_1,j_2),M>&=&\sum_{m_1+m_2=M}
    \underbrace{<j_1~m_1;j_2~m_2|j_1~j_2;j_{12}(j_1,j_2),M>}_{C_{m_1m_2}}|j_1~m_1>\otimes|j_2~m_2>
\ea
Here $C_{m_1m_2} \in \mathbbm R$ denotes the expansion coefficients, the so called Clebsh-Gordan 
coefficients.
On the right hand side $|j_1~m_1;j_2~m_2>=|j_1~m_1>\otimes~|j_2~m_2>$

If we change the order of coupling then $C_{m_1m_2}$ changes its sign:
\ba \label{CGC signchange}    
  <j_1~m_1;j_2~m_2|j_{12}(j_1,j_2)~M=m_1+m_2>
  &=&(-1)^{j_{12}-j_1-j_2}<j_2~m_2;j_1~m_1|j_{12}(j_2,j_1)~M=m_1+m_2>\nonumber\\
  &=&(-1)^{-j_{12}+j_1+j_2}<j_2~m_2;j_1~m_1|j_{12}(j_2,j_1)~M=m_1+m_2>\nonumber     
\ea
As $\mp j_{12}\pm (j_1+j_2)$ is an integer number we are allowed to switch the signs in the exponent of
the factor ($-1$).
The coupled states again form an orthonormal basis:
\ba \label{CGC completeness}
  <j_1~j_2;j_{12}(j_1,j_2),M|j_1~j_2;j_{12}(j_1,j_2),M> &\stackrel{!}{=}&1 \\
  \label{Orthogonality of coupled states}
  <j_1~j_2;\widetilde{j_{12}}(j_1,j_2),\widetilde{M}|j_1~j_2;j_{12}(j_1,j_2),M> &\stackrel{!}{=}&
           \delta_{\widetilde{j_{12}},j_{12}}\delta_{\widetilde{M}M} 
\ea 

In (\ref{Orthogonality of coupled states}) $\delta_{\widetilde{j_{12}},j_{12}}$ comes from the orthogonality
of the $\mathcal{H}^{(2j_{12}+1)}$ in (\ref{Hilbert space decay}), $\delta_{\widetilde{M}M}$ is caused by
the othogonality of the single $|j_1~m_1>,|j_2~m_2>$ (\ref{conditions in single HS})- since always
$M=m_1+m_2$.    

Normalization of the recoupled states (\ref{CGC completeness}) implies according to
(\ref{conditions in single HS}):
\be
      \sum_{m_1+m_2=M}|<j_1~j_2;j_{12}(j_1,j_2),M|j_1~j_2;j_{12}(j_1,j_2),M>|^2=
      \sum_{m_1+m_2=M}|C_{m_1m_2}|^2=1
\ee

Furthermore the $Clebsh-Gordan-coefficients~are~all~real$, which is not obvious, but a result of two
conventions one usually requires \cite{Edmonds}:
\begin{itemize}
  \item[1)] $|j_1~j_2; j_{12}(j_1,j_2)=j_1+j_2~M=j_1+j_2> = |j_1~m_1=j_1> \otimes~ |j_2~m_2=j_2> $
  \item[2)] All matrix elements of $J_3^{(D_1)}$, which are nondiagonal in $|j_1~j_2;j_{12}(j_1,j_2)~M>$ are
            real and nonnegative.
\end{itemize}

The maximal set of simultanuosly diagonalizeable (that is commuting) $2\cdot 2$ operators (\ref{operators in single HS 1})
of the single Hilbertspaces $\mathcal{H}^{(D_1)},\mathcal{H}^{(D_2)}$  is then in $\mathcal{H}^{(D)}$
replaced \footnote{We are a bit sloppy in using this notation, correctly we would have to write:\\
$(J^{(D)})^2=(J^{(D_1)}\otimes \mathbbm{1}_{\mathcal{H}^{(D_2)}}+
             \mathbbm{1}_{\mathcal{H}^{(D_1)}} \otimes J^{(D_2)})^2 
   \hspace{5mm} J_3^{(D)}=J_3^{(D_1)} \otimes \mathbbm{1}_{\mathcal{H}^{(D_2)}}+
                          \mathbbm{1}_{\mathcal{H}^{(D_1)}} \otimes J_3^{(D_2)} $ ~~~~and \\
$                       (J^{(D_1)})^2 =(J^{(D_1)})^2 \otimes \mathbbm{1}_{\mathcal{H}^{(D_2)}}
   \hspace{5mm}         (J^{(D_2)})^2 =\mathbbm{1}_{\mathcal{H}^{(D_1)}} \otimes (J^{(D_2)})^2 $}
by 4 operators: total angular momentum $(J^{(D)})^2$, total projection quantum number
$J_3^{(D)}$, single total angular momenta $(J^{(D_1))^2}$, $(J^{(D_2)})^2$:

\be
  (J^{(D)})^2=(J^{(D_1)}+J^{(D_2)})^2 \hspace{5mm} J_3^{(D)}=J_3^{(D_1)}+J_3^{(D_2)} \hspace{5mm} 
                        (J^{(D_1)})^2 \hspace{5mm} (J^{(D_2)})^2
\ee
which are simultanously digonal in the new basis manifested through: 
\ba \label{new operators from recoupling of two angular momenta}
   (J^{(D)})^2~|j_1~j_2;j_{12}(j_1,j_2),M>&=&j_{12}(j_{12}+1)~|j_1~j_2;j_{12}(j_1,j_2),M>\nonumber\\
   J_3^{(D)}  ~|j_1~j_2;j_{12}(j_1,j_2),M>&=&               M~|j_1~j_2;j_{12}(j_1,j_2),M>\nonumber\\
   (J^{(D_1)})^2~|j_1~j_2;j_{12}(j_1,j_2),M>&=&j_1(j_1+1)~|j_1~j_2;j_{12}(j_1,j_2),M>\nonumber\\
   (J^{(D_2)})^2~|j_1~j_2;j_{12}(j_1,j_2),M>&=&j_2(j_2+1)~|j_1~j_2;j_{12}(j_1,j_2),M>
\ea

\subsection{Recoupling of $Three$ Angular Momenta - $6j$-Symbols}
 
 In this way we can expand the recoupling of three angular momenta in terms of CGC.
 \ba
   |j_{12}(j_1,j_2),j(j_{12},j_3)>
   &=&|j_{12}(j_1,j_2)~j_3;j(j_{12},j_3)~M=m_1+m_2+m_3>\nonumber\\ \nonumber\\
   &=&\sum_{m_{12}~m_3}<j_{12}~m_{12};j_3~m_3 |j_{12}~j_3;j~m_{12}+m_3>\cdot\nonumber\\
   & & ~~~~~~~~~~~~~|j_{12}~m_{12}>|j_3~m_3>\nonumber\\ \nonumber\\         
   &=&\sum_{m_{12}~m_3}<j_{12}~m_{12};j_3~m_3 |j_{12}~j_3;j~m_{12}+m_3>\cdot\nonumber\\
   & &\sum_{m_1~m_2}<j_1~m_1;j_2~m_2|j_1~j_2;j_{12}~m_{12}=m_1+m_2>\cdot\nonumber\\
   & & ~~~~~~~~~~~~~|j_1~m_1>|j_2~m_2>|j_3~m_3>\nonumber\\ \nonumber\\
   &=&\sum_{m_1~m_2}<j_{12}~m_1+m_2;j_3~M-m_1-m_2 |j_{12}~j_3;j~M>\cdot\nonumber\\
   & & ~~~~~~~~<j_1~m_1;j_2~m_2|j_1~j_2;j_{12}~m_1+m_2>\cdot\nonumber\\
   & & ~~~~~~~~~~~~~|j_1~m_1>|j_2~m_2>|j_3~M-m_1-m_2>
 \ea

As we can see, as we couple angular momenta successively, the order of coupling plays an
important role. Different orders of coupling will lead to different phases of the wavefunctions 
(see (\ref{CGC signchange})~).
Concerning this it would be nice to have a transformation connecting different ways of recoupling.
This tranformation between two different ways of coupling 3 angular momenta $j_1,j_2,j_3$ to a
resulting $j$ defines the $6j$-symbols, see appendix \ref{6 j}.

\subsection{Recoupling of $n$ Angular Momenta - $3nj$-Symbols}\label{introduce recoupling scheme}

As mentioned before the case of successive coupling of 3 angular momenta to a resulting $j$
can be generalized. For this purpose let us first comment on the generalization principle before we go into detailed
definitions. 

Theorem \ref{Clebsh Gordan theorem} can be applied to a tensor product of $n$
representations $\pi_{j_1}\otimes \pi_{j_2} \otimes \ldots \otimes \pi_{j_n}$ by reducing out step by step
every pair of representations. 
This procedure has to be carried out until all tensor products are reduced out.
One then ends up with a direct sum of representations each of them having a weight corresponding 
to an allowed value of the total angular momentum the $n$ single angular momenta 
$j_1,j_2,\ldots,j_n$ can couple to.   

But there is an arbitrariness in how one couples the $n$ angular momenta together, that is, the order 
by which $\pi_{j_1}\otimes \pi_{j_2} \otimes \ldots \otimes \pi_{j_n}$ is reduced out (by applying 
\ref{Clebsh Gordan theorem}) matters.
\\
\\
Let us now have a system of $n$ angular momenta.
First we fix a labelling of these momenta, such that we have $j_1,j_2,\ldots,j_n$. 
Again the first choice would be a tensor basis
$|\vec{j}~\vec{m}>$ of all single angular momentum states $|j_k~m_k>$, $k=1 \ldots n$ defined by:

\be \label{tensor basis n angular momenta}
  |\vec{j}~\vec{m}>=|(j_1,j_2,\ldots,j_n)~(m_1,m_2,\ldots,m_n)>:=
                   \bigotimes\limits_{k=1}^n |j_k~m_k>
\ee
with the maximal set of $2n$ commuting operators $(J_I)^2,J_I^3$, $(I=1,\ldots,n)$.

Now we proceed as in section \ref{couple TWO} finding the commuting operators according to
(\ref{new operators from recoupling of two angular momenta}), that is a basis in which the total angular
momentum $(J_{tot})^2=(J)^2=(J_1+J_2+\ldots+J_n)^2$ is diagonal (quantum number $j$)
together witch the total magnetic 
quantum number $J_{tot}^3=J^3=J_1^3+J_2^3+\ldots+J_n^3$ (quantum number $M$).

As $(J)^2 ~and~ J^3$ are 2 operators, we need $2(n-1)$ more quantum numbers of operators commuting with each
other and with $(J)^2 ~and~ J^3$ to have again a maximal set.
We choose therefore the $n$ operators $(J_I)^2$, $I=1,\ldots,n$ of total single angular momentum (quantum
numbers $(j_1,\ldots,j_n):=\vec{j})$. So we
are left with the task of finding addtional $n-2$ operators commuting with the remaining ones.
For this pupose we define:

\begin{Definition}{Recoupling Scheme} \label{Def Recoupling scheme}\\
   A recoupling scheme $|\vec{g}(IJ)~\vec{j}~j~m>$ is an orthonormal basis, diagonalizing besides $(J)^2$,
   $J^3$, $(J_I)^2$ \linebreak($I=1,\ldots,n$) the squares of the additional $n-2$ operators
   $G_2,G_3,\ldots ,G_{n-1}$ defined as\footnote{Note, that formally $G_n:=G_{n-1}+J_n=J_{total}$}: \\
   \begin{sloppypar}
   $ G_1:=J_I,~G_2:=G_1+J_J,~G_3:=G_2+J_1,~G_4:=G_3+J_2,~\ldots,
     ~G_I:=G_{I-1}+J_{I-2},$ \linebreak
     $G_{I+1}:=G_I+J_{I-1},~G_{I+2}:=G_{I+1}+J_{I+1},~G_{I+3}:=G_{I+2}+J_{I+2},
     ~\ldots,~G_J:=G_{J-1}+J_{J-1},$ \linebreak
     $~G_{J+1}:=G_J+J_{J+1},~G_{J+2}:=G_{J+1}+J_{J+2},~\ldots,~
     G_{n-1}:=G_{n-2}+J_{n-1}$\\
    \end{sloppypar}
    The vector $ \vec{g}(IJ):=\big(g_2(j_I,j_J),g_3(g_2,j_1),\ldots,g_{I+1}(g_I,j_{I-1}),
    g_{I+2}(g_{I+1},j_{I+1}),\ldots,g_J(g_{j-1},j_{j-1}),\linebreak 
    g_{j+1}(g_J,j_{j+1}),\ldots,g_{n-1}(g_{n-2},j_{n-1})\big)$ carries as quantum numbers the $n-2$
    eigenvalues of the operators\\ $(G_2)^2,~\ldots ,~(G_{n-1})^2$
 \end{Definition}

So we recouple first the angular momenta labelled by $I,J$ where $I<J$ and secondly all the other angular
momenta successively (all labels are with respect to the a fixed label set), by taking into account the
allowed values for each recoupling according to theorem \ref{Clebsh Gordan theorem}. 
\\[1cm]
Let us define
furthermore the so called $standard~recoupling~scheme$ or $standard~basis$:

\begin{Definition}{Standard Basis} \label{Def Standard basis}\\
   A recoupling scheme based on the pair $(I,J)=(1,2)$ with
   \[ G_K=\sum\limits_{L=1}^K J_L \hspace{10cm}\]
   is called standard basis.
\end{Definition}

Using definition \ref{Def Recoupling scheme} with the commutation relations (\ref{CR J}) and the fact, 
that single 
angular momentum operators acting on different single angular momentum Hilbert spaces
commute\footnote{That is $\big[ J_I^i,J_J^j \big]=0$ whenever $I \ne J$}, one can easily check
that for every recoupling scheme
\begin{itemize}
   \item[(i)]the $G_I$'s fulfill the angular momentum algebra (\ref{CR J}).
   \item[(ii)] $(J)^2,~(J_I)^2,~(G_K)^2,~J^3$ commute with each other $\forall~ I,K=1\ldots n$
\end{itemize}
Note, that it is sufficient to prove these two points in the Standard basis $\vec{g}(12)$, 
because every other basis $\vec{g}(IJ)$ is related to it by simply relabelling the $n$ angular momenta.\\

We have thus succeeded in giving an alternative description of an $n$ angular momenta system by all 
possible occurring intermediate recoupling stages $G_I$ instead of using the individual magnetic
quantum numbers. 

Obviously every orthonormal basis spanned by a recoupling scheme $|\vec{g}(IJ)~\vec{j}~j~m>$ 
is singled out by the labelling, namely the index pair $(IJ)$ and therefore not identical, as we have
already seen in the case of the $two$ angular momentum problem. So we are in need of a transformation
connecting the different bases that is expressing one basis, e.g. belonging to the pair $(IJ)$, in terms of
another basis, e.g. belonging to the pair $(KL)$, respectively. This leads to the following.

\begin{Definition}{$3nj$-Symbol} \label{Def 3nj-symbol}\\
   The generalized expansion coefficients of a recoupling scheme in terms of the standard recoupling
   scheme are called $3nj$-symbols:
   \[ |~\vec{g}(IJ)~\vec{j}~j~m> = \sum\limits_{all~\vec{g}'(12)} 
                      \underbrace{<\vec{g}'(12)~\vec{j}~j~m~|~\vec{g}(IJ)~\vec{j}~j~m>}_{3nj-symbol}~
		      |~\vec{g}'(12)~\vec{j}~j~m> 
   \]
   The summation has to be extended over all possible values of the intermediate recouplings \linebreak
   $\vec{g}'(12)=(g_2'(j_1,j_2),g_3'(g_2',j_3),\ldots,g_{n-1}'(g_{n-2}',j_{n-1})$, that
   is all values of each component $g_k'$ allowed by theorem \ref{Clebsh Gordan theorem}. 
\end{Definition} 

In calculations we will supress the quantum numbers $\vec{j},j,m$, since they are identical all the time, and
write abbreviating $<\vec{g}(IJ)~|~\vec{g}'(12)>$. 
Note additionally, the followoing properties of the $3nj$-symbols:
\begin{itemize}
   \item[(i)] They are real, due to the possibility to express them as Clebsh-Gordan-coefficients:
               \[<\vec{g}(IJ)~|~\vec{g}'(12)>~=~<\vec{g}'(12)~|~\vec{g}(IJ)>\]
   \item[(ii)]They are rotationally invariant, i.e. independent of the magnetic quantum numbers $m_k$
              occurring in (\ref{tensor basis n angular momenta}).
\end{itemize}


\pagebreak

\section{Properties of the $6j$-Symbols \label{6 j}}
In this section we will give an overview on the $6j$-symbols because they are the basic structure we will
use in our recoupling calculations, every coupling of $n$ angular momenta can be decomposed into 
them. 
For further details we refer to \cite{Edmonds}, \cite{Varshalovich}.

\subsection{Definition}
   The $6j$-symbol is defined as \cite{Edmonds},p 92:
   \ba 
   \left\{ \begin{array}{ccc} \label{definition 6j symbol}
   j_1 & j_2 & j_{12}\\
   j_3 & j   & j_{23}
   \end{array} \right\} 
   &:=&[(2j_{12}+1)(2j_{23}+1)]^{-\frac{1}{2}}(-1)^{j_1+j_2+j_3+j}
   \nonumber\\
   &&\times <j_{12}(j_1,j_2),j(j_{12},j_3)|j_{23}(j_2,j_3),j(j_1,j_{23})>
   \nonumber\\
   & = &[(2j_{12}+1)(2j_{23}+1)]^{-\frac{1}{2}}(-1)^{j_1+j_2+j_3+j}
   \nonumber\\
   &&\times \sum_{m_1~m_2}<j_1~m_1;j_2~m_2|j_1~j_2~j_{12}~m_1+m_2> \nonumber\\
   &&\times <j_{12}~m_1+m_2;j_3~m-m_1-m_2|j_{12}~j_3~j~m> \nonumber\\
   &&\times <j_2~m_2;j_3~m-m_1-m_2|j_2~j_3~j_{23}~m-m_1> \nonumber\\
   &&\times <j_1~m_1;j_{23}~m-m_1|j_1~j_{23}~j~m> 
   \ea
   The terms under the summation are called Clebsh-Gordon-Coefficients.

\subsection{Explicit Evaluation of the $6j$-Symbols}

   A general formula for the numerical value of th3 $6j$-symbols has been derived by Racah
   \cite{Racah}, \cite{Edmonds}, p.99:
   \begin{samepage}
      \ba \label{A1}
      &&\left\{ \begin{array}{ccc}
      j_1 & j_2 & j_{12}\\
      j_3 & j   & j_{23}
      \end{array} \right\}
      =\Delta(j_1,j_2,j_{12})\Delta(j_1,j,j_{23})\Delta(j_3,j_2,j_{23})
      \Delta(j_3,j,,j_{12})w
      \left\{ \begin{array}{ccc}
      j_1 & j_2 & j_{12}\\
      j_3 & j   & j_{23}
      \end{array} \right\}, \nonumber\\
      &&
      \Delta(a,b,c)=\sqrt{\frac{(a+b-c)!(a-b+c)!(-a+b+c)!}{(a+b+c+1)!}}
      \nonumber\\
      && 
      w
      \left\{ \begin{array}{ccc}
      j_1 & j_2 & j_{12}\\
      j_3 & j   & j_{23}
      \end{array} \right\}
      =\sum_n (-1)^n (n+1)! \times \nonumber\\
      &&\times [(n-j_1-j_2-j_{12})!(n-j_1-j-j_{23})!(n-j_3-j_2-j_{23})!
      (n-j_3-j-j_{12})!]^{-1}\times
      \nonumber\\
      &&\times [(j_1+j_2+j_3+j-n)!(j_2+j_{12}+j+j_{23}-n)!
      (j_{12}+j_1+j_{23}+j_3-n)!]^{-1} 
      \ea
      The sum has to be extended over all positive integer values of n such that no factorial in
      the denominator has a negative argument. That is:\\
      $\max[j_1+j_2+j_{12},j_1+j+j_{23},j_3+j_2+j_{23},j_3+j+j_{12}] \le n \le
       \min[j_1+j_2+j_3+j,j_2+j_{12}+j+j_{23},j_{12}+j_1+j_{23}+j_3]$
   \end{samepage}	

   \begin{samepage}
      \begin{description} \label{integer conditions}
	\item[Remark]From (\ref{A1}) we are provided with some additional requirements, the
	arguments of the $6j$-symbols have to fulfill: Certain sums or differences of them
	have to be integer for beeing proper($\equiv$integer) arguments for the factorials:\\
	\\
        from $\Delta(a,b,c)$ one gets:
	  \begin{itemize}
	     \item a,b,c have to fulfill the triangle inequalities:
		$(a+b-c)\ge 0$,~ 
		$(a-b+c)\ge 0$,~ 
		$(-a+b+c) \ge 0$,
	     \item $(\pm a \pm b \pm c)$ has to be an integer number
	  \end{itemize}
	 from the w-coefficient one gets:
	  \begin{itemize}
	     \item  $j_1+j_2+j_3+j,~j_2+j_{12}+j+j_{23},~j_{12}+j_1+j_{23}+j_3$ ~are 
	      integer numbers.
	  \end{itemize} 
      \end{description}
      The following (trivial but important) relations are frequently used in calculations involving 
      $6j$-symbols: 
      \ba \label{integer exponents}
          (-1)^{z} & = & (-1)^{-z} ~~~\forall z \in \mathbb{Z} \nonumber\\
	  (-1)^{2z} & = & 1 ~~~~~~~~~~~\forall z \in \mathbb{Z} \nonumber\\
          (-1)^{3k} & = & (-1)^{-k}~~~\forall k=\frac{z}{2} ~~~with ~z \in \mathbb{Z}
      \ea
   \end{samepage}

\subsection{Symmetry Properties} 
    The $6j$-symbols are invariant 
    \begin{itemize}
      \item under any permutation of the columns:
      \ba \label{symmetry1}
      \lefteqn{\left\{ \begin{array}{ccc}
      j_1 & j_2 & j_3\\
      j_4 & j_5 & j_6
      \end{array} \right\} 
      =
      \left\{ \begin{array}{ccc}
      j_2 & j_3 & j_1\\
      j_5 & j_6 & j_4
      \end{array} \right\} 
      =
      \left\{ \begin{array}{ccc}
      j_3 & j_1 & j_2\\
      j_6 & j_4 & j_5
      \end{array} \right\} =} \nonumber\\
      & &
      =
      \left\{ \begin{array}{ccc}
      j_2 & j_1 & j_3\\
      j_5 & j_4 & j_6
      \end{array} \right\} 
      =
      \left\{ \begin{array}{ccc}
      j_1 & j_3 & j_2\\
      j_4 & j_6 & j_5
      \end{array} \right\} 
      =
      \left\{ \begin{array}{ccc}
      j_3 & j_2 & j_1\\
      j_6 & j_5 & j_4
      \end{array} \right\} 
      \ea
    \item under interchange of the upper and lower arguments of two columns at the same time.
          E.g.
      \begin{equation} \label{symmetry2}
        \left\{ \begin{array}{ccc}
	j_1 & j_2 & j_3\\
	j_4 & j_5 & j_6
	\end{array} \right\} 
	=
	\left\{ \begin{array}{ccc}
	j_1 & j_5 & j_6\\
	j_4 & j_2 & j_3
	\end{array} \right\} 	  
      \end{equation}
    \end{itemize}

 \subsection{Orthogonality and Sum Rules}
    \begin{description}
       \item[Orthogonality Relations]
	  \ba \label{orthogonality relation}
	     \sum_{j_{23}}(2j_{12}+1)(2j_{12}^{'}+1)	
	        \left\{ \begin{array}{ccc}
		j_1 & j_2 & j_{12}\\
		j_3 & j & j_{23}
		\end{array} \right\}
  		\left\{ \begin{array}{ccc}
		j_1 & j_2 & j_{12}^{'} \\
		j_3 & j & j_{23}
		\end{array} \right\} 
		& = &
		\delta_{j_{12} j_{12}^{'}}
 	  \ea
       \item[Composition Relation]
	  \ba
	     \sum_{j_{23}}(-1)^{j_{23}+j_{31}+j_{12}}(2j_{23}+1)	
	        \left\{ \begin{array}{ccc}
		j_1 & j_2 & j_{12}\\
		j_3 & j & j_{23}
		\end{array} \right\}
  		\left\{ \begin{array}{ccc}
		j_2 & j_3 & j_{23} \\
		j_1 & j & j_{31}
		\end{array} \right\} 
		& = &
  		\left\{ \begin{array}{ccc}
		j_3 & j_1 & j_{31} \\
		j_2 & j & j_{12}
		\end{array} \right\} 
	  \ea
       \item[Sum Rule of Elliot and Biedenharn]
	   \ba \label{EBI}
	     \lefteqn{	
	        \left\{ \begin{array}{ccc}
		j_1 & j_2 & j_{12}\\
		j_3 & j_{123} & j_{23}
		\end{array} \right\}
  		\left\{ \begin{array}{ccc}
		j_{23} & j_1 & j_{123} \\
		j_4 & j & j_{14}
		\end{array} \right\} =} \nonumber\\
		& &
		=
	        (-1)^{j_1+j_2+j_3+j_4+j_{12}+j_{23}+j_{14}+j_{123}+j}
	        \nonumber\\
	        &&\times \sum_{j_{124}}
	        (-1)^{j_{124}}~(2j_{124}+1)	
	        \left\{ \begin{array}{ccc}
		j_3 & j_2 & j_{23}\\
		j_{14} & j & j_{124}
		\end{array} \right\}
  		\left\{ \begin{array}{ccc}
		j_2 & j_1 & j_{12} \\
		j_4 & j_{124} & j_{14}
		\end{array} \right\} 
		\left\{ \begin{array}{ccc}
		j_3 & j_{12} & j_{123} \\
		j_4 & j & j_{124}
		\end{array} \right\} 
	      \ea   
    \end{description}

\pagebreak
\section{Comment on the Smallest Non-Vanishing Eigenvalue \label{diskussion 4-vertex}}

In this section we will briefly summarize, what can be done to obtain a lower bound of the spectrum of the
matrices occurring when expressing the Volume Operator on a recoupling scheme basis at the gauge invariant
4-vertex. This is mainly done to illustrate the remarkable symmetries 
in that case.
The idea is to obtain a lower bound of the non-zero-eigenvalues by applying theorem (\ref{Gersgorin discs}), 
on the inverse matrix, giving an upper bound for its eigenvalues and therefore a lower bound for the non-zero
volume spectrum.\\

The general form for the gauge-invariant 4-vertex was obtained in (\ref{general matrix 4 vertex})
\setlength{\arraycolsep}{1.2mm}
\be
  A=\left( \begin{array}{cccccc}
             0      & -a_1    & 0      & \cdots  & \cdots  & 0         \\
	     a_1    &  0      & -a_2   &         &         & \vdots    \\
	     0      & a_2     &  0     & \ddots  &         & \vdots    \\
	     \vdots &         & \ddots & \ddots  & \ddots  & \vdots    \\
	     \vdots &         &        & \ddots  & \ddots  & -a_{n-1}  \\
	     0      & \cdots  & \cdots & \cdots  & a_{n-1} & 0 
           \end{array} \right)
\ee
\setlength{\arraycolsep}{0.7mm}
We have explicitly discussed the 0-eigenvalues contained in the spectrum of $A$ in section \ref{Eigenvektor fuer 0}.
We know, that only in the odd-dimensional case the matrix $A$ is singular containing one 0-eigenvalue with
the according eigenvector
\be
  \lo{(n)}\Psi:=x\cdot \left[
  1~,0~,\frac{a_1}{a_2}~,0~,\frac{a_1a_3}{a_2a_4}~,0,~\ldots~,
  \frac{a_1a_3\cdot\ldots\cdot a_{n-2}}{a_2a_4\cdot\ldots\cdot a_{n-1}}\right]
\ee
where $n:=\dim A$ and $x$ an arbitrary scaling factor. We will denote the k$^{th}$ element of $\lo{(n)}\Psi$ by
$\lo{(n)}\Psi_k$. 
For technical reasons we will set $x=\frac{y}{\lo{(n)}\Psi_n}$ in the following to obtain:
\be
  \lo{(n)}\Xi:=y\cdot \left[\frac{a_2a_4\cdot\ldots\cdot a_{n-1}}{a_1a_3\cdot\ldots\cdot a_{n-2}}~,0~,
  \ldots~,\frac{a_2a_4}{a_1a_3}~,0~,\frac{a_2}{a_1}~,0~,1\right]
\ee

\subsection{Even Dimension $n$ of $A$}
    Since $A$ is regular in that case we can invert it to find:
    \setlength{\arraycolsep}{1.2mm}
    \be \label{inverse matrix gerade dimension}
     A^{-1}=\left( \begin{array}{ccccccccccc}
        	0 & \frac{\Xi_{n-1}}{a_1} & 0& \frac{\Xi_{n-3}}{a_3}&0 &\frac{\Xi_{n-5}}{a_5}&0&\cdots&\cdots
		&\frac{\Xi_1}{a_{n-1}}\\
		
		-\frac{\Xi_{n-1}}{a_1}& 0 & 0 & 0 & 0&0&0&\cdots&\cdots& 0    \\
		
		0&0&0&\frac{\Xi_{n-1}}{a_3}&0&\frac{\Xi_{n-3}}{a_5}&0&\cdots&\cdots&\frac{\Xi_3}{a_{n-1}}\\
		
		-\frac{\Xi_{n-3}}{a_3}      & 0     &  -\frac{\Xi_{n-1}}{a_3}&0&0&0  &0 & \cdots &\cdots& 0\\
		0 &    0     & 0 & 0 &0 &  \frac{\Xi_{n-1}}{a_5} & 0&\cdots&\cdots&\frac{\Xi_5}{a_{n-1}}    \\
		-\frac{\Xi_{n-5}}{a_5}&0&-\frac{\Xi_{n-3}}{a_5}&0&-\frac{\Xi_{n-1}}{a_5}&0&0&\cdots&\cdots&0\\
		0&0&0&0&0&0&0&\ddots&&\vdots \\
		\vdots & \vdots &\vdots&\vdots&\vdots&\vdots& \ddots&\ddots&\ddots&\vdots  \\
		\vdots & \vdots &\vdots&\vdots&\vdots&\vdots& &\ddots&0&\frac{\Xi_{n-1}}{a_{n-1}}  \\

	-\frac{\Xi_1}{a_{n-1}} &0 &-\frac{\Xi_3}{a_{n-1}}&0& -\frac{\Xi_5}{a_{n-1}}&0 &\cdots &\cdots
	&-\frac{\Xi_{n-1}}{a_{n-1}}& 0\\ 
              \end{array} \right)
    \ee
    \setlength{\arraycolsep}{0.7mm}
    where we used the  components $\lo{(n-1)}\Xi_k:=\Xi_k$ of the 0-eigenvector $\lo{(n-1)}\Xi$ of the $(n-1)$-odd
    dimensional case with scaling factor $y=1$.

\subsection{Odd Dimension of $A$}
    Since $A$ is not regular we have to project out its nullspace $\lo{(n)}\Xi$ first (with arbitrary
    prefactor $y$). That can be done by
    applying a similarity transformation $W$ on A to obtain $R:=W^{-1}AW$:
    \setlength{\arraycolsep}{1.2mm}
    \be
      \begin{array}{ccc}
       W=\left( \begin{array}{cccccccc}
        	 1  & 0 & 0  & 0  &\cdots  & \cdots  & 0&\Xi_1\\
		 0  & 1 & 0  & 0  &\cdots  & \cdots  & 0&\Xi_2
		     \\
		 0      & 0  &  1 & 0 &  &         & \vdots&\vdots    \\
		 0      & 0  &  0     & 1
		 &\  &         & \vdots &\vdots   \\
		 \vdots &         & & &\ddots  &  & \vdots &\vdots   \\
		 \vdots &         &        & &  &  & 1&\Xi_{n-1} \\
		 0      & \cdots  & \cdots & \cdots&\cdots  & \cdots &0&\Xi_n 
               \end{array} \right) 
	       
	       &~~~~&
    W^{-1}=\left( \begin{array}{cccccccc}
        	 1  & 0 & 0  & 0  &\cdots  & \cdots  & 0&-\frac{\Xi_1}{\Xi_n}\\
		 0  & 1 & 0  & 0  &\cdots  & \cdots  & 0&-\frac{\Xi_2}{\Xi_n}
		     \\
		 0      & 0  &  1 & 0 &  &         & \vdots&\vdots    \\
		 0      & 0  &  0     & 1
		 &\  &         & \vdots &\vdots   \\
		 \vdots &         & & &\ddots  &  & \vdots &\vdots   \\
		 \vdots &         &        & &  &  & 1&-\frac{\Xi_{n-1}}{\Xi_n} \\
		 0      & \cdots  & \cdots & \cdots&\cdots  & \cdots &0&\frac{1}{\Xi_n} 
               \end{array} \right)

       \end{array}
    \ee
    \setlength{\arraycolsep}{0.7mm}    
Now one can check, that (all even components of $\lo{(n)}\Xi$ vanish, again $y=1$):

   \setlength{\arraycolsep}{1.2mm}
   \be
     R:=W^{-1}AW=\left( \begin{array}{ccccccc|c}
        	 0  & -a_1 & 0  & 0  &\cdots  & \cdots  & -\frac{\Xi_1}{\Xi_n}a_{n-1}&0\\
		 a_1  & 0 & -a_2  & 0  &\cdots  & \cdots  & 0&0
		     \\
		 0      & a_2  &  0 & -a_3 &\cdots  &\cdots & -\frac{\Xi_3}{\Xi_n}a_{n-1}&0    \\
		 0      & 0  &  a_3     & 0& \ddots &        & \vdots &\vdots   \\
		 \vdots &  &&\ddots &\ddots  & \ddots & \vdots &\vdots   \\
		 \vdots &         & & &\ddots  & \ddots & -a_{n-2}-\frac{\Xi_{n-1}}{\Xi_n}a_{n-1}&0    \\
		 \vdots &         &        & &  &a_{n-2}  & 0&0\\ \hline
		 0      & \cdots  & \cdots & \cdots&\cdots  & 0&\frac{a_{n-1}}{\Xi_n}&0 
               \end{array} \right) 
   \ee
   \setlength{\arraycolsep}{0.7mm}
To discuss the non-zero-spectrum of $A$ it is sufficient to discuss the now regular submatrix $\tilde{R}$ 
being the submatrix of $R$ with n$^{th}$ row and column deleted. One obatins:

    \setlength{\arraycolsep}{1.2mm}
    \be
     \tilde{R}^{-1}=\left( \begin{array}{ccccccccccc}
        	0 & \frac{\Xi_{n-1}}{a_1} & 0& \frac{\Xi_{n-3}}{a_3}&0 &\frac{\Xi_{n-5}}{a_5}&0&\cdots&\cdots
		&\frac{\Xi_1}{a_{n-1}}\\
	        M_{21}& 0 & M_{23} & 0 & M_{25}&0&M_{27}&\cdots&M_{2~n-1}& 0    \\
		0&0&0&\frac{\Xi_{n-1}}{a_3}&0&\frac{\Xi_{n-3}}{a_5}&0&\cdots&\cdots&\frac{\Xi_3}{a_{n-1}}\\
		
	        M_{41}& 0 & M_{43} & 0 & M_{45}&0&M_{47}&\cdots&M_{4~n-1}& 0    \\
		0 &    0     & 0 & 0 &0 &  \frac{\Xi_{n-1}}{a_5} & 0&\cdots&\cdots&\frac{\Xi_5}{a_{n-1}}    \\
	        M_{61}& 0 & M_{63} & 0 & M_{65}&0&M_{67}&\cdots&M_{6~n-1}& 0    \\
		0&0&0&0&0&0&0&\ddots&&\vdots \\
		\vdots & \vdots &\vdots&\vdots&\vdots&\vdots& \ddots&\ddots&\ddots&\vdots  \\
		\vdots & \vdots &\vdots&\vdots&\vdots&\vdots& &\ddots&0&\frac{\Xi_{n-1}}{a_{n-1}}  \\

	M_{n-1~1} &0 &M_{n-1~3}&0& M_{n-1~5}&0 &\cdots &\cdots&M_{n-1~n-2}& 0\\ 
              \end{array} \right)
    \ee
    \setlength{\arraycolsep}{0.7mm}

Here $M_{ij}=\frac{\det{\tilde{R}_{(ij)}}}{\det{\tilde{R}}}$ and $\tilde{R}_{(ij)}$
is a shortcut for the submatrix of $\tilde{R}$ one obtains by deleting row $i$ and column $j$. Basically
this is the definition for the matrix element of the inverse matrix. Since unfortunately the $M_{ij}$ are
hard to control (but of order $\frac{1}{a}$) we are unable to give an 
explicit upper bound for the spectrum
of $\tilde{R}^{-1}$ and therefore a lower bound on the spectrum of $A$ according to 
 theorem (\ref{Gersgorin discs}). It is remarkable, however, that half of 
the sructure of 
(\ref{inverse matrix gerade dimension}) is being reproduced by the odd 
dimensional case.
  
\end{appendix}
%

%
%
%
%


\pagebreak



\addcontentsline{toc}{chapter}{Bibliography}
\begin{thebibliography}{99}
%
%
%

\bibitem{1} C.\ Rovelli, L.\ Smolin, ``Discreteness of volume and
area in quantum gravity'' Nucl. Phys. B {\bf 442} (1995) 593, Erratum :
Nucl. Phys. B {\bf 456} (1995) 734

\bibitem{Geometry II}
A.~Ashtekar,~J.~Lewandowski\\  ``Quantum Theory of Gravity II: Volume Operators'',\\\
[arXiv:~gr-qc/9711031]


\bibitem{0}
C. Rovelli,\\ 
``Quantum Gravity'', Cambridge University Press 2004, at press.\\
``Loop Quantum Gravity", Living Rev. Rel. {\bf 1} (1998) 1,\\\
[gr-qc/9710008]\\
T.~Thiemann,\\
``Modern Canonical Quantum General Relativiy'',\\\
to appear in Cambridge University Press,\\\ [arXiv:~gr-qc/0110034].\\
``Lectures on Loop Quantum Gravity'', Lecture Notes in
Physics, {\bf 631} (2003) 41 -- 135,\\\ 
[gr-qc/0210094]\\
A. Ashtekar, J. Lewandowski,\\ 
``Background Independent Quantum Gravity: A Status Report'',\\\
[gr-qc/0404018]


\bibitem{TT:QSD I}
T.~Thiemann,\\
``Anomaly-free Formulation of non-perturbative,
four-dimensional Lorentzian Quantum Gravity", Physics Letters {\bf B380}
(1996) 257-264.\\\ 
[gr-qc/9606088]

\bibitem{TT:QSD II}
T.~Thiemann,\\
``Quantum Spin Dynamics (QSD)",
Class. Quantum Grav. {\bf 15} (1998) 839-73, [gr-qc/9606089];
``II. The Kernel of the Wheeler-DeWitt Constraint Operator",
Class. Quantum Grav. {\bf 15} (1998) 875-905, [gr-qc/9606090];
``III.
Quantum Constraint Algebra and Physical Scalar Product in Quantum General
Relativity", Class. Quantum Grav. {\bf 15} (1998) 1207-1247,
[gr-qc/9705017];
``IV. 2+1 Euclidean Quantum Gravity as a model to test 3+1
Lorentzian Quantum Gravity", Class. Quantum Grav. {\bf 15} (1998)
1249-1280, [gr-qc/9705018];
``V. Quantum Gravity as the Natural Regulator of the Hamiltonian
Constraint
of Matter Quantum Field Theories",
Class. Quantum Grav. {\bf 15} (1998) 1281-1314, [gr-qc/9705019];
``VI. Quantum Poincar\'e Algebra and a Quantum Positivity of Energy
Theorem for Canonical Quantum Gravity",
Class. Quantum Grav. {\bf 15} (1998) 1463-1485, [gr-qc/9705020]

\bibitem{2}
T.~Thiemann,\\
``The Phoenix Project: Master Constraint Programme for Loop Quantum 
Gravity'',\\\
[gr-qc/0405080]

\bibitem{3} R. Loll, Phys. Rev. Lett. {\bf 75} (1995) 3048

\bibitem{dePietri}
R. De Pietri,\\ ``Spin Networks and Recoupling in Loop Quantum Gravity'', 
Nucl. Phys. Proc. Suppl. {\bf 57} (1997) 251\\\
[arXiv:~gr-qc/9701041]

\bibitem{dePietriII}
R. De Pietri,~C. Rovelli,\\
``Geometry Eigenvalues and Scalar Product from 
Recoupling Theory in Loop Quantum
Gravity'', Phys.Rev. {\bf D54} (1996) 2664\\\
[arXiv:~gr-qc/9602023 v2]

\bibitem{TT:vopelm}
T.~Thiemann,\\
``Closed formula for the matrix elements of the volume operator in
canonical quantum gravity'', J. Math. Phys. {\bf 39} (1998) 3347-3371\\\
[arXiv:~gr-qc/9606091].

\bibitem{4}
J.~Brunnemann,\\
``Spectral Analysis of the Volume Operator in Canonical Quantum General 
Relativity'', Diploma Thesis, Humboldt Universit\"at zu Berlin,
December 2002

\bibitem{4a} M. Bojowald, H. A. Morales -- Tecotl,\\
``Cosmological Applications of Loop Quantum Gravity'',\\\
[gr-qc/0306008]

\bibitem{5}
A.~Perez,\\
``Spin Foam Models for Quantum Gravity'', Class. Quant. Grav. 
{\bf 20} (2003) R43\\\
[gr-qc/0301113]

\bibitem{Edmonds}
A.~R.~Edmonds\\
``Angular Momentum in Quantum Mechanics,''
Princeton University Press, Fourth printing 1996

\bibitem{Varshalovich}
D.~A.~Varshalovich, A.~N.~Moskalev, V.~K.~Khersonskii,~\\
``Quantum Theory of Angular Momentum'',~~
World Scientific, 1988

\bibitem{Sexl Urbandtke}
R.U.~Sexl~H.K.~Urbandtke, \\``Relativit\"at, Gruppen, Teilchen'', Springer Verlag Wien/New York 1982,
Zweite, erweiterte Auflage

\bibitem{Carroll}
S.M.~Carroll, \\``Lecture Notes on General Relativity'', \\\ [arXiv:~gr-qc/9712019]

\bibitem{Wald}
Robert~M.~Wald,~\\
``General Relativity'',
The University of Chicago Press,1984

\bibitem{Racah}
G.~Racah, ``Theory of Complex Spectra. II'',\\ Phys. Rev.62 (1942), published in \cite{Biedenharn/van Dam}

\bibitem{Biedenharn/van Dam}
L.C.~Biedenharn~H. van Dam, \\``Quantum Theory of Angular Momentum'' (a collection of reprints and original
papers edited by the authors), Academic Press New York and London 1965

\bibitem{Horn}
Alfred~Horn~\\
``Eigenvalues of sums of Hermitian matrices'',
Pacific J. Math.12 (1962),225-241

\bibitem{Fulton}
William~Fulton~\\
``Eigenvalues, invariant factors, highest weights and Schubert
calculus'',\\\
[arXiv:~math.AG/9908012]

\bibitem{Holz}
D.E. Holz~H. Orland~A. Zee,\\ ``On the Remarkable Spectrum of a Non-Hermitean Random Matrix Model'', \\\
[arXiv:math-ph/0204015]



\bibitem{Seifert}
M.~Seifert,\\ ``Angle and Volume Studies in Quantized Space'',\\\
 [arXiv:~gr-qc/0108047]

\bibitem{Dyson}
F.J.~Dyson, \\``The Dynamics of a Disordered Linear Chain'', Phys.~Rev. 92, 1331~(1953)


\bibitem{Gantmacher}
F.~R.~Gantmacher\\
``Matrizentheorie'',
VEB Deutscher Verlag der Wissenschaften, Berlin 1986

\bibitem{Zurmuehl}
R.~Zurm\"uhl,\\ ``Matrizen'', Springer Verlag Berlin/G\"ottingen/Heidelberg~1964,~4.Auflage

\bibitem{Marcus Minc}
M.~Marcus~H.~Minc,\\ ``A Survey of Matrix Theory and Matrix Inequalities'', Dover Publications, Inc. New
York 1992 
 
 
\bibitem{HU-script}
H.~Grassmann\\
``Algebra und Geometrie - Vorlesungsscript''\\\
[http://www.irm-mathematik.hu-berlin.de/$\sim$hgrass]


\bibitem{Dirac constrained system}
P.A.M.~Dirac, ``Lectures on Quantum Mechanics'',\\ 
Academic Press INC. New York/London, Second printing 1967

\bibitem{Nolting}
W.~Nolting,\\``Grundkurs Theoretische Physik: 5 Quantenmechanik,~Teil 2: Methoden und Anwendungen'', 
Friedrich Vieweg \& Sohn Verlgsgesellschaft mbH, Braunschweig/Wiesbaden, 1997

\bibitem{Bogoljubov}
N.N.~Bogoljubov,~D.V.~Sirkov, \\``Quantenfelder'', Physikverlag Weinheim / VEB Deutscher Verlag der
Wissenschaften, Berlin 1984


\bibitem{Wintner}
A.~Wintner,\\  ``Spektraltheorie der unendlichen Matrizen'', Verlag von S. Hirzel in Leipzig 1929




\bibitem{TTM}
Teubner-Taschenbuch der Mathematik, Bd. 1+2, B.G. Teubner
Stuttgart, Leipzig 1996

\bibitem{6}
A. Ashtekar, J. Lewandowski\\
``Relation between Polymer and Fock Excitations'', 
Class. Quant. Grav. {\bf 18} (2001) L117-L128\\\
[gr-qc/0107043]\\
T. Thiemann\\
``Complexifier Coherent States for Quantum General Relativity'',
[gr-qc/0206037]

\bibitem{7}
J. Brunnemann, T. Thiemann,\\ 
``Towards the Cosmological Sector of Loop Quantum Gravity'', in preparation

\end{thebibliography}
\end{document}